\documentclass[twoside,12pt]{article}
\usepackage{epsfig}
\usepackage{graphicx}
\usepackage{epstopdf}
\usepackage{amssymb,amsmath}
\usepackage[super,compress]{cite}
\usepackage{wrapfig}
\usepackage{float}

\usepackage{xcolor}
\definecolor{lcolor}{rgb}{0.,0.0,0.}
\definecolor{citcolor}{rgb}{0,0.,0.5}
\usepackage[breaklinks,colorlinks,urlcolor=blue,citecolor=blue,linkcolor=blue]{hyperref}
\usepackage{mciteplus}

\newcommand{\be}{\begin{equation}}
\newcommand{\ee}{\end{equation}}
\newcommand{\bea}{\begin{eqnarray}}
\newcommand{\eea}{\end{eqnarray}}

\newcommand{\chisq}{\chi^2/\mathrm{d.o.f.}}

\def\eq#1{{Eq.~(\ref{#1})}}
\def\fig#1{{Fig.~\ref{#1}}}

\begin{document}

\title{ \vspace{1cm} Gluon saturation and initial conditions \\ for relativistic heavy ion collisions}
\author{J.~L.\ Albacete,$^{1}$ C.\ Marquet,$^2$ \\
\\
$^1$CAFPE and Departamento de F\'{\i}sica Te\'orica y del Cosmos,\\ Universidad de Granada, E-18071 Granada, Spain\\
$^2$Centre de Physique Th\'eorique, \'Ecole Polytechnique,\\CNRS, 91128 Palaiseau, France}
\maketitle
\begin{abstract} 
We present an overview of theoretical aspects of the phenomenon of gluon saturation in high energy scattering in Quantum Chromo Dynamics. Then we review the state-of-the-art of saturation-based phenomenological approaches to the study and characterisation of the initial state of ultra-relativistic heavy ion collisions performed at RHIC and the LHC. Our review focuses mostly in the Color Glass Condensate effective theory, although we shall also discuss other approaches in parallel. 
\end{abstract}
\tableofcontents
\section{Introduction}

The main goal of the experimental programs on ultra-relativistic heavy-ion collisions (HIC) at Brookhaven's relativistic heavy-ion collider (RHIC) and at CERN's large hadron collider (LHC) is to study QCD matter under extreme conditions. As the RHIC experiments were announcing what is now considered the prime ingredient of RHIC's legacy -- the discovery of a new state of matter of extremely high density and temperature called the Quark Gluon Plasma (QGP) -- a part of the community was studying the QCD description of the initial nuclei themselves, and what it implied for hadronic scattering in high-energy limit. At that time, the importance of a detailed understanding of the physics of the initial state was already recognized, but overall, perhaps under-appreciated. Rather, signals related to the expected thermal properties of the QGP were considered as the most promising observables for discovery. 

As in any other hadronic collision process, a precise understanding of the partonic composition of the projectile nuclei at all the relevant scales is indeed crucial. A typical lead-lead collision at the LHC gives rise approximately 10000 detected particles in the final state. Further $\sim 99\%$ of such particles carry a relatively small transverse momentum compared to the centre of mass collision energy, 2.76 TeV at the LHC, $p_t\lesssim 1\div2$ GeV. This simple observation, together with a ballpark analyses of the collision kinematics, leads immediately to the conclusion that most of the produced particles in a high-energy heavy ion collision are originated from partons (mostly gluons) that carry a small fraction $x$ of the light-cone (equivalently, longitudinal) momentum of their parent nucleon. Typical values for RHIC and LHC energies at central rapidities are $x\sim 10^{-2}$ and $x\sim 10^{-4}$, although even smaller values are reached at more forward rapidities\footnote{This values arise from simple $2\to 1$ kinematics neglecting fragmentation effects: $ x\sim m_{t}e^{-y}/\sqrt{s}$, with $\sqrt{s}$ the nucleon-nucleon centre of mass energy, $m_{t}$ and $y$ the transverse mass and rapidity of the produced particle.}.

Therefore, a detailed understanding of the --abundant-- small-$x$ gluons in the wave function of the colliding nuclei is mandatory for a proper characterization of the QCD medium produced in HIC.  As the goal had become to turn the physics of the QGP into a quantitative science, it was realized that bulk observables in HIC were as much sensitive to features of the initial state as they are to properties of the quark-gluon plasma itself, making it impossible to precisely extract the medium transport coefficients, such as the viscosity over entropy density ratio, without a proper quantitative understanding of the physics of the initial state. This observation was confirmed by the first lead-lead results from the LHC.

Most recently, the results of proton-lead collisions at the LHC, which were thought of by many as a mere cold-matter control experiment, have instead put into question our interpretation of RHIC and LHC results on heavy-ion collisions: high-multiplicity proton-nucleus collisions seem to produce the same QGP signals as nucleus-nucleus do, without any sign of the systematic effects the smaller system size would imply. This has brought even more importance into the physics of the initial state as it is believed that the solution to this puzzle lays in the understanding of high-multiplicity proton-nucleus (and proton-proton) collisions.

As a matter of fact, one of the major challenges since the discovery of QCD has been to understand hadronic Fock-space wave functions in the high-energy
limit. It was realized long ago that, due to the soft singularity 
 of the splitting function of non-Abelian gauge bosons, hadronic wave
functions would contain a large number of gluons with a small
(light-cone) momentum fraction $x$~\cite{Gribov:1984tu}. This
expectation was confirmed by $e+p$ deep-inelastic scattering
experiments at HERA (see below) and is accounted for by modern PDF parameterizations. 
At the same time it also became clear that such fast growth of the gluon densities in hadron towards small-$x$ could not continue indefinitely or, otherwise, unitarity of the theory would be violated. 
Rather, it was proposed that at sufficiently small-$x$, a novel non-linear dynamics of the soft color fields would emerge, taming the non-abelian avalanche of soft gluons towards the small-$x$ domain and restoring unitarity.  This new phenomenon is commonly referred to as {\it saturation} of gluon distributions at low $x$ and manifests itself in non-linear, density dependent terms in the QCD
evolution equations.

The onset of gluon saturation effects is controlled by a dynamically generated transverse momentum scale, the saturation scale $Q_s(x)$.
Although saturation effects are expected to be an intrinsic feature of any hadronic wave function at sufficiently high energies, the saturation momentum $Q_s(x)$ is boosted further in an ultra-relativistic heavy nuclei due to the superposition of the gluon fields of its constituent nucleons induced by Lorentz contraction in the longitudinal direction. In other words,  the density of large-$x$ ``valence'' charges per unit
transverse area increases in proportion to the thickness $\sim
A^{1/3}$ of a nucleus~\cite{McLerran:1993ni,McLerran:1993ka,Kovchegov:1996ty}. These
represent the sources for the small-$x$ soft gluon fields. Thus,
non-linear color field dynamics should be operative in nuclear wave functions at higher transverse momentum scales than for nucleons.

While the need for non-linear corrections to QCD evolution equations in the high-energy regime is well established from a theoretical point of view, the relevant question for present analyses of HIC data is to what extent such corrections are present at available collision energies and, if so, what is the best theoretical framework for their description.
Concerning the first of these two questions, one important lesson learned from experimental data collected in Au+Au and Pb+Pb collisions at RHIC and the LHC, respectively, is that bulk
particle production in ion-ion collisions is very different from a
simple superposition of nucleon-nucleon collisions. This is evident,
for instance, from the measured charged particle multiplicities which
exhibit a strong deviation from scaling with the number of independent
nucleon-nucleon collisions: $\frac{dN^{AA}}{d\eta}(\eta=0)\ll N_{coll}
\frac{dN^{NN}}{d\eta}(\eta=0)$. One is led to the conclusion that
strong coherence effects among the constituent nucleons, or the
relevant degrees of freedom at the sub-nucleon level, must be present
during the collision process. In a QCD description, coherence effects can be related to the presence
of large gluon densities both prior to the collision and during the
collision process itself\footnote{We note that the diagrammatical interpretation of the collision process is generally gauge dependent. Therefore only  gauge invariant quantities --and not other subset of diagrams-- can be properly identified to happen before or after the collision.} and, hence, to the phenomenon of gluon saturation.

Although different approaches have been developed to study initial state effects in HIC, the main focus of this review shall be on the Color Glass Condensate (CGC) effective theory of QCD for high-energy QCD scattering. The CGC has emerged as the best candidate to approximate QCD in the saturation regime, both in terms of practical applicability and of phenomenological success. The CGC is based on three main physical ingredients. First, high gluon densities correspond to strong classical fields, which permit ab-initio first principles
calculation of hadronic and nuclear wave functions at small-$x$ through classical techniques. Next, quantum corrections are incorporated
via non-linear renormalization group equations such as the B-JIMWLK hierarchy\footnote{These acronym refers to the authors I.~Balitsky, J.~Jalilian-Marian,  E.~Iancu, L.~McLerran, H.~Weigert, A.~Leonidov, and A.~Kovner, rearranged in a {\it gym walk} by A.~H.~Mueller.} or, in the large-$N_c$ limit, the Balitsky-Kovchegov (BK) equation that describe the evolution of the hadron wave function towards small-$x$. The non-linear, density dependent
terms in the CGC evolution equations are ultimately related to unitarity of the theory and, in the appropriate frame and gauge, can be interpreted as due to gluon recombination processes that tame or saturate the growth of gluon
densities for modes with transverse momenta below the saturation scale $Q_s(x)$. Finally, the presence of strong color fields $\mathcal{A}\sim 1/g$ leads to breakdown of standard perturbative techniques whose description of particle production processes is based on a series expansion in powers of the strong coupling $g$ and in inverse power of a necessarily hard scale $Q^2$. Terms of order $g\mathcal{A}\sim \mathcal{O}(1)$ or $Q_s^2/Q^2$ --often referred to in the literature as {\it higher twists}-- need to be resummed to all orders.

The practical importance of initial state studies are many.  In the soft sector, say particles with small transverse
momentum $p_t\lesssim 1$~GeV, the initial conditions determine the
bulk features of multi-particle production in heavy ion collisions such
as $dN/dy$ or $dE_T/dy$ as a function of the relevant variables in the collision: collision energy, rapidity, transverse momentum and the
collision geometry (distribution of produced gluons in the transverse
plane). A very important application of the CGC formalism is
to provide such initial conditions for the subsequent evolution of the system. 
This input affects significantly the transport coefficients such as shear
viscosity extracted from hydrodynamic analysis of heavy-ion
collisions.  A complete description of the initial state of the collision must also account for the fluctuations of the deposited energy density. Event-by-event experimental analyses of data on angular correlations of the produced particles exhibit a much richer structure than event-averaged analyses. The standard interpretation is that fluctuations of the initial energy density or "hot spots" are transported by collective flow to the final state of observed particles. Current analyses of angular correlations in data have been able to determine the harmonic coefficients in the Fourier decomposition of angular correlations of this up $v_6$. This high moments of the angular distribution are sensitive to small distance scales thus providing an additional lever arm to constraint the QGP transport properties. Clearly, phenomenological works should parallel the precision of experimental measurements. This has triggered an intense theoretical and phenomenological activity  for a proper description of all possible sources of fluctuations, specially those  happening at small, sub-nucleonic length scales, responsible for the fine structure of the initial transverse energy-density patterns.

In the ``hard probes'' sector, i.e.\ for particles with a
--perturbatively-- large transverse momentum that do not thermalize
but are used as tomographic probes, a proper distinction of initial
state effects from those due to the QGP (``final state effects'') is
vital for a quantitative characterization of the matter produced in
heavy ion collisions as they may sometimes lead to qualitatively
similar phenomena in observables of interest like, for instance, the nuclear modification factors $R_{AA}$.
Although it is frequently believed that saturation effects are at work only for transverse momenta below the saturation scale $k_t\lesssim Q_s(x)$, the non-linear and non-local character of the CGC evolution equations extend the influence to unitarity correction at higher transverse momenta. Parametric estimates indicate that such {\it extended scaling} window reaches $k_t\sim Q_s^2/\Lambda_{QCD}$.

Last but not least, another primary goal of studies of the initial
state of HIC is to provide proof that (local) thermalization of the system
actually happens over the time scales estimated from hydrodynamical
simulations, $\tau_{therm}\sim 0.5\div1$~fm/c.  Explaining such swift transition between a coherent quantum mechanical state (the nuclear wave functions) to a thermal state (the QGP)  --highly incoherent by definition-- is, arguably, one of the most fundamental open problems in the field of heavy ions.

This review is structured as follows: In section (\ref{sec:three}) we introduce the main theoretical ideas behind the phenomenon of gluon saturation and the basics of the CGC effective theory at a rather heuristic level. We then focus in the discussion of present phenomenological 
analyses of data sensitive to small-$x$ dynamics from a variety of colliding systems with increasing degree of complexity. Section 2. is dedicated to searches of saturation physics in deeply inelastic electron-proton scattering (DIS) data from HERA and to the construction of models for the nuclear wave function. Next, in section 3. we discuss the analysis of exclusive features of single and double inclusive transverse momentum spectra in dilute-dense (i.e, proton-nucleus) scattering.  Finally, in section 4. we focus on the bulk features of multi particle production processes and its correlations in proton-proton (p+p), proton-nucleus (p+A) and nucleus-nucleus collisions.

\subsection{Three views of saturation}
\label{sec:three}

The problem of high energy QCD scattering is a multi-faceted many-body problem that affords several technical formulations and physical interpretations. Progress in the field and insight on particular aspects of the saturation regime has been attained through different technical approaches. Although this variety of languages is certainly an enriching feature, it can also be confusing to the newcomer to the field or to accidental conference bystanders. Below we present a brief survey of the most popular interpretations of saturation phenomenon and their inter-relation at a rather heuristic level.  Detailed reviews on the technical aspects of the CGC formalisms can be found, e.g in refs.~ \cite{Jalilian-Marian:2005jf,Weigert:2005us,Gelis:2010nm} or the book\cite{Kovchegov:2012mbw}.

\subsubsection*{Emission vs recombination}

\begin{figure}[tb]
\epsfysize=10.0cm
\begin{center}
\includegraphics[width=1\textwidth]{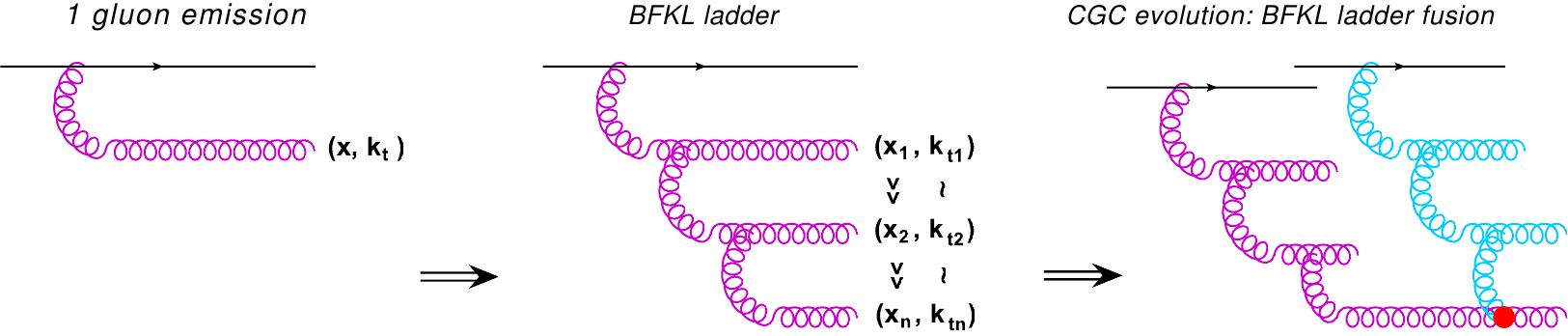}
\caption{Schematic representation of diagrams relevant for small-$x$ evolution: 1 gluon emission, BFKL ladders and BFKL ladder fusion in CGC non-linear evolution.}
\label{fig:ladder}
\end{center}
\end{figure}

Perturbative QCD evolution equations describe the change of hadronic wave functions with the resolution scale at which they are probed.  More precisely,  the object of interest for this discussion is the intrinsic $x$ and $k_{\perp}$ contributions to the infinite momentum hadronic have function in the light-cone gauge. This is given by the Weizsacker-Williams unitegrated gluon distribution (ugd) $\phi^{WW}(x,k_{\perp}$) that literally counts the number of gluons per unit phase space in the the hadron wave function carrying a fraction of the the light-cone momentum $x=k^+/P^+$ and transverse momentum $k_\perp$\footnote{Light-cone coordinates for an arbitrary 4-vector $a^\mu$ are defined as $a^{\pm}=\frac{a^0\pm a^3}{\sqrt{2}}$, $a^i=a^i$ for $i=1,2$.}. Its relation to the usual integrated gluon distribution $xG(x,Q^2)$ is\footnote{For a detailed discussion of the definition and physics interpretation of the different UGD appearing in the CGC formalism see {\protect\cite{Dominguez:2011wm}} }: 
\be
xG(x,Q^2)=\int^{Q^2}d^2k_{\perp}\,\phi^{WW}(x,k_{\perp})=\int^{Q^2}d^2k_{\perp}\,\frac{dN}{dYd^2k_{\perp}}
\ee
where we have introduced the rapidity $Y\equiv\ln(1/x)$. The basic mechanism that controls the evolution of hadronic wave functions is the branching of partons. Let us consider the simple case of a single energetic parton within a hadron (quark or gluon) as the initial state for the evolution. At lowest order, the bremsstrahlung probability for the emission of a gluon carrying a fraction of its light-cone momentum $x$ and a relative transverse momentum $k_{\perp}$ is, for small values of $x$,  given by

\be
dP\simeq C_{R}\frac{\alpha_{s} }{\pi^{2}}\frac{d^{2}k_{\perp}}{k_{\perp}}\frac{dx}{x}\,
\label{branching}
\ee
where $C_R= N_c$ for gluons and $C_R=(N_c^2-1)/2N_c$ for quarks. \eq{branching} exhibits the well known soft singularity of QCD splitting functions, i.e, it diverges for $x\to0$. The integrated probability for one gluon emission is then $P_{1-\text{emission}}\sim \alpha_{s}\ln(1/x)$. Clearly, for small enough values of $x$ the smallness of the strong coupling is compensated by the large logarithm $\ln(1/x)$ and the probability of one emission gets large, of order one. This suggests that additional emissions are also likely to happen. Indeed, imposing strong longitudinal momentum ordering to subsequent emissions, $x_1\gg x_2\gg\dots\gg x_n$, to the successive emissions in the shower represented in Fig (\ref{fig:ladder}), the integrated probability for the emission of $n$ gluons becomes $P_{n-\text{emissions}}\sim \left(\alpha_{s}\ln\frac{1}{x}\right)^n\sim \mathcal{O}(1)$. Perturbative QCD evolution equations like DGLAP or BFKL provide a compact tool for resuming such large logarithms to all orders in the form of differential equations for the scale dependence of the appropriate parton distributions.  Using very compact notation, the BFKL equation and its solution to leading order accuracy in $\alpha_s\ln (1/x)$ can be written as  
\be
\frac{\partial \phi(x,k_{\perp})}{\partial \ln(1/x)}=\mathcal{K}\otimes \phi(x,k_{\perp})\, \Longrightarrow \quad \phi^{BFKL}(x,k_{\perp})\sim x^{-{4N_c\ln 2\over\pi}\,\alpha_s}
\label{dglap-bfkl}
\ee
where $\otimes$ denotes integral convolution and $\mathcal{K}$ is the BFKL kernel (see e.g \cite{Forshaw:1997dc} for detailed expressions).

We see how the BFKL equation leads to an exponential (in rapidity $Y=\ln (1/x$)) growth of the unintegrated gluon distributions at small-$x$. 
This is due to the linear character of the BFKL equation, which only accounts for radiative process like the ones depicted in the left and centre diagrams of Fig \ref{fig:ladder}. The probability of a new emission at a next step is proportional to the enhanced gluon charge built up in the previous steps, leading to a chain reaction responsible of the exponential increase. A similar growth is predicted in the double logarithmic approximation (DLA) of DGLAP evolution. One implicit assumption in both BFKL and DGLAP approaches is that the hadron under consideration remains a dilute parton system at every step of the evolution. This is reflected in the fact that no interference effects between emitters are taken into consideration. However, this hypothesis necessarily breaks down when the gluon densities become large enough. The emission process is then coherent, appearing destructive interference terms which account for gluon recombination processes, as sketched in the right diagram of \fig{fig:ladder}. Therefore, in the high-density regime the probability of emission of a new gluon should depend somehow on the pre-existing charge density, contrary to what happens in BFKL or DGLAP. This would lead to the appearance of non-linear, density-dependent terms in the evolution equations that tame or {\it saturate} the strong growth of gluon densities at small-$x$.

Historically, the first step in that direction dates back to the early eighties. Gribov, Levin and Ryskin \cite{Gribov:1981ac} developed a picture of parton recombination, introduced the concept of saturation and proposed the first non-linear perturbative QCD evolution equation. Later 
Mueller and Qiu\cite{Mueller:1986wy} formulated a related equation also including non-linear gluon fusion terms. The resulting GRL-MQ equation for a low density picture of a spherical proton of radius $R_h$, reads

\be
{\partial^2xG(x,Q^2)\over\partial\ln(1/x)\partial\ln Q^2}={\alpha_sN_c\over\pi}xG(x,Q^2)-{4\alpha_s^2N_c\over3C_FR_h^2}{1\over Q^2}[xG(x,Q^2)]^2.
\label{glrmq}
\ee
The GLR-MQ equation is recovered in the double logarithmic limit of the BK equation (discussed below). Hence, it is valid near the boundary of the saturation region where it provides the first high-density correction to the evolution, but not deep inside the saturation region where the resummation of all-twists (as in the BK equation) is needed.
The very presence of a non-linear term in \eq{glrmq} already implies the emergence of a dimensionful scale, the saturation scale $Q_s$, which signals the transverse momentum scale at which the linear (emission) and non-linear (recombination) terms become parametrically of the same order. The onset of saturation effects also affords a neat geometrical interpretation as due to the overlap of gluons in the transverse plane: The {\it packing factor} $\kappa$, proportional to the transverse gluon density times the typical gluon-gluon interaction cross section, becomes of order one in the saturation regime, yielding the following parametric estimate for $Q_s$:
\be
\kappa= \rho^g\cdot \sigma^{gg}\sim\frac{xG(x,Q^2)}{\pi R_h^2}\cdot \left.{\alpha_s\, \over Q^2}\right|_{Q^2=Q_s^2}\sim 1\quad\Rightarrow Q_s^2(x)\approx {\alpha_s\over \pi R_h^2}\,xG(x,Q_s^2)\sim x^{-\lambda}\,.
\label{kappa}
\ee
Although saturation effects are universal, i.e they govern the small-$x$ dynamics of all hadronic wave function at sufficiently small-$x$, for fixed values of $x$ the saturation scale of a nucleus is enhanced with respect to the nucleon one by the so called {\it oomph} factor $\sim A^{1/3}$ due to the larger ab-initio gluon densities in nuclei.
\be
\frac{xG_A(x,Q^2)}{\pi R_A^2} \sim A^{1/3}\,\frac{xG_N(x,Q^2)}{\pi R_N^2}\quad \Longrightarrow\quad  Q_{sA}^2\sim A^{1/3}\,Q_{s,N}^2\,.
\ee

 The DLA limit $\alpha_s\ln(Q^2)\ln(1/x) \gg 1$ in which the GLR-MQ equation was derived is of little interest in phenomenological applications, since available experimental data from collider experiments correlate the small-$x$ and the small-$Q^2$ domain. The non-linear extensions of the BFKL equation, applicable for small-$x$ and moderate values of $Q^2$ are of more relevance. Actually, the BK equation that we shall discuss in detail later can be written, at leading order, as the non-linear extension of BFKL equation:
\be
\frac{\partial \phi(x,k_{\perp}^2)}{\partial \ln(1/x)}=\mathcal{K}\otimes \phi(x,k_{\perp}^2)-\bar{\alpha_s}\phi^2(x,k_{\perp}^2)
\ee

In more general terms the saturation regime sets in when the kinetic energy ($\sim \partial^2$) becomes of the order of the interaction energy ($\sim g^2A^2$). Taking the gluon distribution as a proxy for the gluon field two-point function, the r.h.s of \eq{kappa} yields $Q^2\sim\partial^2\sim \alpha_s\langle AA\rangle\sim \alpha_s\,xG(x,Q^2)$.   
The kinetic picture that emerges from this interpretation as a balance between emission and recombination terms has been reinforced by the realization by Munier and Peschanski~\cite{Munier:2003vc,Munier:2003sj} that non-linear QCD evolution equations are similar to reaction-diffusion processes studied in statistical systems. The latter define a wide universality class that also contains some specific population evolution models.

\subsubsection*{Strong Classical fields. The MV model}
The saturation phenomenon affords yet another interpretation in terms of strong classical fields which, actually,  play a prominent role in the Color Glass Condensate formulation of high-energy QCD scattering.
The key observation is that the presence of large gluon densities is completely equivalent to the presence of strong gluon fields:  
\be
\phi(k_{\perp}\lesssim Q_s)\sim \frac{1}{\alpha_s}\Rightarrow \mathcal{A}\sim \frac{1}{g}
\ee

This naturally defines a classical scenario: The occupation numbers are much larger than the commutators between creation and annihilation operators. Thus, the quantum fluctuations are just a small correction to the strong background field:
\be
N= a^{\dagger}a \sim {1\over \alpha_s} >> [a^\dagger,a]\sim 1
\ee
The classicality condition can also be understood exploring the relation of time scales of the source "valence" color charges and the soft gluons they radiate off.  Consider the case of a nucleus moving in the positive $x^3$ direction with a large light-cone momentum $P^+$. In the infinite momentum frame (IMF) the valence partons carrying a large fraction of light-cone momentum are Lorentz contracted into a thin {\it pancake} of size $\sim 2R/\gamma$ with $\gamma\sim P^+/m_N$ and $m_N$ the nucleon mass. 
Using the uncertainty principle to estimate the longitudinal and (light-cone) temporal extent of fluctuations carrying a momentum fraction $x$ and with transverse momentum $k_\perp$ one gets:
$
\Delta x^-\sim {1}/({xP^+}) \,;\quad \Delta x^+ \approx {2xP^+}/{k_\perp^2}
$ respectively.
Thus, small-$x$, {\it wee} partons are delocalized over much larger distances and are much shorter lived than fast partons. They see the large-$x$ valence as a frozen, infinitely thin (in the $x^-$ direction) source of color charge. This also applies to the space-time structure of the BFKL-like gluon cascades discussed in the previous section and represented in \fig{fig:ladder}. For a cylindrical nucleus and at a transverse resolution scale $Q^2$, a wee parton "sees" a charge distribution $\mathcal{Q}$ which is on the average color neutral but that has a large typical color charge squared:
\be
\langle \mathcal{Q}^a\rangle =0 \,, \quad \langle \mathcal{Q}^a \mathcal{Q}^b\rangle =\frac{\delta^{ab}}{Q^2}\frac{g^2 N_c A}{\pi R_A^2}\,.
\label{charge}
\ee
The r.h.s of \eq{charge} is obtained assuming that each nucleon contributes a factor $\sim g^2N_c$ to the total net charge squared of the nucleus. Since $R_A\sim A^{1/3}R_{proton}$, one gets $[\mathcal{Q}^a, \mathcal{Q}^b]=if^{abc}\mathcal{Q}^c\ll\mathcal{Q}^2\sim g^2 A^{1/3}/Q^2$ and the charges can be treated as commuting objects, i.e as a classical charge distribution. 
Based on these observations, McLerran and Venugopalan  \cite{McLerran:1993ni,McLerran:1993ka,McLerran:1994vd} (MV) pioneered the use of classical techniques to describe the small-$x$ structure function of a large nucleus from a first-principles calculation via classical techniques.

The MV model relies on a separation of
degrees of freedom into soft, small-$x$ gluons $\mathcal{A}(x)$ --treated as dynamical
gauge fields-- and high-$x$, static valence degrees of freedom (in
terms of light-cone time) characterized by a charge density $\rho(x)$.  The  fast partons can either emit or absorb soft gluons, but in a first approximation they do not deviate from their light-cone trajectories $x^-=0$ (eikonal approximation), so that they are integrated out and no longer considered as dynamical modes.  Thus they generate a color current only in the + direction and localized at $x^-\approx0$, $J_a^{\mu}=\delta^{\mu+}\delta(x^-)\rho_a(x^-,x_{\perp})$. The classical small-$x$ gluon field can be obtained by solving the classical Yang-Mills equation of motion in the presence of such external current
\be 
D_{\mu},\mathcal{F}^{\mu\nu}=J^{\nu}\,
\ee
which, with the appropriate boundary conditions and in the Lorenz gauge $\partial^\mu \mathcal{A}^\mu=0$, admits the following solution: 
\be
\mathcal{A}^{a\mu}= \delta^{\mu+}\,b^a(x)\,,\quad \text{with} \quad -\nabla_\perp^2 b^a(x)=\rho^a(x)\,.
\label{soleom}
\ee
The valence charges are treated as an stochastic variable. All the information about them is encoded in a functional probability distribution $W_\Lambda[\rho]$ that gives the probability of having a certain configuration of the source in the hadron at a scale $\Lambda^+$. In the MV model it is assumed to be given by a local Gaussian probability distribution:
\be
W_{MV}[\rho]= \mathcal{C} \exp\left[ -\int dx^-d^2x_\perp \frac{\rho^a(x^-,x_\perp)\rho^a(x^-,x_\perp)}{2\mu^2(x^-)} \right]
\label{WMV}
\ee
where $\mathcal{C}$ is a normalization constant and $\mu^2(x^-)$ is the per nucleon charge density, related to the saturation scale\footnote{Note that although we are defining the saturation scale in different ways throughout the paper all the definitions are equivalent. A more precise quantitative definition of the saturation will be provided in section 3 in terms of the solution of evolution equations.} as
\be
Q_{sA}^2 =   \frac{g^2C_F}{2\pi}\,\mu_A^2\,,
\label{eq:qsmu}
\ee
where $\mu_A^2$ is the net charge density per unit transverse area, obtained after integrating over the longitudinal extent of the valence charge distribution: $\mu_A^2 = \int dx^-\mu^2(x^-)$. 
The Gaussian probability distribution \eq{WMV} leads to trivial correlators between charges:
\be
\langle \rho^a(x_{\perp},x^-) \rho^b(y_\perp,y^-)\rangle = \delta^{ab}\delta^2(x_\perp-y_\perp)\delta(x^--y^-)\,\mu_A^2
\ee
In order to compute any physical observable sensitive to the small-{\it x} gluons, $\mathcal{O}[A_{\mu}^a]$, one must average over all possible configurations of the sources:
\be
\langle \mathcal{O}[A^\mu] \rangle=\int [d\rho] W_{MV}[\rho] \mathcal{O}[A^\mu[\rho]]\,, 
\ee 
with the relation between the $\rho$'s and the gluon fields given by \eq{soleom}. The fact that the averaging is performed at the level of the observable (i.e, at the level of the amplitude squared and not of the amplitude itself) reflects the fact that at very high energy the hadron is probed in a {\it frozen} configuration and, hence, interferences between different possible field configurations are neglected.  The MV model was later re-derived by Kovchegov in ref.~\cite{Kovchegov:1999yj}. There, rather than the classicality condition, the basic assumption was that the external probe is able to resolve the nucleons within the nucleus. Therefore the color charges at different $x^-$ are treated as belonging to different nucleons. Again, this allows to neglect all the higher order correlators of $\rho$ and leads to the same Gaussian weight functional as in \eq{WMV}, provided the net charge distribution $\mu_A^2$ is large enough.

\noindent The unintegrated gluon distribution functions found in the MV model with a Gaussian ansatz for the statistical weight $W_{\Lambda}$ saturate i.e. they show a power-like, perturbative behavior for large values of the transverse momentum, while they show a much milder logarithmic behavior for small transverse momentum (smaller than the saturation momentum):
\be
\phi^{MV}(x,k_{\perp})\sim\left\{
\begin{array}{cc} 
 \ln(k_{\perp}^2/Q_s^2)& \mbox{for}\quad k_{\perp}\ll Q_s, \\ 
Q_s^2/ k_{\perp}^2 & \mbox{for}\quad k_{\perp}\gg Q_s.
\end{array}
\right.
\label{phimv}
\ee

Notice that the MV model relies in the existence of an ab-initio large gluon density, with no dynamical content explaining the generation thereof, which requires the use quantum evolution equations. Therefore the MV model is well suited to describe the initial condition for the small-$x$ evolution of a large nucleus and that is precisely its main utility in phenomenological applications.
We finish by recalling that the MV set up describes a non-perturbative system in the following sense: The only dimensionfull scale appearing in the problem is the saturation scale which, in turn, is assumed to be perturbatively large $\mu_A^2 \sim A^{1/3} \Lambda_{QCD}^2$ such that $\alpha_s(Q_s^2)\ll1$ and weak coupling methods apply, same as in standard perturbation theory.  However, a standard perturbative expansion is not only a expansion in powers of the coupling but also in powers of the charge density which is large in the MV model. 
More specifically, in the saturation regime the interaction terms in the QCD Lagrangian cannot be expanded out in the covariant derivative $D^\mu= \partial^\mu-ig\mathcal{A}^\mu$ because $ g\mathcal{A}\sim\mathcal{O}(1)$. On the contrary, interesting non-linear saturation effects are expected when the density is large, of order $\alpha_s^{-1}$ and observables of interest, as the gluon distribution itself, exhibit a non-analytic dependence on the strong coupling characteristic of non-perturbative systems.

\subsubsection*{Eikonal approximation, Wilson lines, dipoles etc}

\begin{figure}[tb]
\epsfysize=4.0cm
\begin{center}
\epsfig{file=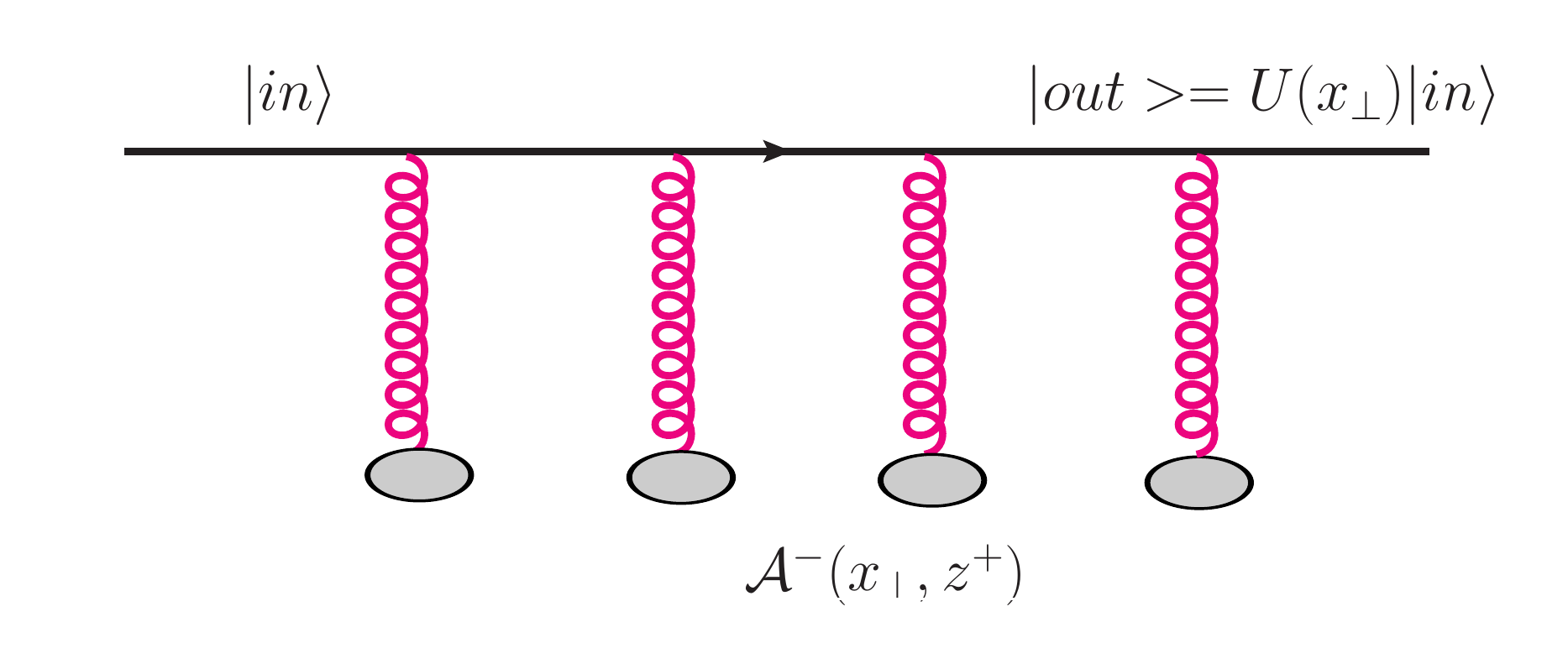,scale=0.45}
\includegraphics[width=0.49\textwidth]{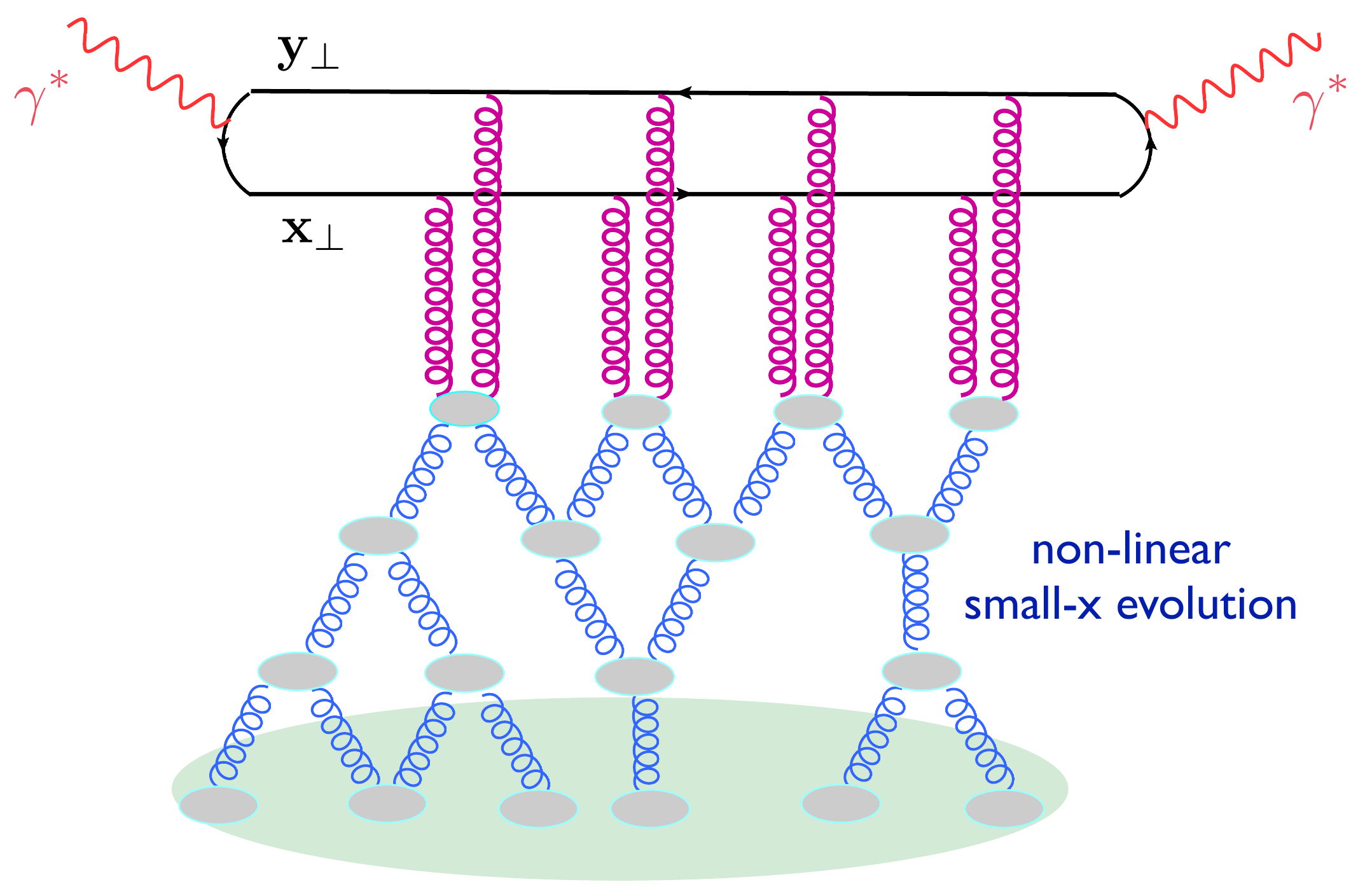}
\caption{Left: Wilson line for right moving quark. Right: Sketch of an electron-nucleus collision in the dipole model including multiple scatterings and non-linear small-$x$ evolution. }
\label{fig:wilson}
\end{center}
\end{figure}

Rather than quarks or gluons or the strong fields they are associated to, the most convenient degrees of freedom to study high-energy QCD scattering are correlators of Wilson lines.
So far we have only discussed the gluon distribution or strong color fields as {\it intrinsic} properties of high energy hadrons or nuclei. However, those definitions and the following physics discussion, are somewhat ambiguous since they are linked to a particular frame or gauge choice (light cone gauge in our case). An efficient way of getting rid off  such ambiguities is to set up a collision problem in order test the strong color of a target field with in terms of gauge and Lorentz invariant  observables. 
Thus, Wilson lines describe the propagation of an external parton (quark or gluon) projectile in the background of the strong field of the target. Let us consider the propagation of a fast right-moving quark scattering on a left-moving nucleus in the high-energy limit, as shown in \fig{fig:wilson}.  The quark will multiply scatter in the strong color field of the target. Under the eikonal approximation --valid at high energies-- the recoil of the propagating parton as it multiply scatters off the background field can be neglected. In transverse coordinate space this implies that the transverse position of the parton projectile remains fixed during its propagation. In this limit, 
\begin{eqnarray}
U( x_{\perp}) & = & 1 +ig\int dz^+\mathcal{A}^{-}(x_{\perp},z^{+})+\frac{(ig)^{2}}{2}\int dz^+ \int^{z^+} dz^{'+}\mathcal{A^{-}}(x_{\perp},z^{+})\mathcal{A^{-}}(x_{\perp},z^{'+}) +\dots \nonumber \\ & = & \mathcal{P}\exp\left[ ig\int dz^{+}\mathcal{A}^{-}(x_{\perp},z^{+})\right]\quad : \quad\in SU(N_c)
\label{wilson}
\end{eqnarray}
with $\mathcal{P}$ denoting the ordering of the color matrices in the exponential w.r.t. their $x^+$ arguments (i.e, when the exponential is expanded the fields with higher values of $x^+$ are to the left). 
Wilson lines are unitary matrices in $SU(N_{c})$, i.e. pure phases. The parton projectile simply picks a color phase in its propagation through the strong color fields. Analogous expressions are obtained for a projectile gluon by simply changing the representation of the external color fields in \eq{wilson}, i.e: $\mathcal{A}\equiv A^a(t^a)_{bc}\rightarrow A^a(T^a)_{bc}=A^af^{abc}$ where the $t$ and $T$'s denote the $SU(3)$ matrices in the fundamental and adjoint representation respectively. 4-gluon vertices are suppressed at high-energies, hence the similarity between quark and gluon scattering.  The simplest invariant object that can be built from Wilson lines is the scattering matrix for a quark-antiquark color dipole of transverse size $r_\perp=|x_\perp-y_\perp|$.
\be
S(r_\perp,Y)=\frac{1}{N_c}\langle\text{tr}\left\{ U(x_{\perp})U^{\dagger}(y_{\perp})\right\}\rangle_Y
\ee
where, again, the average should be performed over all the possible configurations for the color fields of the target. Using the Gaussian ansatz of the MV model \eq{WMV} one gets the Glauber-Muller\cite{Mueller:1989st} formula for the dipole scattering matrix:
\be
S(r_\perp,Y)=\exp\left[-\frac{r_\perp^2 Q_s^2(Y)}{4} \right];\quad \quad\text{with}\quad Q_s^2(Y)\equiv \frac{g^2 C_F\mu^2}{2\pi}\ln(1/r_\perp \Lambda)\,. 
\label{dip}
\ee
$Q_s^2$ is the saturation scale at rapidity of the sources $Y$ and $\Lambda$ is an infrared cutoff of the order of the nucleon scale. The infra-red divergence occurs because the assumption of a truly local Gaussian
distribution ignores the fact that color neutralization occurs on distance scales
smaller than the nucleon size: two $\rho$'s can only be uncorrelated if they are at transverse coordinates separated by at least the distance scale of color neutralization. It should be noted that although each partonic component in the projectile scatters eikonally, i.e it is unmodified up to a color phase, the whole projectile state --a color dipole in our case-- does not. On the contrary, each quark line picks a different color phase and the initial colorless state will get decohered --or colored-- though the propagation giving rise to a inelastic final state or, in other words, absorption.    

This simple result offers an additional aspect of saturation physics. Color dipoles with a transverse size much larger that the inverse saturation scale will be fully absorbed, whereas much smaller dipoles will propagate unaltered through the background field generated by such collection of random charges 
\begin{eqnarray}
S(r_\perp)\approx 0 \quad \text{for}\quad r_\perp\gg r_s\equiv1/Q_s\\
S(r_\perp)\approx 1 \quad \text{for}\quad r_\perp\ll r_s\equiv1/Q_s
\label{s}
\end{eqnarray}

The above result show how multiple scatterings are crucial for preserving unitarity of the theory: The S-matrix has the meaning of a probability and, therefore, is bounded between $[0,1]$. Expanding \eq{dip} one recovers the leading twist result $1-S(r_{\perp})\sim r_\perp^2$ which,  if extrapolated to larger values of $r_\perp$ would quickly lead to unitarity violation $S>1$. Rather, higher order terms in the expansion lead to the exponential form of \eq{dip} which obviously preserves unitarity.
Other relevant feature of \eq{dip}, is that, modulo the logarithmic term absorbed in the definition of the saturation scale, all the dependence of the dipole S-matrix on its natural variables $r_\perp$ and $Y$ is contained in dimensionless parameter $l_s(Y)\equiv r_\perp^2Q_s^2(Y)$ that characterizes the change between the dilute and dense regimes. This feature is known as {\it geometric scaling} and has played an important role in the discussion of saturation effects in HERA data on e+p structure functions, which exhibit an analogous scaling law.  Both features, unitarity and geometric scaling can be regarded as a hallmarks of saturation effects. Importantly, those two features are preserved by the solution of the non-linear evolution equations that encode the quantum corrections to the semiclassical picture presented in this section.   
Finally, \eq{dip} indicates that, deep in the saturation regime color neutralization happens at the scale of the saturation scale $Q_s$ instead of $\lambda_{QCD}$. This kind of nonlinearity --eikonalization of the gauge boson fields-- appears also in Abelian theories such as QED. The non-linearities intrinsic to non-Abelian theories appear at the level of the evolution and renormalisation of the gluon fields of the target, as we shall see later.  

As we shall see later, quantum corrections bring in modifications to the dipole S-matrix in \eq{dip}. They result not only in an increased saturation scale, but they also affect the specific functional form of the dipole S-matrix. In more general terms, the coherent emission of new gluons encoded in the renormalisation group equations generates correlations between the effective valence charges not taken into account in the Gaussian MV model.

Similar correlators of Wilson lines enter in the calculation of multi-particle production processes at small-$x$. The building blocks of the the calculation of physical processes are correlators of at least two Wilson lines of the kind $\langle\text{tr}\left\{U(x_{1\perp})U^{\dagger}(x_{2\perp})\right\}\rangle$, $ \langle\text{tr}\left\{U(x_{1\perp})U^{\dagger}(x_{2\perp})\dots U(x_{n\perp})\right\}\rangle$ etc or products of such correlators.  Actually, most of the recent theoretical advances in the CGC focus on the calculation of higher $n$-point correlators.

\subsection{The Color Glass Condensate}

The Color Glass Condensate effective theory for the small-$x$ degrees of freedom in high-energy QCD scattering provides a solid calculational framework to study saturation effects.  The starting point in the formulation of the Color Glass Condensate effective theory is analogous to the MV model: it relies on a separation of degrees of freedom into valence (static) and soft (dynamic) d.o.f. This separation is performed at some arbitrary light-come momentum $\Lambda^+$.
Valence charges are those with light cone momentum $p^+>\Lambda^+$ and are treated as a random variable over which one has to integrate to calculate physical observables. In turn the dynamical modes are the strong color fields with momentum $k^+<\Lambda^+$, and hence sit at small values of  $x=k^+/P^+$, with $P^+$ the total hadron or nucleus momentum.

However, the MV model is purely classical.  Quantum corrections to this classical approximation are large if one tries to describe modes with a much smaller light-come momentum than the scale at which the theory is defined 
$\Lambda^+$. For $x'=\lambda x$, they are of order $\ln 1/\lambda$ and, therefore, become large for $\lambda\ll 1$. This is due to the fact that the separation scale between hard and soft modes is totally arbitrary, and the interactions do not disappear as we move away from this scale. In other words, a semi-soft gluon with light-cone momentum fraction $x$ initially emitted from a valence source sitting at larger $x$ can itself be a source for the radiation of even softer gluons with light-cone momentum fraction $x'\ll x$, as in the BFKL ladders illustrated in \fig{fig:ladder}.    
These corrections can, however, be resumed by means of a Wilsonian renormalization group procedure in which quantum fluctuations inside the momentum strip $\Lambda^{+}>k^+>\Lambda^{'+}$ are integrated out and incorporated in the effective theory by renormalizing the color sources $\rho_a$ and its correlations or, equivalently, the statistical weight $W_{\Lambda^+}[\rho]$. In this way the classical + quantum calculation can be reproduced by a purely classical calculation, but with a modified statistical weight $W_{\Lambda^{'+}}[\rho]$, whose variation with the scale $Y\!=\!\Lambda^{+}/\Lambda^{'+}$ is given by a Renormalization Group Equation, the B-JIMWLK equations. This is a remarkable, highly non-trivial and very powerful feature of the CGC formalism

In order to discuss quantum corrections and evolution equations let us consider the change of the dipole scattering amplitude off a dense target with increasing collision energy. In more generality the discussion that follows is valid for a projectile composed of an arbitrary number of quarks and gluons as long as the total color charge is small, i.e for any dilute projectile. This problem can be approached in different frames yielding complementary, but strictly equivalent, pictures on non-linear QCD evolution.

\subsubsection*{Two views of evolution}

{\bf The JIMWLK equation}. One first option to study the rapidity evolution of dipole-nucleus scattering is to boost the dense target, leaving the dipole projectile unchanged. The boost will open up the phase space for additional gluon radiation. However such emission happens coherently from the strong pre-existing strong color field of the target, characterized as a static background field. 
The precise diagrammatic content of this derivation is quite involved. Shortly, it accounts for the possibility of two BFKL-like cascades as those described in \fig{fig:ladder} to merge into a single cascade. Thus,  this approach makes explicit the interpretation of evolution as a balance between emission and recombination terms. The resulting JIMWLK equation can be written in a Hamiltonian form for the weight functional:

\be
\frac{\partial W_Y[\rho]}{\partial Y}= H\, W_Y[\rho]\,,\quad \text{with}\quad 
H=  {1\over 2}\int_{x_\perp y_\perp}\frac{\delta}{\delta\rho^a(x_\perp)}\, \chi^{ab}(x_\perp, y_\perp)[\rho]\, \frac{\delta}{\delta\rho^b(y_\perp)}\,.
\label{jimwlk}
\ee  
The kernel of the evolution $\chi[\rho]$ is positive definite non-linear function of $\rho$ and is highly non-local in both transverse and longitudinal coordinates. This non-local character of small-$x$ evolution or, more in general, non-linear coherence effects, is often overlooked in phenomenological analyses. However, it is of primary importance since it interconnects the physics of semihard momentum modes with the one of softer modes via diffusion. Thus, it poses simultaneous constraints to both soft and hard sectors which must then be described at the same time, i.e within the same model for coherence effects.  

Eq. (\ref{jimwlk}) was first derived by Jalilian-Marian, Kovner, McLerran and Weigert in \cite{JalilianMarian:1996xn}, and has been further analyzed and discussed in many works \cite{Jalilian-Marian:1997gr,Jalilian-Marian:1998cb,Kovner:1999bj,Iancu:2000hn,Iancu:2001ad,Ferreiro:2001qy}. In terms of the Regge field theory, the JIMWLK equation includes triple pomeron vertex and effectively resum the fan diagrams of Pomeron Calculus. In the low density, or weak field, limit this equation linearizes and reduces to the BFKL equation, as shown in ref. \cite{Jalilian-Marian:1997jx}.\\

\begin{figure}[htb]
\begin{center}
\includegraphics[width=1\textwidth]{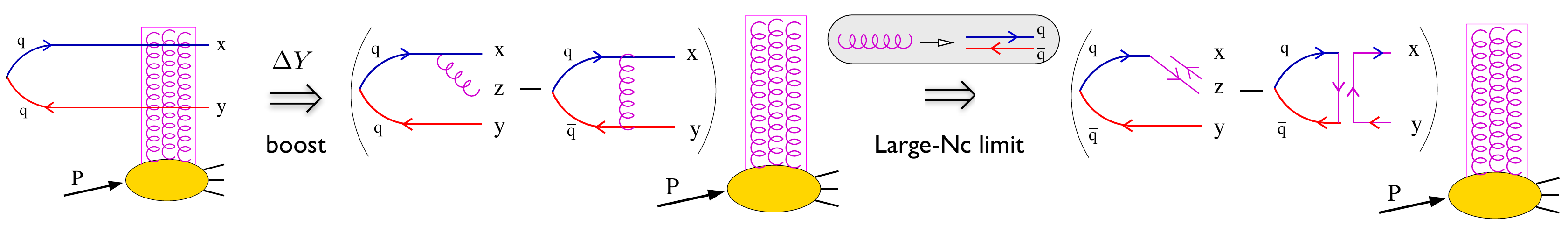}
\end{center}
\vspace*{-0.25cm}
\caption[a]{Schematic representation of the derivation of the BK equation.}
 \label{fig:bk}
\end{figure}

\noindent{\bf The Balitsky hierarchy}. An alternative formalism to the JIMWLK approach was developed by Balitsky from the operator product expansion for high-energy scattering \cite{Balitsky:1995ub}. In this approach the target is not evolving, and the evolution is achieved by boosting the projectile dipole. The additional energy provided by the boost will induce the emission of small-$x$ gluons at some transverse position ${\bf z}$ from either the quark or antiquark legs of the dipole (see \fig{fig:bk}), calculated at leading order in the resummation parameter $\alpha_s \ln(1/x)$. The newly created $q\bar{q}g$ system then scatters eikonaly with the target with an enhanced amplitude at the scale $Y=\ln(x'/x)$. The resulting equation for the dipole evolution reads

\begin{eqnarray}
\frac{d}{dY}\langle \text{tr}\left\{U(x_\perp)U^{\dagger}(y_\perp) \right\}\rangle_Y & =& \frac{1}{\pi^2}\int d^2{\bf z}\, \mathcal{K}_{\bf x y z}\left( \langle [ \tilde U(z_\perp)]^{ab} \,\text{tr} \left\{ t^a U(x_\perp)t^bU^{\dagger}(y_\perp)\right\}\rangle_Y \right. \nonumber \\
& & \left. -C_F\langle \text{tr}\left\{U(x_\perp)U^{\dagger}(y_\perp) \right\}\rangle_Y \large\right)\,
\label{bal-eq}
\end{eqnarray}
where $\tilde U$ denotes a Wilson line in the adjoint representation and
\be
\mathcal{K}_{\bf x y z}=\alpha_s \,\frac{(\bf{x}-\bf{y})^2}{(\bf{x}-\bf{z})^2(\bf{z}-\bf{y})^2}
\label{bkker}
\ee 
is the evolution kernel at leading logarithmic accuracy in $\alpha_s \ln (1/x)$, with $\alpha_s$ fixed.
It is evident from the r.h.s. of \eq{bal-eq} that the evolved $q\bar{q}g$ system contains a richer color structure than the original dipole. In particular, it contains a double trace operator (see \eq{fierzU}) which average value at the evolved scale $Y=\ln(x'/x)$ should be known in order to evaluate the change of scattering amplitude before and after the boost. In turn, the rapidity evolution of some initial double-trace operator would generate more complicated color structures under the small-$x$ gluon radiation that drives the evolution and so on. At the end of the day one gets a infinite hierarchy of coupled differential evolution for the rapidity evolution of expectation values for different $n$-point correlators of Wilson lines, known as the Balitsky hierarchy. \\

\noindent{\bf The Balitsky-Kovchegov equation.}
The infinite set of coupled equations given by the JIMWLK-Balitsky hierarchy is very difficult to handle in practice. Their solutions are not known analytically nor easy to find numerically. A convenient analysis tool is provided by the Balitsky-Kovchegov (BK) equation\cite{Kovchegov:1999ua,Balitsky:1996ub}, which can be obtained as the large-$N_c$ and mean field limit of the Balitsky hierarchy, as follows. The Fierz identity of SU(3) algebra 
\be 
 \sum_{a}t^a_{ij}\,t^a_{kl}={1\over 2}\delta_{il}\,\delta_{jk}-{1\over 2N_c}\delta_{ij}\,\delta_{kl}
\label{fierz}
\ee
allows to write the following relation for the double trace operator in the r.h.s of \eq{bal-eq} 
\be 
\tilde U(z_\perp)^{ab} \,\text{tr} \left\{t^a U(x_\perp)t^bU^{\dagger}(y_\perp)\right\}= \text{tr}\left\{U(x_\perp)U^{\dagger}(z_\perp)\right\}\,\text{tr}\left\{U(z_\perp)U^{\dagger}(y_\perp)\right\}-\frac{1}{N_c}\text{tr}\left\{U(x_\perp)U^{\dagger}(y_\perp)\right\}
\label{fierzU}
\ee
Taking now the large-$N_c$ limit one can neglect the last term in \eq{fierzU}, which simplifies enormously the color configurations generated under the evolution. In particular, the system created after a first step of the evolution can now be regarded as an ensemble of dipoles: the pre-existing one and two new dipoles, the one formed by the parent quark and the antiquark line of the gluon $({\bf x}, {\bf z})$, and other with the antiquark and the quark line of the gluon, $({\bf y},{\bf z})$. Within this picture further evolution can be interpreted as a dipole branching process, and the large-$N_c$ corresponds to replacing gluon lines by zero size $q\bar{q}$ lines. Further, assuming that the average of the product  of two dipole scattering matrices over the field configurations of the target factorizes $\langle S_{xz} S_{zy} \rangle_Y\rightarrow \langle S_{xz}\rangle_Y \langle S_{zy} \rangle_Y$, i.e taking the mean field limit, one gets a closed evolution equation for the rapidity evolution of the color dipole scattering amplitude off a dense hadronic target, the BK equation:

\be
\frac{\partial \mathcal{N}_{{\bf x y};Y}}{\partial Y}=\frac{N_c}{2\pi^2} \int d^2{\bf z}\, \mathcal{K}_{\bf xyz}\left[\mathcal{N}_{{\bf x z};Y}+\mathcal{N}_{{\bf z y};Y} - \mathcal{N}_{{\bf x y};Y} -\mathcal{N}_{{\bf x z};Y}\mathcal{N}_{{\bf z y};Y}  \right]\,,
\label{bk}
\ee
where we have introduced the dipole scattering amplitude $\mathcal{N}_{\bf x y}(Y)\equiv \mathcal{N}({\bf r}=\frac{\bf x-y}{2},{\bf b}=\frac{\bf x+y}{2};Y)= 1-\langle S(\bf {x ,y})\rangle_Y$ and the evolution kernel $\mathcal{K}_{\bf xyz}$ is given by \eq{bkker}. 

At leading logarithmic accuracy the BK equation can be Fourier transformed to momentum space, thus providing a straightforward non-linear extension of the BFKL equation for the unintegrated gluon distribution~\cite{Braun:2000wr}:

\be
{\partial \phi({{\bf k}},Y)\over \partial Y}={\alpha_sN_c\over \pi^2}k^2\int {d^2{\bf q}\over ({{\bf k}}-{\bf q})^2}\left[{\phi({\bf q},Y)\over q^2}-{\phi({{\bf k}},Y)\over q^2+({{\bf k}}-{\bf q})^2}\right]- {\alpha_sN_c\over\pi}\phi^2({{\bf k}},Y).
\label{bkk}
\ee

In this derivation,  the non-linear term in the r.h.s. of \eq{bk} arises as due to simultaneous scattering of the two newly created dipoles, it can be related to gluon recombination terms by means of the  equivalence between the JIMWKL equations and the Balitsly hierarchy. This term prevents the dipole amplitude to growing above unity, thus ensuring unitarity of the theory. 
Despite its approximate nature, the BK equation has become the starting point for the study of unitarity effects in high energy evolution of hadron structure. Their solutions have been investigated in many numerical and analytical works and, as we shall see, as nowadays it provides the main phenomenological tool to explore saturation effects in phenomenological analyses of experimental data.

\subsection{Recent developments and new directions}

The results presented in the introduction above constitute the basic elements of saturation physics and are well known since the foundation of the CGC effective theory. However, the physics input and technical tools of the CGC effective been extended and improved in many directions through an intense theoretical activity that continuous nowadays.
Here he briefly review some of the most recent developments and new lines of research in the field, not delving into the technical details of the different works. Some of these works have had a decisive role in improving the description of experimental data and will be explicitly discussed in the following sections. \\

\noindent {\bf Evolution equations at next-to-leading order}. Briefly, it can be said that the CGC is now entering the next-to-leading (NLO) order era. Actually, the kernel of the BK and B-JIMWLK evolution equations are now known to NLO accuracy in the resummation parameter $\alpha_s\ln (1/x)$ \cite{Balitsky:2008zza} or, also, to running coupling accuracy through the resummation of a partial subset of NLO diagrams~\cite{Kovchegov:2006vj,Gardi:2006rp,Balitsky:2006wa,Albacete:2007yr} to all orders\footnote{Hence, the resulting equation includes subsets of diagrams at all orders: NNLO, NNNLO etc.}. These works relied in the use of the BLM method\cite{BLM} to set the scale of the running coupling. 
Two recent works have derived the full NLO corrections to the JIMWLK Hamiltonian~\cite{Kovner:2013ona} and the NLO hierarchy of the evolution equations for Wilson-line operators or Balitsky hierarchy~\cite{Balitsky:2013fea}. These advances are of great relevance in order to bring closer theoretical predictions and data. It was known from long ago that the BK-JIMWLK evolution 
at the degree of accuracy of their original derivation (LO in $\alpha_s\ln (1/x)$ with $\alpha_s$ fixed) is $Q_s^2\approx Q_0^2(x_0/x)^\lambda$ with $\lambda \approx 4.8 N_c\alpha_s$\cite{Albacete:2004gw}. This growth is too fast to be reconciled with the energy dependence observed in HIC or DIS data. Indeed, the different phenomenological analysis of DIS data discussed below indicate an optimal value $\lambda\approx 0.2\div 0.3$. Clearly, these values can only be attained in a LO evolution scheme for unrealistically small values of the strong coupling. In turn, once the physics input in the evolution kernel is enriched by either taking into consideration running coupling effects (which correspond to an all-order resummation of a partial subset of NLO diagrams) or through the use of collinearly improved NLO BFKL evolution in the presence of saturation boundaries as done in \cite{Triantafyllopoulos:2002nz}, a noticeably smaller value of the exponent $\lambda$ is obtained, rendering the theory compatible with experimental data.\\

\noindent {\bf Production processes at next-to-leading order}. The efforts for furnishing non-linear QCD evolution equations with a richer physics input are also being paralleled by the calculation of particle production processes at NLO accuracy. A full NLO generalization of the dipole factorization formula for the calculation of structure functions in deep inelastic scattering (DIS) at small-$x$~\cite{Balitsky:2010ze,Beuf:2011xd} and its representation as a $k_t$-factorization~\cite{Balitsky:2012bs} formula are now available. Concerning hadronic collisions, full NLO~\cite{Chirilli:2011km} and inelastic~\cite{Altinoluk:2011qy} contributions to hybrid calculation of single inclusive particle production in dilute-dense scattering are now available, while running coupling corrections to the $k_t$-factorization formula have been calculated in ref.~\cite{Horowitz:2010yg}. Finally, a proof of factorization of multi-particle production processes in nucleus-nucleus collisions at NLO has been given in refs.~\cite{Gelis:2008ad,Gelis:2008rw}. \\

\noindent {\bf Beyond the 2-point function. Approximate solution of B-JIMWLK equations}. So far most searches of saturation effects have been based on the analyses of inclusive data from either DIS or hadronic collisions. However, these analyses do not allow to extract definitive conclusions on the presence or relevance of non-linear QCD dynamics in data. This has motivated the exploration of more exclusive observables that offer a deeper insight into small-$x$ regime and, thereby, more discriminating power among different physical scenarios. These studies make necessary going beyond the BK equation which applicability is limited to observables sensitive to only the 2-point (dipole) amplitude, like structure functions in DIS or single inclusive hadron cross sections in pA collisions. Thus, large attention has been devoted lately to the study of di-hadron azimuthal correlations in d+Au collisions or to the ridge-like correlations appearing in RHIC and LHC A+A collisions and also in p+p and p+Pb high multiplicity collisions at the LHC. The calculation of these correlations involve multi-particle correlations beyond the 2-point function and, therefore, demand the resolution of the full B-JIMWLK equations. An even more urgent case is the calculation of inclusive gluon production in nucleus-nucleus collisions. The solutions of the corresponding Classical Yang-Mills equations are well known numerically, but the quantum evolution of the sources given by B-JIMWLK evolution --needed for instance to extrapolate from RHIC to LHC energies-- has not been implemented to date in phenomenological studies.  

Both the theoretical interest and phenomenological relevance of this problem have triggered intense activity in the search for approximate analytical solutions to the B-JIMWLK equations. The most popular approach in this quest has been to look for approximate mean field Gaussian solutions of the B-JIMWLK equations, thus extending the Gaussian MV model to arbitrarily small values of $x$, albeit with a non-local kernel. The main advantage of this approach is that Gaussian distributions allow to extract arbitrary $n$-point correlators in terms of the {\it only} 2-point function through a systematic and straightforward application of Wick's theorem. Kovchegov and Weigert showed in \cite{Kovchegov:2008mk}  that a truncation of JIMWLK evolution in the form of a minimal Gaussian generalization of the BK equation captures some of the remaining $1/N_c$ contributions leading to an even better agreement with JIMWLK evolution. The validity of this approach has been further justified in \cite{Iancu:2011nj} and, finally, Gaussian analytical solutions have been tested against numerical solutions of the full B-JIMWLK equations in \cite{Dumitru:2011vk, Alvioli:2012ba}, showing a very good agreement  between the two for the 4 and 6-point functions for some specific transverse configurations. 
Furthermore, our knowledge of exclusive particle production and multi-particle correlations (di-hadrons, hadron-photon etc) has been advanced significantly through a series of recent works that establish the precise relation between the $n$-point functions of the nuclear wave function and the observables of interest~\cite{JalilianMarian:2004da,Dominguez:2011wm}. Recent studies also shed light on the rise of double parton scattering in the CGC framework\cite{Lappi:2012xe}, a phenomenon poorly investigated to date.\\

\noindent {\bf Symmetrization of B-JIMWLK equations}. The B-JIMWLK evolution equations are tailored to describe the situation where a small perturbative projectile scatters off a large dense target.
They neglect important effects --variably referred to as ÒPomeron loopsÓ~\cite{Mueller:2004sea}-- ÒfluctuationsÓ~\cite{Iancu:2004es} or Òwave function saturation effectsÓ~\cite{Kovner:2005jc} that become important in less asymmetric situations, i.e those where both the projectile and target are in the high or low density regime.
Pomeron loops or gluon number fluctuations are of higher order in $\alpha_s$ and are linear in the source density since they result from the splitting of a single gluon. Hence, they are parametrically sub-leading in the high-density regime (remember that recombination processes are enhanced by powers of the gluon density $\sim \alpha_s\phi$). However, they can strongly influence the approach towards saturation. For instance, these fluctuations reduce the (average) speed of the saturation front and also make the value of $Q_s$ a fluctuating quantity itself, washing out the geometric scaling property of the solutions of B-JIMWLK equations. 
It has been argued that the onset of Pomeron loop effects may be considerably delayed in evolution rapidity once the running of the coupling is taken into account~\cite{Dumitru:2007ew}. This observation rises the hope that they maybe safely discarded in phenomenological applications at LHC energies. However, their proper incorporation into the high-energy QCD formalism and, thereby, reaching a full generalization of the B-JIMWLK equations to to arbitrarily dense colliding systems is still an important theoretical open question subject to many studies.\\

\noindent {\bf The Glasma and the thermalization problem}. Other important line of research in the CGC focuses in the characterisation of the system produced right after a heavy ion collision and its evolution towards thermalization. This pre-equilibrium state, dubbed as Glasma, is far from being a thermal state or a Quark Gluon Plasma.
Rather, it is highly coherent, anisotropic state which properties can be investigated through the solution of the classical Yang Mills equation of motion until proper times $\tau\lesssim 1/Q_s$. It is known from time ago that the classical solutions are unstable under small changes in the initial conditions. This instabilities grow exponentially with the proper time $e^\tau$, quickly overwhelming the leading order contribution to the energy-momentum tensor. Recently, a first-principle calculation of the fluctuations spectrum has been performed in\cite{Epelbaum:2013waa} by solving the Yang-Mills equations linearized around the classical background with appropriate initial conditions.
Further, first NLO-resummed results in the Color Glass Condensate framework for the energy-momentum tensor shortly after a heavy ion collision were presented in ref.\cite{Gelis:2013rba}. There, it was shown that for sufficiently large --but realistic-- values of the strong coupling a regime of nearly ideal hydrodynamical expansion is reached shortly after the collision. 
While further studies may be needed in order to determine the equilibration time and system properties as a function of the natural scales of the problem (collision energy, initial geometry etc), this result provides a first compelling indication of short-time thermalization dynamics within the CGC formalism, hence driven by weakly coupled dynamics.  \\



All in all, substantial and continuos progress is being made towards the ultimate CGC goal of providing a theoretically solid and phenomenologically complete description of the full space-time development of a heavy ion collision until the thermalization time.
However, it is worth remarking that the significant theoretical progress reported above has not been yet fully exploited in phenomenological applications, which leaves a large margin for improvement in the analysis of the both RHIC and LHC data.

\subsection{Other approaches}

Aside of the CCG formalism, other approaches have been developed for the description of initial state effects of heavy ion collisions.
These other approaches typically start from standard leading twist perturbation theory and introduce modifications to it to account for nuclear effects, either in the form of shadowing weights or higher twists corrections. Importantly, all these works  include strong coherence effects among the constituent nucleons, or the relevant degrees of freedom at the sub-nucleon level, at different stages of the collisions process. 
On physical grounds, coherence phenomena are related to the presence of high gluon densities in the wave function of the colliding nuclei at small values of Bjorken-$x$. Briefly, one can identify coherence effects at the level of the wave function and also at the level of primary particle production, as sketched in \fig{coher1}. To the first category belong the nuclear shadowing (in a partonic language) or the percolation and string fusion (in non-perturbative approaches). In both cases, when different constituents, whichever the degrees of freedom chosen are, overlap in phase space according to some geometric criterium recombination of such constituents happen, thus reducing the total number of scattering centers --gluons-- entering the collision process. To the second category correspond the inclusion  of multiple scatterings or the presence of energy dependent momentum cutoffs in event generators. 

All these ingredients are akin, at least at a conceptual level, to those dynamically built in the CGC --saturation and multiple scatterings-- although they are formulated in very different ways. Thus, one can conclude that coherence effects are essential for the description of experimental data. The pertinent debate now is which theoretical framework is the best suited for their description: whether the CGC (at its present degree of accuracy) or others. Below we make a brief and incomplete survey of some popular approaches alternative to the CGC one. A rather exhaustive compilation can be found in e.g ref.~{\protect\cite{Abreu:2007kv}. 
\begin{figure}[htb]
\begin{center}
\includegraphics[width=0.27\textwidth]{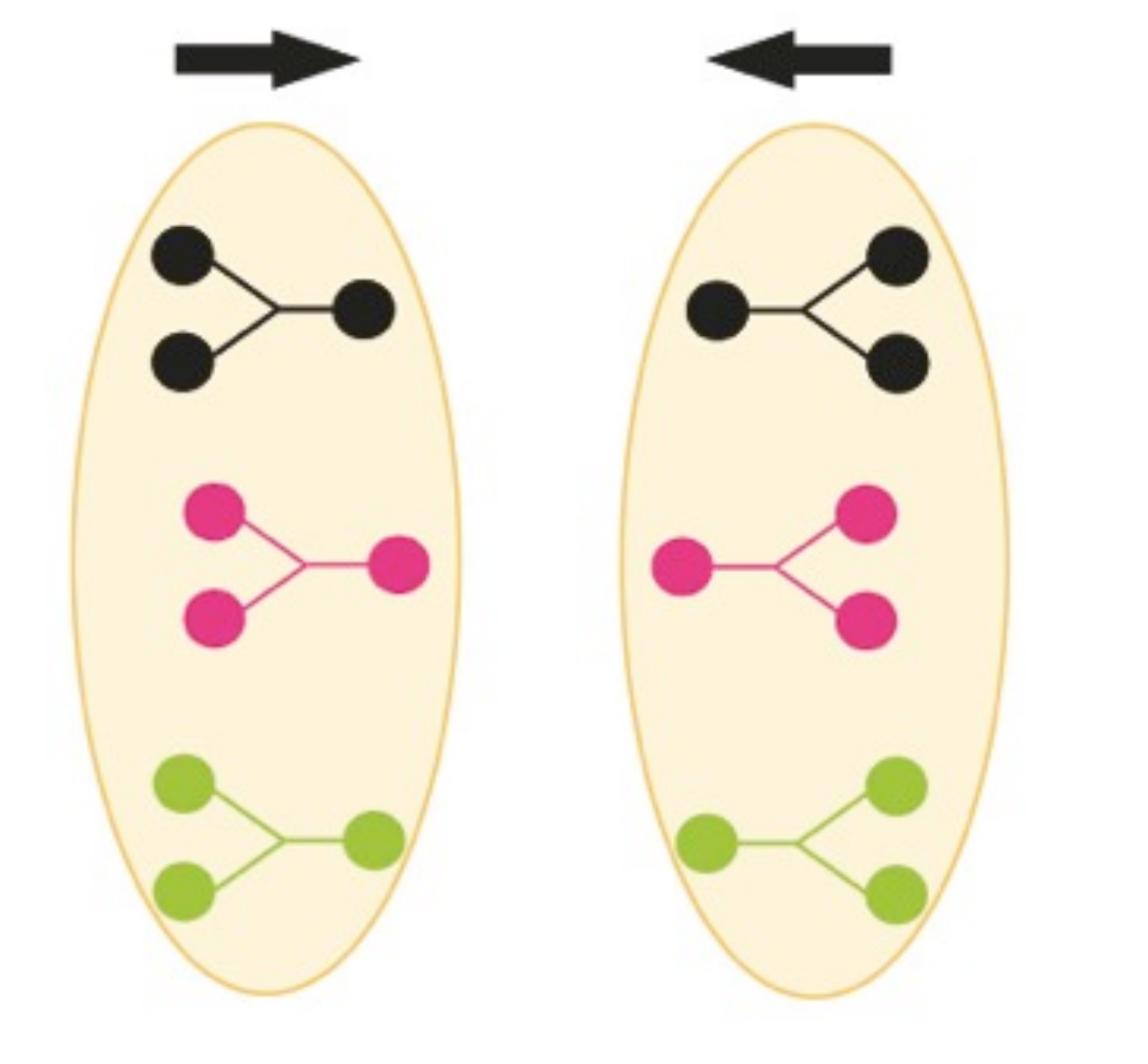}
\hspace*{1.5cm}
\includegraphics[width=0.39\textwidth]{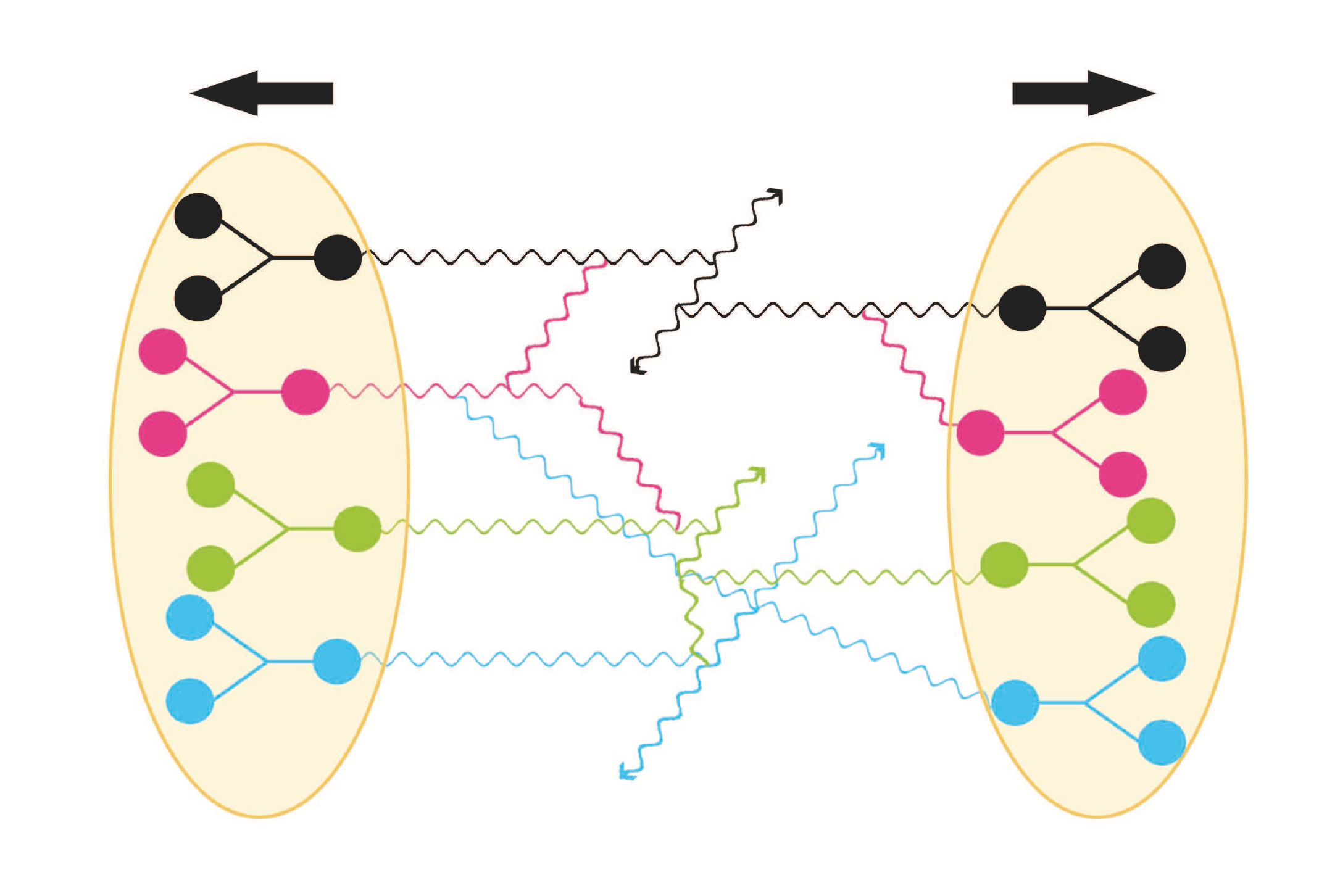}
\end{center}
\vspace*{-0.15cm}
\caption[a]{Left: Schematic representation of coherence effects in HIC at the level of the nuclear wave functions (left) and of particle production processes (right). }
\label{coher1}
\end{figure}

A minimal modification of the leading twist formalism is provided by the nuclear parton distribution approach (nPDF)\cite{Eskola:2009uj,Hirai:2001np,Kovarik:2013sya}. The starting hypothesis is that leading twist collinear factorization is valid to describe particle production with perturbatively high transverse momentum in nuclear reactions, $p_t\!>\!1\div 2$ GeV. All the nuclear effects are then absorbed into the nuclear PDF's, made proportional to the nucleon PDF's through scale dependent factors: $f_{A,i}(x,Q^2)= R_{A,i}(x,Q^2)\,f_{N,i}(x,Q^2)$. The $Q^2$-dependence of nPDF's is described by DGLAP evolution, while all the information on the $x$-dependence is encoded in the initial conditions for DGLAP evolution and extracted from global fits to experimental data. In the shadowing region $x\lesssim 10^{-2}$ all nPDF parameterizations exhibit a suppression of the gluon nuclear PDF, $R_{A,g}<1$. The lack of available small-$x$ data on nuclear reactions to constrain the nPDF fits is reflected in the large uncertainty bands for the nuclear gluon distributions shown in \fig{coher}. While the nPDF shadowing is of phenomenological origin and lacks of a clear dynamical input, many other dynamical models for shadowing are available in the literature (see e. the review in ref.~\cite{Armesto:2006ph}).  
 
One step further is taken by the calculation of power corrections to the leading-twist approximation, either in the coherent limit of the high-twist formalism~\cite{Qiu:1988dn,Qiu:2003vd} or through an all orders resummation of incoherent multiple scatterings in {\it Glauber-like} formalisms~\cite{Accardi:2003jh}.  Performing the complete resummation including energy-momentum conservation is a challenging task. Sometimes, a detour of the strict calculation is taken by resorting to unintegrated parton distributions which include information about the intrinsic transverse momentum of the partons $k_0$  and the average transverse momentum gained during the interaction with the nucleus $\Delta k:$ $f_{A,i}(x,Q^2) \rightarrow F_{A,i} (x,Q^2 ,\langle k_0 \rangle + \langle \Delta k \rangle (\sqrt{s},b,p_t))$. While the intrinsic $k_0$ is adjusted in p+p collisions, the gained transverse momentum is let to depend on the collision energy, centrality and $p_t$ of the detected hadron. 

Monte Carlo generators like HIJING\cite{Wang:1991hta} or HYDJET\cite{Lokhtin:2008xi} aim at a full characterization of the collision event, both in the hard and soft sectors. They are two component models where particle production below some transverse momentum cutoff $p_t< p_{t0}$ is regarded as non-perturbative and described by means of soft cross-sections based on string models. In turn particle production above the cutoff $p_t>p_{t0}$ proceeds through independent minijet production from different nucleon-nucleon collisions and is described within standard LO perturbation theory including shadowing corrections. In practice, phase-space arguments motivate the implementation of energy-dependent cut-offs $p_{t0}(\sqrt{s})$ in order to ensure the hypothesis of independent particle production from different sources. \fig{coher} (right) shows the energy dependence of $p_{t0}(\sqrt{s})$ in some p+p and A+A event generators. Its strong rise with increasing energy --stronger than the typical one for the nuclear saturation scale-- signals the increasing importance of collective effects in particle production processes. The EPOS event generator implements multiple scatterings corresponding  Gribov-Regge theory and perturbative QCD and explicitly accounts for energy-momentum conservation and, in its latest version, also for parton saturation effects \cite{Werner:2013vba}.

\begin{figure}[htb]
\begin{center}
\includegraphics[width=0.49\textwidth]{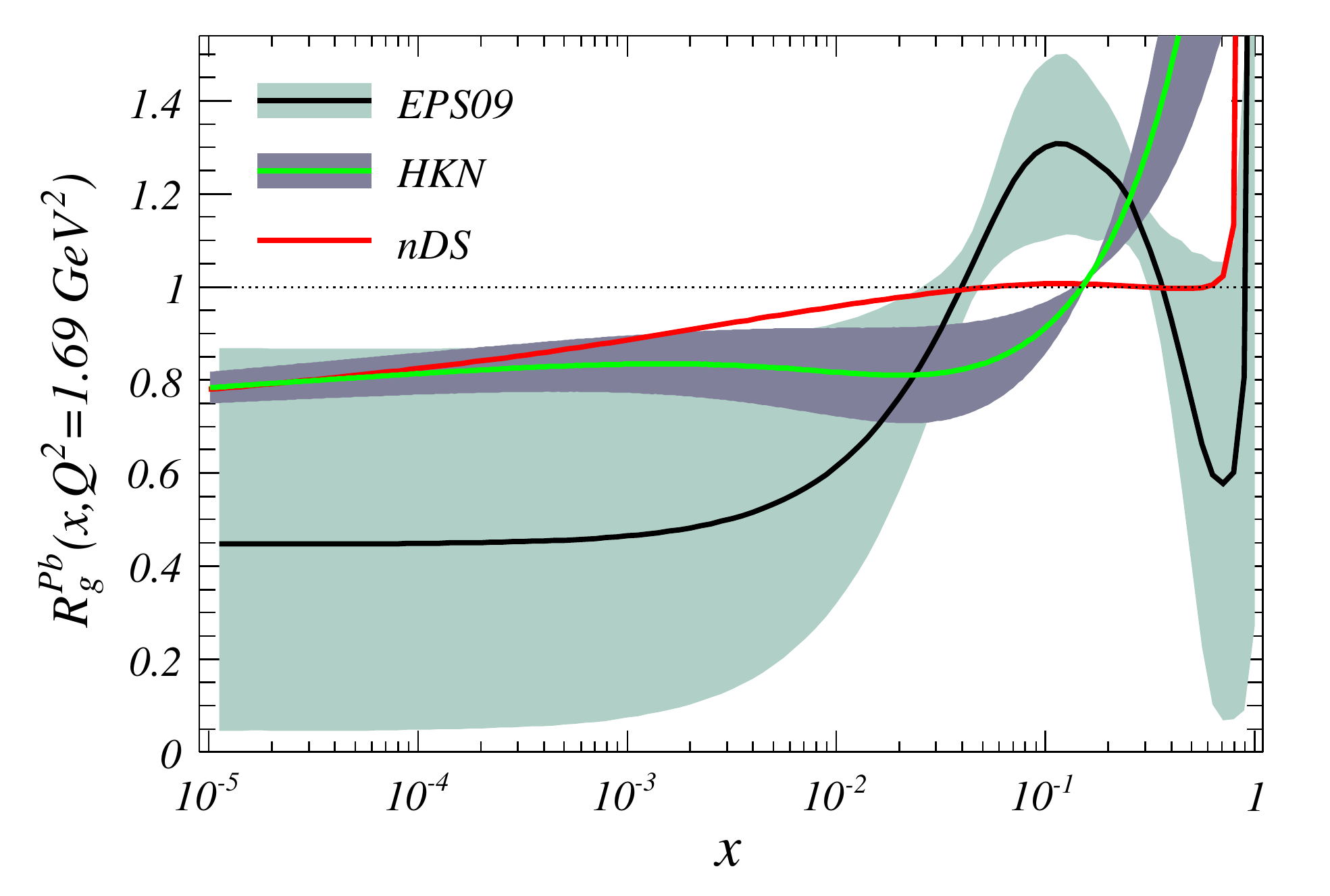}
\includegraphics[width=0.46\textwidth]{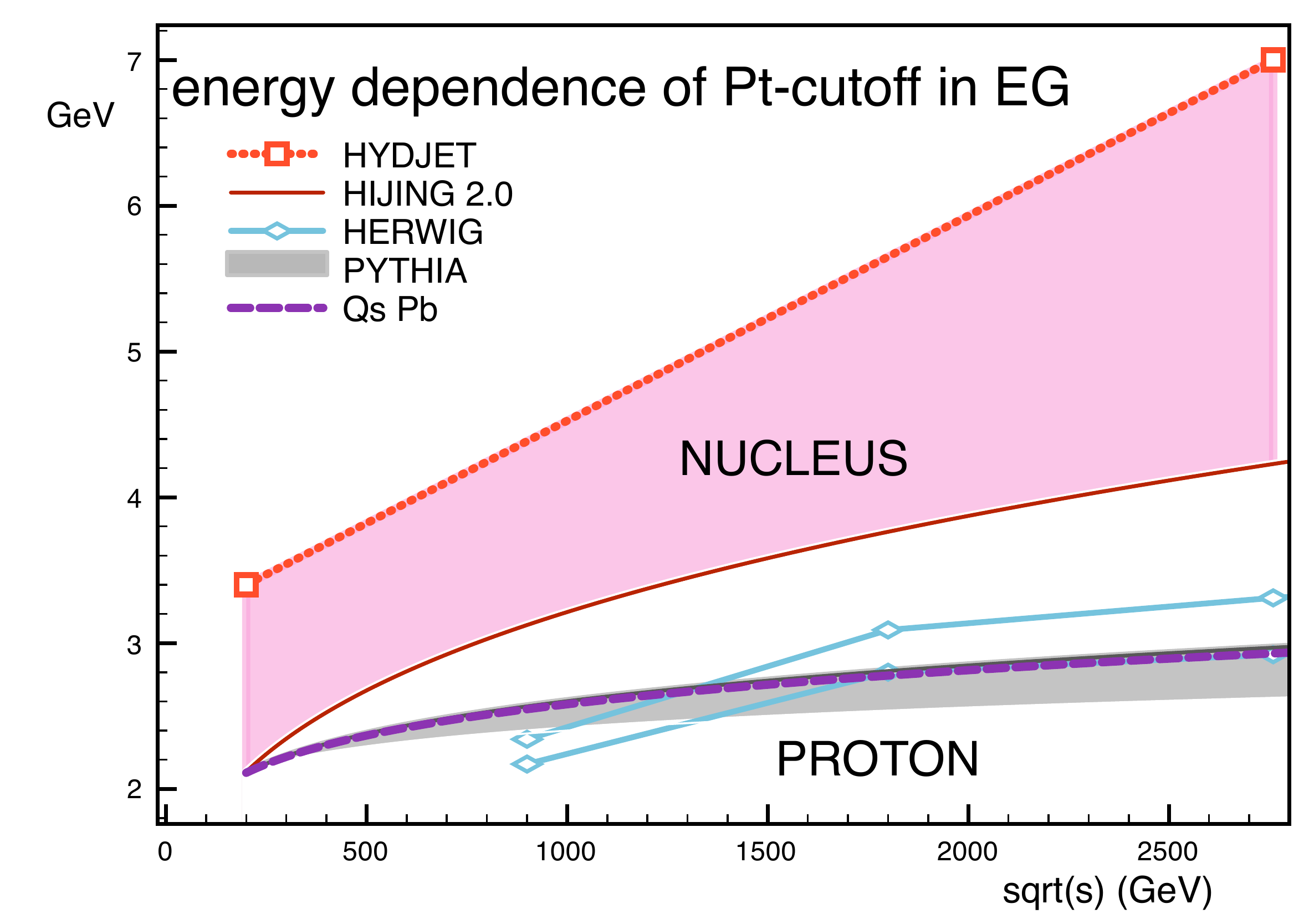}
\end{center}
\vspace*{-0.15cm}
\caption[a]{Left: Different nPDF parametrizations for the gluon distribution at the scale $Q^2=1.69$ GeV$^2$ (figure from ref.~\cite{Salgado:2011wc}). Right: Collision energy dependence of the transverse momentum cutoff in MC event generators for proton-proton collisions (two different PYTHIA tunes) and heavy ion collisions (HYDJET and HIJING). Also shown the characteristic value of the saturation scale $Q_{s}\sim \sqrt{s}^{0.15}\,$. }
\label{coher}
\end{figure}

\section{Deep Inelastic Scattering}

\subsection{e+p collisions}

Despite the fact that CGC effects are expected to be enhanced in
nuclei versus protons, which is due to the larger valence charge densities per unit transverse area in nuclei, 
so far the most exhaustive searches for the gluon saturation phenomenon have been
performed using data on proton reactions. This is mainly due to the
large body of high quality Deep Inelastic Scattering (DIS) data on electron-proton collisions at small-$x$ collected in HERA. This is in sharp contrast with extreme paucity of small-$x$ data in electron-nucleus collisions, specially for large nuclei with only 8 experimental data points for a lead nucleus at $x\!<\!10^{-2}$ available to date.
Further, e+p collisions provide the cleanest experimental and theoretical ground to directly study the partonic content of hadrons. For these reasons most --if not all-- phenomenological approaches for heavy ion
collisions borrow empiric information from the analysis of e+p data,
either to parametrize the Bjorken-$x$ or energy dependence of the
saturation scale or to also constrain more exclusive features
of the unintegrated gluon distribution of the proton. Here we briefly review the different saturation-based approaches for the description of e+p data and discuss how knowledge gained in these analyses is used to build up models for the nuclear wave function. 

All CGC-based models for the description of HERA data rely in the dipole-model formulation of DIS\cite{Nikolaev:1990ja}. There, and using light-cone perturbation theory, the total virtual photon-proton cross section can be written as the convolution of the light-cone wave function squared for a virtual photon to fluctuate into a quark-antiquark dipole, $\left|\Psi_{T,L}\right|^2$, and the quark-antiquark dipole cross section off the hadronic target. For transverse and longitudinal polarizations of the virtual photon one writes: 
\begin{equation}
  \sigma_{T,L}^{\gamma^*-h}(x,Q^2)=\sum_f\int_0^1 dz\int d^2{\bf r}\,\vert
  \Psi_{T,L}^f(e_f,m_f,z,Q^2,{\bf r})\vert^2\,
  \sigma^{q\bar{q}}(r,x)\,,
\label{dm}
\end{equation}
where, $e_f$, $m_f$ and $z$ refer to the quark electric charge, its mass and the fraction of longitudinal momentum of the photon carried by it. According to the optical theorem, the dipole cross section is given by the integral over impact parameter of the (imaginary part of) dipole-hadron scattering amplitude:
\be
\sigma^{q\bar{q}}(r,x)=2\int d^2b \,{\cal N}(x,r,b)\,.
\ee
The dipole amplitude encodes all the information about the strong interactions --along with all $x$-dependence-- of the scattering process. Its modelling constitutes the main difference amongst different CGC approaches.
\vspace*{0.2cm}

$\bullet$ {\bf The GBW model} ~\cite{Golec-Biernat:1998js} provided the first analysis of DIS data in terms of saturation physics. It relies in the following simple parametrisation of the dipole cross section:
\be 
\sigma^{dip}=\sigma_0 \left[1-\exp\left(-\frac{r^2Q_{s\rm N}^2(x)}{4}\right)\right]\,.
\label{gbw}
\ee
This functional form is very similar to the MV model, up to logarithm corrections. 
A good description of data on structure functions was obtained with the proton saturation scale given by 
$Q_{s\rm N}^{2}(x)=Q_0^2\left(\frac{x_{0}}{x}\right)^{\lambda}
$
with $Q_0^2=1$ GeV$^2$, $\lambda=0.288$ and $x_{0}=3\cdot10^{-4}$. This empiric  parametrisation found a
theoretical justification in the analysis of the collinearly improved LO and NLO BFKL evolution in the presence of
saturation boundaries\cite{Triantafyllopoulos:2002nz}.  However, the GBW model leads to an exponential fall-off of the UGD in transverse momentum space incompatible with both more exclusive experimental data and with the well established power-law behaviour dictated by perturbative QCD calculations. 

$\bullet$ {\bf IIM \& b-CGC}\cite{Iancu:2003ge,Watt:2007nr}: This other analytic model aimed at furnishing the GBW with known features of BFKL dynamics. Later it was updated to also include explicit impact parameter dependence\cite{Watt:2007nr}. This latest incarnation of the model was dubbed b-CGC model. The dipole scattering amplitude reads
\be
\mathcal{N}(x,r,b)=\left\{ \begin{array}{cc}  \mathcal{N}_0 \left(\frac{rQ_{s\rm N} }{2} \right)^{(\gamma_s+\frac{1}{\kappa\lambda Y}\ln\frac{2}{rQ_{s\rm N}}   )}&:\quad rQ_{s\rm N}\le 2  \\ 1-e^{-A\ln^2(B\, r\, Q_{s\rm N})} & :\quad  rQ_{s\rm N}>2   \end{array}\right.
\label{IIM}
\ee   
with $Y=\ln(x_0/x)$ and $\kappa=\chi''(\gamma_s)/\chi'(\gamma_s)$ where $\chi$ is the LO BFKL kernel. The coefficients $A$ and $B$ are determined from the continuity condition at the matching point $rQ_{s\rm N}=2$. 
In its b-dependent version, the saturation scale is given by: 
\be
Q_{s\rm N}^2(x,b)=\left( \frac{x_0}{x} \right)^\lambda \exp\left[-\left(b^2/2B_{CGC} \right)^{\frac{1}{\gamma_s}}\right]\,.
\ee
The remaining constants are left as fit parameters (see ref.~\cite{Watt:2007nr} for explicit values). However, the effective anomalous dimension in this model $\gamma_s +\frac{1}{\kappa\lambda Y}\ln\frac{2}{rQ_{s\rm N}}$ with $\gamma_s\approx 0.6 \div 0.7$ yields a very hard proton UGD at high transverse momentum and is disfavoured by the analyses of inclusive hadron production in d+Au collision at RHIC or p+p collisions at the LHC. This is contrast with the rcBK approach discussed below, where values of the  anomalous dimension $\gamma\!>\!1$ in the initial condition for the evolution  lead to a better agreement with data.


$\bullet$ {\bf IP-Sat}: Other popular saturation approach for the description of HERA data at small-$x$ is the Impact Parameter Saturation model (IP-Sat), first proposed by Teaney and Kowalski in \cite{Kowalski:2003hm} and recently updated in\cite{Rezaeian:2012ji} 
to include the much more precise data on reduced cross-sections 
from the combined analysis by the H1 and ZEUS 
collaborations\cite{Aaron:2009aa}. The main assumption in the IP-Sat model is that the dipole-nucleon cross section is well described by the Glauber-Mueller resummation of  two-gluon rescatterings:
\be
\mathcal{N}(x,r,b)=1-\exp\left[ -\frac{\pi^2r^2}{2N_c}\,\alpha_s(\mu^2)xg(x,\mu^2)T_g(b)\right]\,,\quad \text{with} \quad T_G(b)=\frac{1}{2\pi B_G}\exp\left( -\frac{b^2}{2B_G} \right)\,,
\label{ipsat}
\ee
where $ xg(x,\mu^2)$ is the standard gluon distribution of the proton evaluated a a scale $\mu^2=\mu_0^2+1/r^2$ and 
$\langle b^2\rangle = 2B_G$ the average squared gluonic radius of the proton. In the IP-Sat model the evolution of the dipole amplitude is due to the $Q^2$-evolution of the gluon distribution $xg(x,\mu^2)$ according to LO DGLAP quarkless evolution. 
The eikonalized form in \eq{ipsat} account for higher twist corrections important at small-$x$ and ensures a smooth matching with the perturbative QCD dipole expression at large $Q^2$ for a given $x$ (at the LO degree of accuracy). 
The initial condition for DGLAP evolution at the initial scale is taken of the form $xg(x,\mu_0^2)= A_gx^{-\lambda_g}(1-x)^{5.6}$. Table (\ref{t-1}) shows the fit parameters obtained in\cite{Rezaeian:2012ji} after fitting all available data with $x<10^{-2}$ and  $Q^2\in\left[0.75,650\right]$ GeV$^2$.

\begin{table}
  \centering
  \begin{tabular}{cccc|ccc|c}
    \hline\hline
    Data & $B_G/\mathrm{GeV}^2$ & $m_{u,d,s}$/GeV & $m_c$/GeV & $\mu_0^2/\mathrm{GeV}^2$ & $A_g$ & $\lambda_g$ & $\chisq$ \\ \hline
     $\sigma_r$ & 4 & $\approx 0$ & $1.27$ & $1.51$ & $2.308 $ & 0.058& $298.89/259 =1.15$ \\ \hline
      $\sigma_r$ & 4 & $\approx 0$ & $1.4$ & $1.428$ & $2.373 $ & 0.052& $316.61/259 =1.22$ \\ \hline\hline
  \end{tabular}
  \caption{IP-Sat parameters from the fits to HERA data on reduced cross section in the kinematic range $Q^2\in [0.75, 650]\,\text{GeV}^2$ and $x\le 0.01$ \cite{Rezaeian:2012ji}. 
  }
  \label{t-1}
\end{table}
\begin{figure}[htb]
\begin{center}
\includegraphics[width=0.35\textwidth]{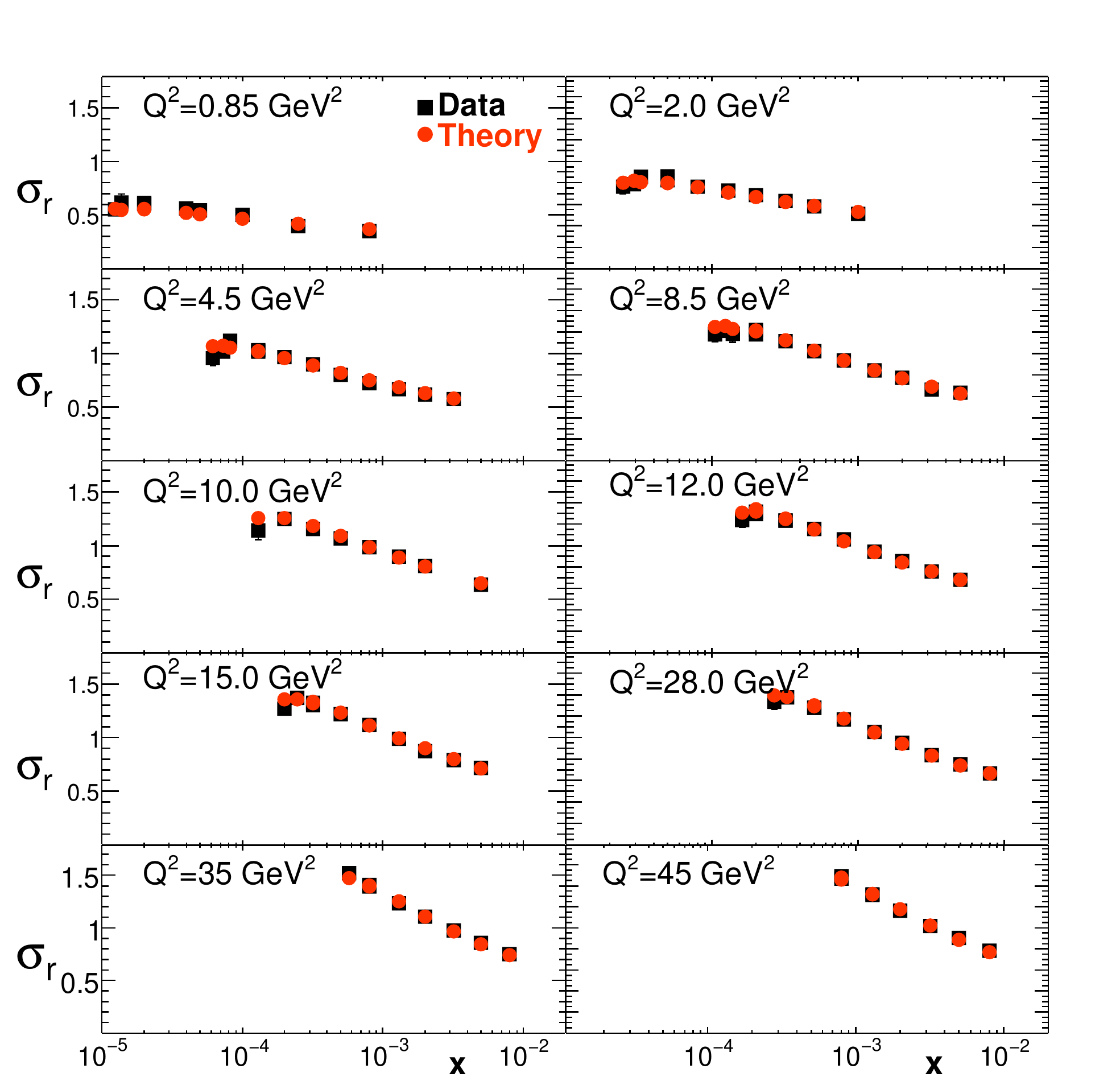}
\includegraphics[width=0.45\textwidth]{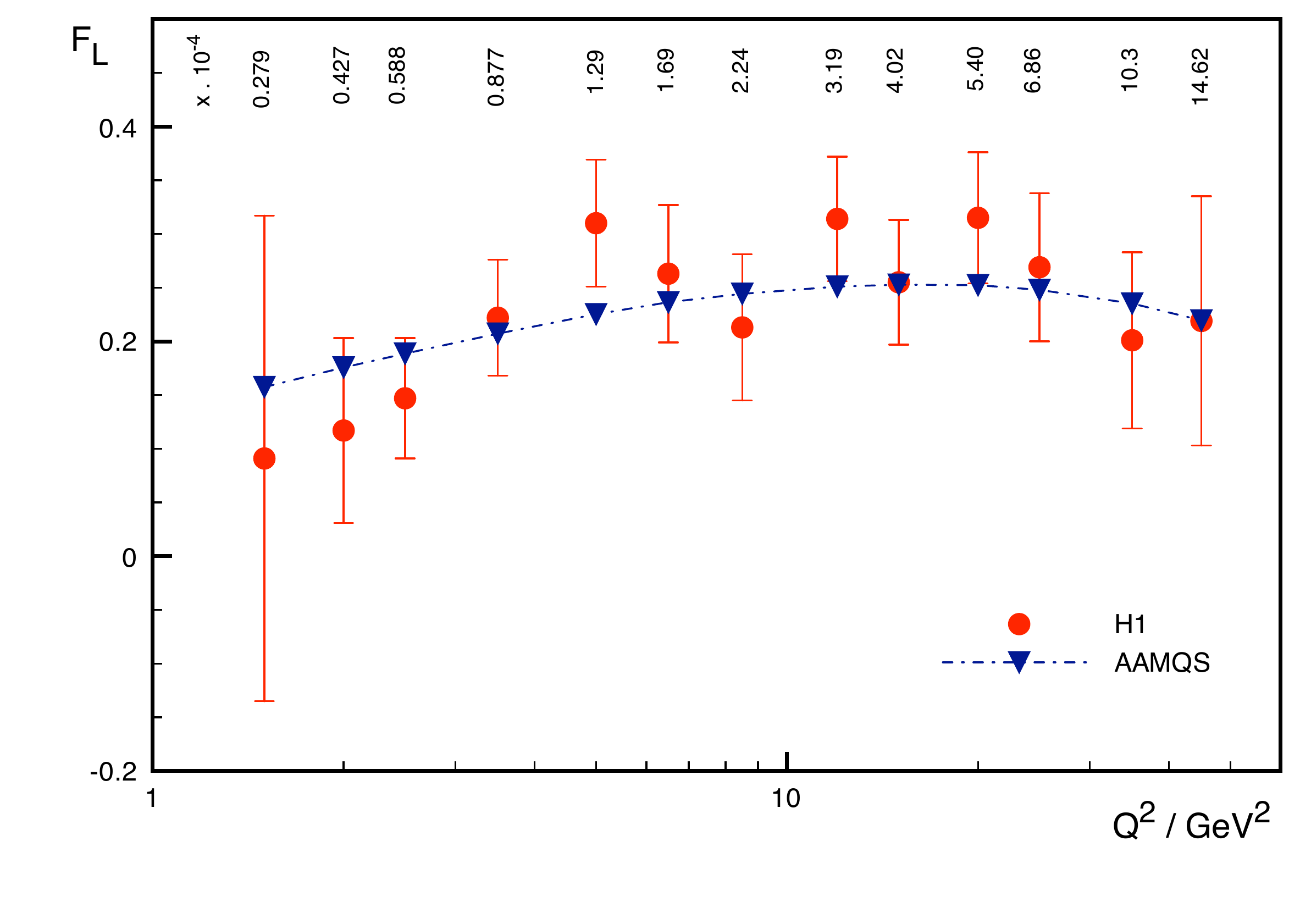}
\end{center}
\vspace*{-0.15cm}
\caption[a]{Left: rcBK fits to HERA data on the reduced cross section
  in e+p scattering at small-$x$ from~{\protect\cite{Albacete:2010sy}}. Right: Comparison of the latest $F_L$ data~{\protect\cite{Collaboration:2010ry}} (red dots) with the AAMQS calculation (blue triangles).}
 \label{fig:ep-ea}
\end{figure}

$\bullet$ {\bf rcBK}: An important step forward towards a theory driven analysis of the data was brought about by the AAMQS global
fits\cite{Albacete:2009fh,Albacete:2010sy}. The AAMQS approach relies in the use of the translationally invariant running coupling BK evolution equation (Eqs. (\ref{bkker}) and (\ref{bk})) to describe the small-$x$
dependence of the dipole scattering amplitude. Although several prescriptions have been proposed in the literature 
for the kernel of the BK equation at running coupling accuracy, it was shown in ref.~\cite{Albacete:2007yr} that Balitsky's prescription~\cite{Balitsky:2006wa} minimizes the role of higher-order, {\it conformal} corrections. Under such prescription the running coupling kernel reads
\be
\mathcal{K}^{run}_{\bf xyz}=\alpha_s(r^2)\,\left[\frac{r^2}{r_1^2r_2^2}+\frac{1}{r_1^2}\left(\frac{\alpha_s(r_1^2)}{\alpha_s(r_2^2)} -1\right) +\frac{1}{r_2^2}\left(\frac{\alpha_s(r_2^2)}{\alpha_s(r_1^2)} -1\right)  \right]
\label{kerbal}
\ee 
where $r_1\equiv|{\bf x-z}|$and  $r_2\equiv|{\bf z-y}|$. Using the kernel \eq{kerbal} in \eq{bk} instead of the fixed coupling kernel \eq{bkker} one gets the running coupling BK equation as used in the AAMQS fits. The
free parameters in the AAMQS fits mainly correspond to the free parameters in the initial conditions for the
evolution, as follows:
\begin{equation}
\mathcal{N}(r,x\!=\!x_0=0.01,b)=
1-\exp\left[-\frac{\left(r^2\,Q_{s0{\rm N}}^2\right)^{\gamma}}{4}\,
  \ln\left(\frac{1}{\Lambda\,r}+e\right)\right]\ ,
\label{ic}
\end{equation}
where, $\Lambda=0.241$~GeV, $Q_{s0{\rm N}}$ is the proton saturation
scale at the initial $x_{0}$ and $\gamma$ is a dimensionless parameter
that controls the steepness of the unintegrated gluon distribution for
momenta above the initial saturation scale $k_t>Q_{s0\rm N}$. 
The AAMQS fits were updated in \cite{Albacete:2010sy} to include the much more precise data from the combined analysis by the H1 and ZEUS collaborations\cite{Aaron:2009aa} as well as the contributions from heavy quarks. With this set up, the AAMQS fits provide a remarkably good description of both data on the inclusive, $F_2$ (\fig{fig:ep-ea} left) and longitudinal, $F_L$ (\fig{fig:ep-ea} right), structure functions.  Although the AAMQS fits clearly favor
values $\gamma\!>\!1$, they do not uniquely determine its optimal value
(and neither does the analysis of forward RHIC data performed in
ref.~\cite{Fujii:2011fh}). Rather, different pairs of $(Q_{s0{\rm
    N}}^{2},\gamma)$-values provide comparably good values of
$\chi^{2}/d.o.f\sim1$, see Table~\ref{tabfits}. The ``degeneracy'' is
due to correlations among parameters and also because HERA data is too
inclusive to constrain exclusive features of the proton UGD. Similar to the GBW model, the AAMQS fits consider the proton to be homogenous in the transverse plane. Equivalently the proton density profile is taken to be a step-function $T_p(b)=\theta(b-b_0)$, which allow the replacement of the integral over impact parameter by a global normalization constant $\int d^2b \rightarrow \sigma_0$, treated as another fit parameter. Another good fit to HERA data on inclusive structure functions has been recently presented in ref.~\cite{Lappi:2013zma}. There main novelty with respect the AAMQS fits is the choice of a new set of initial conditions: the anomalous dimension, $\gamma$ in \eq{ic} is kept to unity in and, in turn, the constant under the logarithm, $e$ is taken as another free parameter. Energy conservation corrections, which slow down further the evolution speed, were incorporated to rcBK fits in ref.~\cite{Kuokkanen:2011je}, providing also a good description of diffractive and inclusive data.

A recent comparative study of the stability of standard DGLAP and rcBK fits to small-$x$ HERA data under changing boundary conditions performed by systematically excluding subsets of data from the fits showed that rcBK fits do provide more robust fits to data than DGLAP ones~\cite{Albacete:2012xq}, thus providing an additional indication for the relevance of non-linear small-$x$ dynamics in the analysis of HERA data.
\begin{table}[htbp]
   \centering
   \begin{tabular}{|c|c|c|c|c|c|c|} 
\hline
   
     Set  & $ Q_{s0{\rm N}}^{2}$ (GeV$^{2}$)  & $\gamma$          \\          
 \hline
        MV    & 0.2 &  1 \\
        MV$^{\gamma_1}$       & 0.168 & 1.119  \\
        MV$^{\gamma_2}$      & 0.157  & 1.101 \\
       \hline
   \end{tabular}
\caption{\small Values of the three set of AMMQS parameters implemented in the rcBK Monte Carlo{\protect \cite{Albacete:2010ad,Albacete:2012xq}}. A complete list of AAMQS fit parameters can be found in {\protect\cite{Albacete:2009fh,Albacete:2010sy}}.}
   \label{tabfits}
\end{table}

All the approaches described above attain an excellent description of inclusive e+p data. 
Arguably, the main advantage of the IP-Sat or b-CGC models is the explicit impact parameter dependence of the amplitude which, unlike the GBW or rcBK approaches, allows the description of HERA data on exclusive diffractive processes. In turn, the rcBK approach is more solidly grounded in the CGC theoretical framework and provides a direct test on non-linear evolution equations, allowing a more detailed characterization of the proton UGD through the parameters $\gamma$ or $e$.  
Ideally, one would like to achieve a unified description of BK evolution and impact parameter dependence. 
Control over the impact parameter dependence is crucial in heavy ion collisions, as this new variable controls the initial size and shape on the interaction area and, hence, the initial energy deposited in the collisions.
However, the impact parameter dependence of the dipole amplitude is controlled by long- range, non-perturbative phenomena rooted in the physics of confinement and thus is not amenable to a perturbative description like the one encoded in the BK or B-JIMWLK equations\cite{Kovner:2001bh}. Attempts to solve the impact parameter dependent BK equation\cite{Berger:2011ew} through the inclusion of a non-perturbative gluon mass in the evolution kernel in order to mimic confinement effects show a strong dependence on the details of such regularization. Such insufficiency of the theory plus the lack of direct empiric information makes unavoidable some degree of modeling for the nuclear wave functions at small-$x$ in order to correctly account for confinement effects. 

However, it should be noted that although the saturation scale can be given an unambiguous definition, the physical object that carries full physical meaning is the hadron unintegrated gluon distribution. In the end, the transition from the dilute to the high-density regime does not occur sharply at some momentum scale --the saturation scale--. Rather, it is a gradual, smooth one that happens over a sizable range of transverse momenta (of the order of a few GeV) and is, therefore, sensitive to the shape of the full hadron UGD.

\begin{figure}[htb]
\begin{center}
\includegraphics[width=0.41\textwidth]{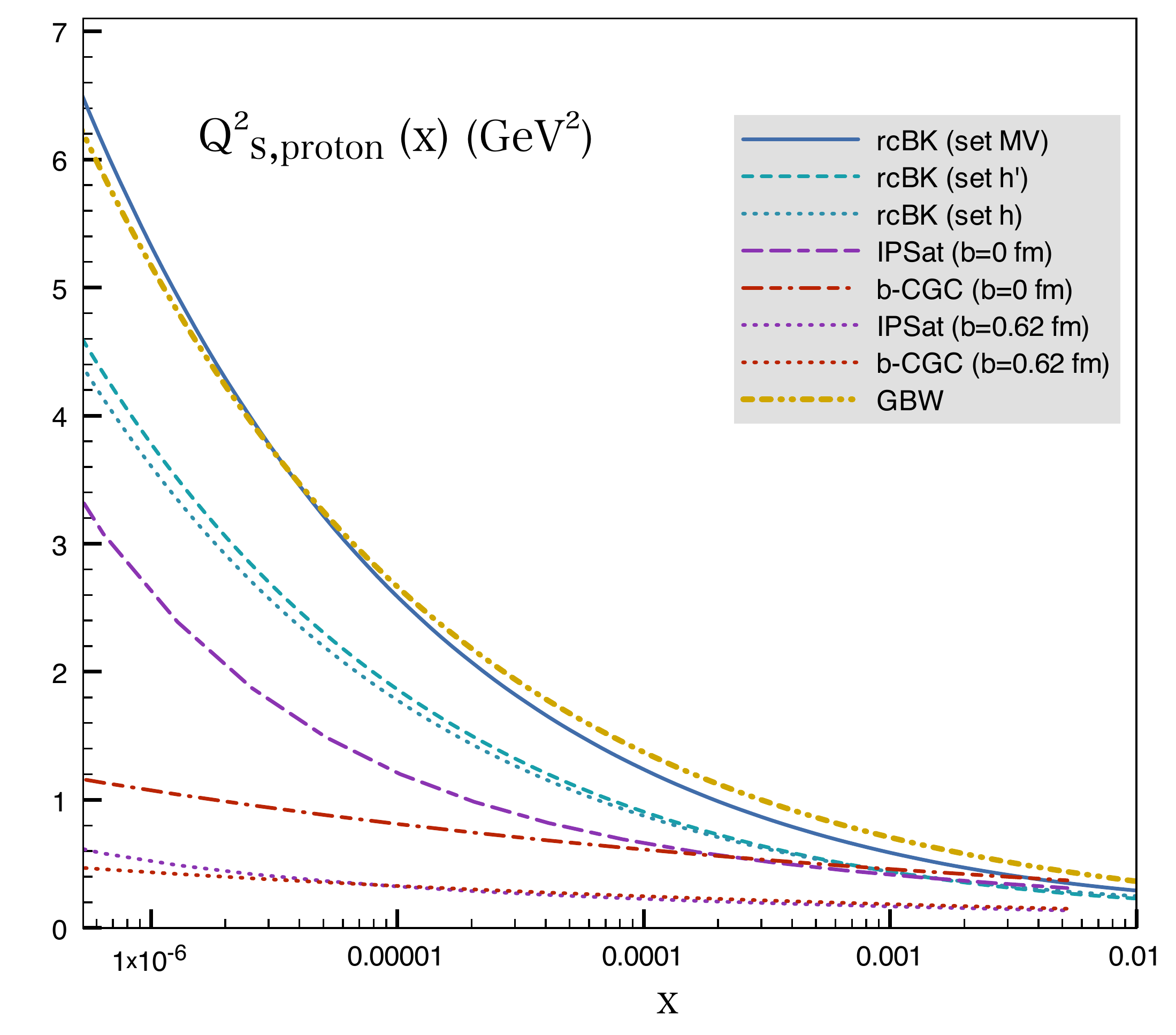}
\includegraphics[width=0.42\textwidth]{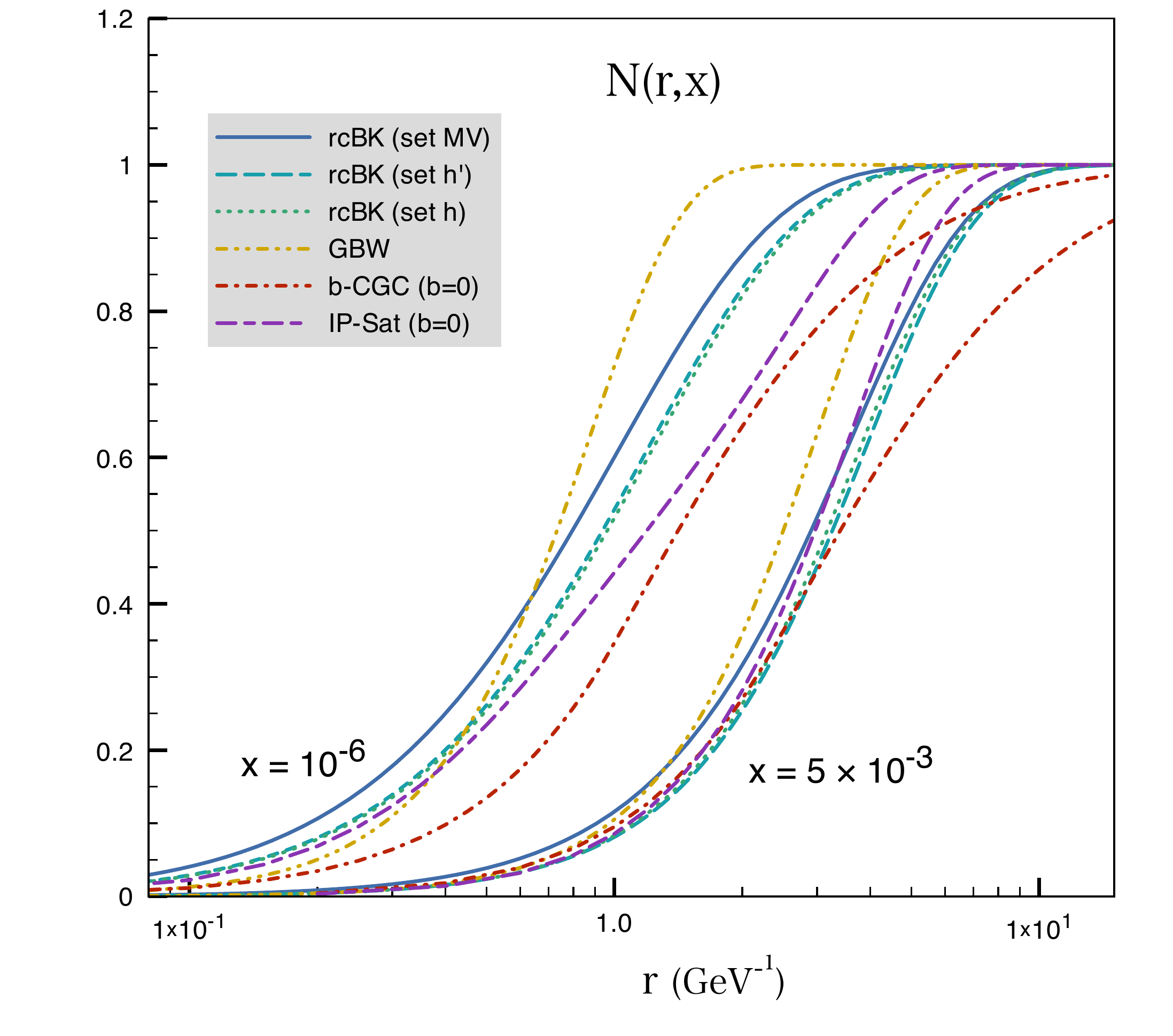}
\end{center}
\vspace*{-0.15cm}
\caption[a]{Left: Proton saturation scale extracted from the different models discussed in this section. Right: Dipole-proton scattering amplitude at two values of Bjorken-$x$: $5\times 10^{-3}$ and $10^{-6}$.}
\label{fig:qsat}
\end{figure}

\subsection{Models for the nucleus wave function} 

Here we briefly review how the models for e+p collisions described above are geometrically extended to build models for the nuclear wave function. We also indicate their relation to some of the most popular models to describe multi particle production in heavy ion collisions discussed later.
\begin{wrapfigure}{r}{0.35\textwidth}
\centering
    \includegraphics[width=0.25\textwidth]{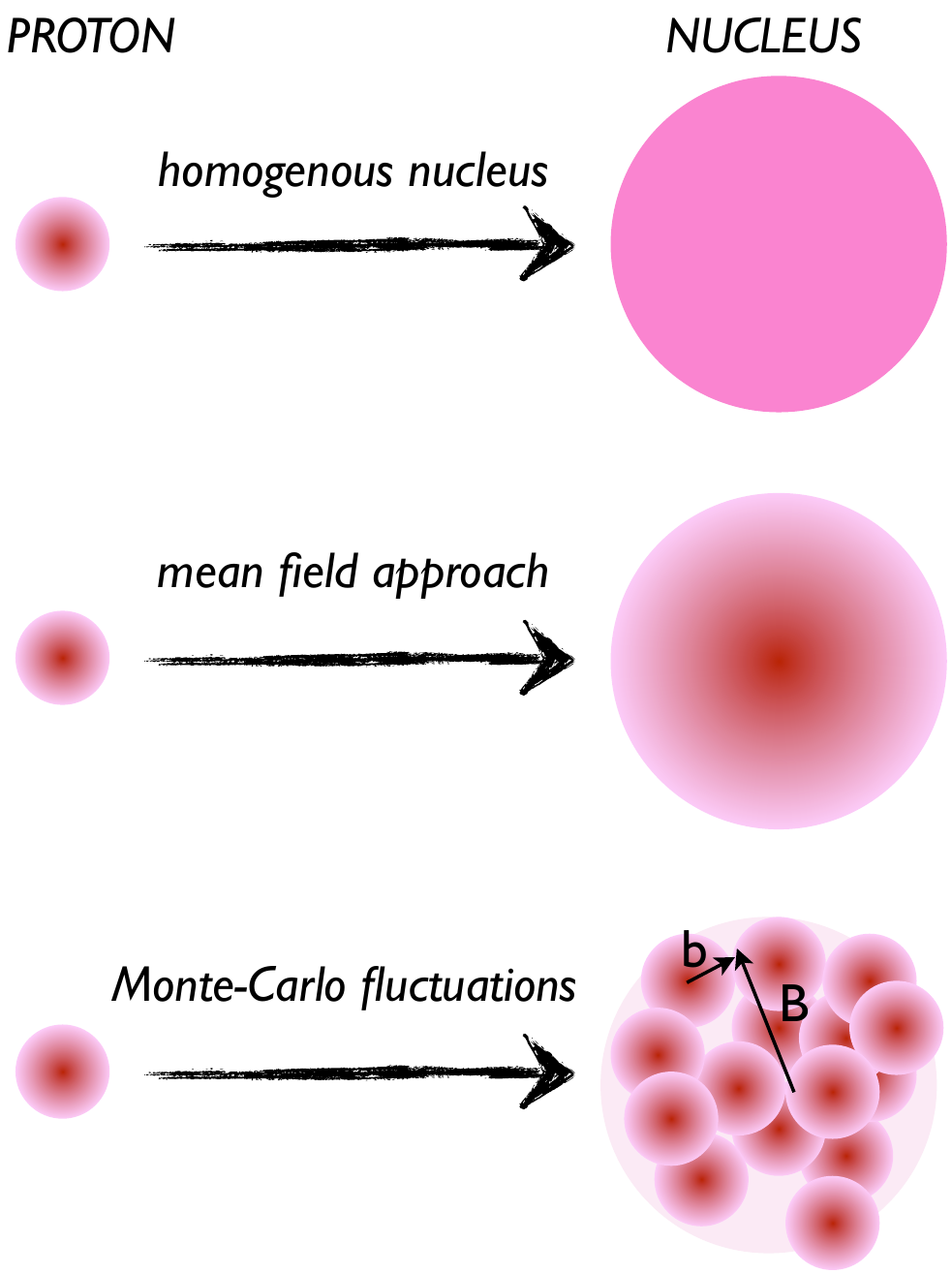}
    \hspace{-20pt}
    \label{fig:ptoA}
  \caption{Geometric extension of proton models to the nuclear case. }
\end{wrapfigure}

With the exception of the b-CGC model, in CGC approaches one normally
relates the b-dependence of the nuclear saturation scale to the proton
saturation scale and the local nuclear density: $Q_{sA}^{2}(B,x)\! =\!
T_{A}(B)\,Q_{s,\rm N}^{2}(x)$. This can be done at different levels of
refinement: assuming a homogeneous nucleus, a nucleus with a
Woods-Saxon density distribution $T_A(B)$ without fluctuations, or, finally,
also accounting for geometry fluctuations, as illustrated in \fig{fig:ptoA}. 
One popular phenomenological approach, first proposed in ref.~\cite{Armesto:2002ny} and recently used in ref.~\cite{Lappi:2013zma}, consists in the use of the 
optical Glauber model to generalize our dipole-proton amplitude (or cross-section) to dipole-nucleus scattering:
\be
\mathcal{N}^A(x,r,{\bf B})= 1-\exp\left[ -\frac{1}{2}AT_A({\bf B})\,\sigma_{dip}^{e+p}(x,r) \right]
\ee
This form is an average of the dipole cross section over the fluctuating positions of the nucleons in the nucleus.
In models that explictely accounting for geometry fluctuations, the position of nucleons in the transverse plane is sampled stochastically through Monte Carlo methods in strict analogy with the Monte Carlo Glauber model. The longitudinal position of the nucleons are integrated out and, in some cases a hard-core nucleon-nucleon repulsive potential to avoid full overlap between nucleons is assumed. Then, for each transverse configuration of the nucleons the local saturation scale of the nucleus is taken to be proportional to the number of overlapping nucleons in that transverse location $B$. 
Below we provide some more details on how some popular models for multi particle production in nuclear reactions 

\begin{itemize}
\item{KLN and MC-KLN}\cite{Kharzeev:2001yq,Kharzeev:2004if,Kharzeev:2007zt}: The works of  Kharzeev, Levin and Nardi pioneered the saturation based modelling of heavy ion collisions. The KLN model assumes a homogeneous nucleus of constant density in the transverse plane. The relation between the nucleon and nuclear saturation scale is taken to be $Q_{s\rm A}(x)^2\sim N_{part}\,Q_{s\rm N}^2(x)$, with the proton saturation scale given by the GBW parametrization \eq{gbw} and $N_{part}$ the number of participants in a nucleus-nucleus collision. In a pA collision $N_{part}\sim A^{1/3}$ but, in general it is not an intrinsic property of the nuclear wave function. Rather, it depends on the colliding object, i.e the other nucleus. In turn, the functional form of the nuclear UGD (equivalently, the dipole-nucleus cross section) is not taken from the GBW model but rather inspired in the expected generic features of saturation effects:
\be
\phi_{KLN}(x,k_\perp)\sim\frac{1}{Q_{sA}^2(x)}\frac{Q_{sA}^2(x)}{\text{max}\left\{ Q_{sA}^2(x),k_\perp^2 \right\}}
\ee
The KLN model was then updated in \cite{Drescher:2006ca,Drescher:2007ax} in order to include geometrical fluctuations, leading to the NC-KLN model. 
 
\item{rcBK and MC-rcBK}:  The MC-rcBK\cite{Albacete:2010ad,Albacete:2012xq} was built as an upgrade of the original MC-KLN model in order to replace the KLN UGD's by the rcBK ones. Thus, the geometric set up runs exactly parallel in both cases. However, there is one main difference between the two: In the MC-rcBK approach a transverse distribution of initial saturation scales is generated (see \eq{ic}). Then, every point in the transverse plane is assumed to evolve independently through translationally invariant rcBK evolution. This is contrast to the original rcBK\cite{Albacete:2007sm} fit to RHIC multiplicities, where a homogeneous nucleus was assumed $Q_{sA}^2\approx A^{1/3}Q_{s\rm N}^2$ and then evolved as a whole entity. Both procedures led to significantly different predictions for the total multiplicities in Pb-Pb collisions at the LHC. This is because initially dilute systems evolve faster than denser ones due to the enhanced role of non-linear terms in the latter case and to scaling violation terms induced by the running coupling. This immediately implies a non-commutativity of geometrical averaging and small-$x$ evolution. This is an important physical effect that is further enhanced by fluctuations and, as we discuss later, may have relevant observable consequences. 
The presence of anomalous dimension $\gamma>1$ in rcBK UGD sets poses an additional difficulty in modeling the nuclear wave function. The {\it natural} prescription $Q_{sA}^{2}(B)\! =\!
T_{A}(B)\,Q_{s,\rm N}^{2}$ may lead to an inconsistent definition of the nuclear modification factor in the limit of large transverse momentum because additivity of the dipole scattering amplitude in the number of nucleons, $N(r;x_0,B) \sim T_A(B)$ at small $r$, is violated. To fix this issue another possibility to relate the nucleon and nuclear saturation scales was considered: $ Q_{sA}^{2}(B)\! =\!
T_{A}(B)^{1/\gamma}\,Q_{s,\rm N}^{2}$   

\item{CYM and IP-Glasma}: Unlike all the previously described approaches, in the CYM and IP-Glasma models the relevant object is not the nuclear UGD. Rather knowledge of the full probability distribution $W_Y[\rho]$ of color charges within the nucleus is needed.  The IP-Glasma model, relies in an extension of the Gaussian MV model (or, equivalently, the Glauber-Mueller model) to arbitrarily small values of $x$, hence effectively incorporating small-$x$ evolution in an ad-hoc way. There, the color charge squared per unit area in \eq{WMV} is replaced by a $x$ and $b$-dependent one $\mu_A^2\rightarrow \mu_A^2(x,x_\perp,b_\perp)$. Similar to MC-KLN or MC-rcBK, the positions of nucleons in the transverse plane $b_{A\perp}$ are stochastically sampled trough a Woods-Saxon distribution. Then,  $\mu_A^2(x,x_\perp,b_\perp)$ is obtained by summing the individual contributions of every nucleon $\mu_p^2(x,b_{p\perp})$, where $b_{p\perp}$ refers to the transverse distance to the center of the contributing nucleon. Finally $\mu_p^2(x,b_{p\perp})$ is made proportional to the saturation scale obtained  from the IP-Sat model through the condition $ \mathcal{N}_{IPSat}(x,r_{s\perp}\equiv 2/Q_{sp},b_{p,\perp})=1-e^{-1/2}$. It should be noted that there is an overall logarithmic uncertainty in the exact numerical factor relating $\mu_p^2(x,b_{p\perp})$ and the saturation scale $ Q_{sp}(x,b_{p\perp})$ stemming from the IP-Sat model throughout the previous condition may depend on details of the calculation\cite{Lappi:2007ku}. This is due to the fact that the correlators of Wilson lines in the MV model do not match exactly with those yielded by the Glauber-Mueller ansatz \eq{ipsat}.

\end{itemize}

\subsection{e+A collisions}

The predictive power of the CGC depends on the level of accuracy of the calculations (leading-order vs. next-to-leading order), but also on the amount of non-perturbative inputs needed (initial conditions to the small-$x$ evolution, impact parameter dependence, and possibly other inputs specific to some observables). Electron-ion (e+A) collisions provide the best option to reduce these uncertainties. We shall describe in the rest of the review how the three CGC-inspired models of the nuclear wave function introduced above are used to describe p+A and A+A collisions. However, one should keep in mind that eventually, e+A collisions will be needed to put final constraints on these models.

Indeed, there are several aspects that proton-nucleus collisions cannot investigate to a satisfactory degree. For instance, this is the case of the impact parameter dependence of the nuclear gluon distribution and of the saturation scale, which has always been the main non-perturbative input in CGC calculations. As described above, what is done in the most advanced CGC models, is to treat the nucleus as a collection of Woods-Saxon distributed CGCs, and to evolve (down in x) the resulting gluon density at different impact parameters independently. Conceptually that is not good enough, and e+A collisions are crucial to address this issue by precisely imaging the transverse structure of nuclei at small-$x$.

The Electron Ion Collider (EIC) and Large Hadron electron Collider (LHeC) are two complementary proposals for the next decade. The EIC has the versatility to study the atomic number dependence while the LHeC can reach lower $x$ values. Complete details about these two proposals are given in the EIC White Paper \cite{Accardi:2012qut} and in the LHeC Conceptual design report \cite{AbelleiraFernandez:2012cc}. The standard deep inelastic scattering (DIS) measurements that have been performed in e+p collisions at HERA can be performed in e+A at these facilities, as summarized in \cite{Marquet:2012tb}.

In particular, hard diffractive events off nuclei would be measured for the first time. Such processes were a surprising QCD feature at HERA: a proton in its rest frame hit by a 25 TeV electron remains intact 10\% of the time. Interestingly enough, they are naturally understood in QCD when non-linear evolution is taken into account, as they are actually subject to strong non-linear effects even for $Q^2$ values significantly bigger than $Q_s^2$ \cite{GolecBiernat:1999qd,Marquet:2007nf}. In e+A collisions the amount of diffractive events are expected to be a smoking gun for parton saturation \cite{Frankfurt:2003gx,Kugeratski:2005ck,Kowalski:2008sa}. 

Another particularly interesting process is diffractive vector meson production (or deeply virtual Compton scattering when the produced vector particle is a photon) which provides access to the spatial distributions (and correlations) of partons in the transverse plane. It is one of the few observables for which using an impact-parameter averaged saturation scale is not enough, already at the proton level \cite{Kowalski:2006hc,Marquet:2007qa,Rezaeian:2013eia}. In e+A collisions, it would provide the transverse imaging needed to constrain the initial state models \cite{Caldwell:2010zza,Lappi:2010dd,Horowitz:2011jx,Toll:2012mb}. Note that there is no such direct access to the spatial gluon distribution in p+A collisions.

Finally, di-hadron production in e+A collisions can be measured without final-state flow ``contamination'' (as seems to be the case in p+A collisions at the LHC, see below). The best experimental evidence of parton saturation so far has been observed at RHIC, looking at the azimuthal angle dependence of the correlation function of forward di-hadrons: the disappearance of the away-side peak in central d+Au collisions compared to p+p collisions (see below). Qualitatively, the effect will be similar in e+A collisions, but that at the quantitative level, the di-hadron production process in e+A collisions involves a different operator definition of the unintegrated gluon distribution, and therefore is complementary to the RHIC measurement in p+A collisions \cite{Dominguez:2011wm}.

\section{Transverse momentum spectra and correlations in p+p and p+A collisions}
\label{sec:3}

Proton nucleus collisions provide a better handle on initial state and
gluon saturation effects than those of heavy ions. This is mainly due
to the absence of strong final state effects induced by the presence of a
QGP. There are two distinct but related approaches to hadron
production in high energy asymmetric collisions such as proton-nucleus or very
forward proton-proton and nucleus-nucleus.  In those
collisions, particle production processes in the central rapidity
region probe the wave function of both projectile and target at small
values of $x$ and can be described in terms of the $k_t$-factorization
formalism where the differential cross-section for inclusive gluon
production in A+B collisions is given by~\cite{Kovchegov:2001sc} \bea
\frac{d\sigma^{A+B\to g}}{dy\, d^2p_{t}\, d^2R} &=& K^{k}\,
\frac{2}{C_F} \frac{1}{p_t^2} \int^{p_t} \frac{d^2k_t}{4}\alpha_s(Q)
\nonumber\\ & & \int d^2b\, \varphi\left(\frac{|p_t+k_t|}{2},x_1;b\right)\,
\varphi\left(\frac{|p_t-k_t|}{2},x_2;R-b\right)~,
\label{kt2}
\eea 
where $y$, $p_{t}$ and $R$ are the rapidity, transverse momentum
and transverse position of the produced gluon,
$x_{1(2)}=(p_{t}/\sqrt{s_{NN}})\exp(\pm y)$ and
$C_F=(N_c^2-1)/2N_c$. The upper bound in the $k_t$-integral in \eq{kt2} is introduced ad-hoc in order to regulate the divergence $p_t\to 0$ that would otherwise yield infinite multiplicities after $p_t$-integration.  At the degree of accuracy of its original derivation, leading logarithmic in $\alpha_s \ln (1/x)$, the strong coupling in \eq{kt2} is fixed. However, in phenomenological applications the coupling is often evaluated at some scale $Q$, the precise value thereof varying among models.
The relevant UGD's for the $k_{t}$-factorization
formula above are given by 
\begin{equation}
\varphi(k,x,{\bf R})=\frac{C_F}{\alpha_s(k)\,(2\pi)^3}\int d^2{\bf r}\
e^{-i{\bf k}\cdot{\bf r}}\,\nabla^2_{\bf r}\,\mathcal{N}_A(r,x,{\bf R})\,
\label{phi}
\end{equation}
where $\mathcal{N}_{A}$ is the dipole amplitude in the adjoint
representation which, in the large-$N_c$-limit, is related to the
dipole amplitude in the fundamental representation by
$\mathcal{N}_A(r,x,{\bf R})=2\,\mathcal{N}_{F}(r,x,{\bf
  R})-\mathcal{N}_{F}^2(r,x,{\bf R})$.

However, at more forward rapidities the proton is probed at larger
values of $x$ while the target nucleus is deeper in the small-$x$
regime. In that context $k_{t}$-factorization fails to grasp the
dominant contribution to the scattering process. Rather, the {\it
  hybrid} formalism proposed in \cite{Dumitru:2005gt} is more
appropriate. In the hybrid formalism the large-$x$ degrees of freedom
of the proton are described in terms of usual parton distribution
functions (pdf's) of collinear factorization with a scale dependence
given by DGLAP evolution equations, while the small-$x$ glue of the
nucleus is still described in terms of its UGD. At leading order,
inclusive particle production is given by:
\bea
\frac{dN^{pA\to hX}}{d\eta\,d^2k}=\frac{K^h}{(2\pi)^2\!\!}\int_{x_F}^1\frac{dz}{z^2}\, \left[\sum_{q}x_1f_{q/p}
(x_1,Q^{2})\tilde{N}_F\left(x_2,\frac{p_t}{z}\right)D_{h/q}(z,Q^{2}) \right.\nonumber\\ +\left. 
x_1f_{g/p}(x_1,Q^{2})\tilde{N}_A\left(x_2,\frac{p_t}{z}\right)D_{h/g}(z,Q^{2})\right] 
\label{hybel}\,,
\eea
where 
\begin{equation}
\tilde{N}_{F(A)}(k,x,{\bf R})=\int d^2{\bf r}\;
e^{-i{\bf k}\cdot{\bf r}}\left[1-\mathcal{N}_{F(A)}(r,x,{\bf R})\right].
\label{phihyb}
\end{equation}
The K-factors $K^k$ and $K^h$ in  Eqs. (\ref{kt2}-\ref{hybel}) are not the result of any calculation. They have been added by hand to account for higher order corrections and, in practice, to adjust the normalisation of theoretical curves to experimental data in phenomenological works. Ideally, they should be equal to unity.

Clearly, the
{\it hybrid} formalism is restricted to production of particles with
high enough transverse momentum, as implied by its use of collinear
factorized pdf's and fragmentation functions. Unfortunately, no
smooth matching between the $k_{t}$-factorization and {\it hybrid}
formalisms is known to date.  Also, their corresponding limits of
applicability --equivalently the precise value of $x$ at which one
should switch from one to the other-- have only been estimated on an
empirical basis (in most analyses it is taken to be $x_0\approx 10^{-2}$).

Running coupling corrections to the $k_t$-factorization formula have been calculated in ref.~\cite{Horowitz:2010yg}, although they have note been yet implemented in phenomenological works. Partial and full next-to-leading corrections to the hybrid formalism have been calculated
in~\cite{Altinoluk:2011qy,Chirilli:2011km,Chirilli:2012jd}. These works showed that the factorized form of \eq{hybel} holds at one loop order, since the arising collinear divergences can be factorized into the splittings of the parton distribution from the incoming nucleon and the fragmentation function for the final state hadron, while the rapidity divergence at one-loop order is factorized into the BK evolution for the dipole gluon distribution of the nucleus. 
These new contributions, dubbed as {\it inelastic} corrections in ref.~\cite{Altinoluk:2011qy}, account for the finite pieces of DGLAP splitting and are parametrically of order $\ln(k_t/Q_s)$, hence expected to be important at high transverse momenta $k_t\gg Q_s$. 
In principle these corrections should bring the calculation of particle production into agreement with the standard perturbative result at large transverse momentum. However, and somewhat unexpectedly, first phenomenological studies\cite{Albacete:2012xq,Stasto:2013cha} indicate that NLO corrections become negative at high $p_t$, and in
fact dominate over the leading order result in some kinematic regions. This suggest that calculations beyond NLO accuracy or performing effective partial resummations at high $p_t$ may be needed in order to stabilize the perturbative series.

\subsection{Single Inclusive}

\subsubsection*{Single inclusive hadron distributions}
Here we discuss the phenomenological works for the description of single inclusive hadron distributions in p+p collisions at Tevatron and LHC energies and in d+Au collisions at RHIC.  We start by discussing the results obtained in the framework of $k_t$-factorization \eq{kt2}. As shown in \fig{ppspectra} mid-rapidity single inclusive hadron distributions measured in p+\={p} and p+p collisions at Tevatron and the LHC respectively are well described within the LO $k_t$-factorization formalism using the rcBK ugd's extracted from fits to HERA data up to intermediate transverse momenta, $p_t\lesssim 10$ GeV. 
The theoretical curves correspond to ref.~\cite{Albacete:2012xq} and are obtained folding the single gluon distribution in \eq{kt2} with the appropriate fragmentation function:
\be
\frac{dN^{A+B\to hX}}{dy\, d^2p_{t}} = \int d^2R \int \frac{dz}{z^2}\,D_g^h\left( z=\frac{p_t}{k_t}, \mu\right)\frac{1}{\sigma_s}\frac{d\sigma^{A+B\to gX}}{dy\, d^2p_{t}\, d^2R}\,
\ee
where $\sigma_s$ is the effective interaction area and $\mu$ the factorisation scale, taken to be proportional to the transverse momentum of the produced hadron.
The different curves in \fig{ppspectra} correspond to the use of different proton ugd sets in \eq{kt2} and to a thorough exploration of uncertainties: factorisation and running coupling scales, choice of fragmentation functions, ugd sets etc.  Importantly, ugd sets with the steeper initial gluon spectrum (UGD sets $h$ and $h'$ in Table 2.) lead to significantly better agreement with the data than the classical MV model initial condition with $\gamma = 1$ (dotted curves in \fig{ppspectra}). This result gives an important consistency check of the rcBK analyses of e+p data, where initial conditions with anomalous dimension $\gamma>1$ provided a better description of data. Further, since in hadronic collisions $x \sim p_t$ for spectra at fixed rapidity, the shape of the spectra does provide a direct test of the rcBK evolution speed. The correct description of p+p spectra is also important because they enter in the calculation of nuclear modification factors, a key observable for the discussion of initial state effects. 
 
Next, we discuss the analyses of RHIC forward data based on the {\it hybrid} formalism \eq{hybel}. Until the start of operation of the LHC this were the only data providing direct access to the nuclear wave function at small-$x$ and have being the subject of multiple analyses, not all of them covered here.  Some recent works have been able to provide a good description of both p+p and d+Au RHIC forward data, as shown in \fig{dAuforward}. rcBK evolved ugd's were employed for the first time in ref.~\cite{Albacete:2010bs}, although under the assumption of homogeneous nuclear density and no clear connection to DIS analyses. Further, a K-factor $K^h=0.4$ was needed to adjust the normalization of the most forward neutral pion distribution at $\eta=4$.
This work was then improved by the use ugd sets constrained by HERA data either under the IP-Sat model\cite{Tribedy:2011aa} or the rcBK \cite{Lappi:2013zma} model, and also to include a more detailed treatment of the nuclear geometry~\cite{Albacete:2012xq}.  Contrary to p+p data at LHC mid-rapidities, forward RHIC spectra can be relatively well described with either MV-like or $\gamma>1$ initial conditions for rcBK evolution both in p+p and d+Au collisions. This probably indicates that, due to the vicinity of RHIC forward data to the kinematic limit $x\to1$, the shape of the spectra in this region is dominated by the scale dependence of the large-$x$ d.o.f of the of the projectile, i.e its parton distribution function, rather that by the exclusive features of the used ugd set.
\vspace*{-0.05cm}
\begin{figure}[htbp]
\begin{center}
\includegraphics[width=0.50\textwidth]{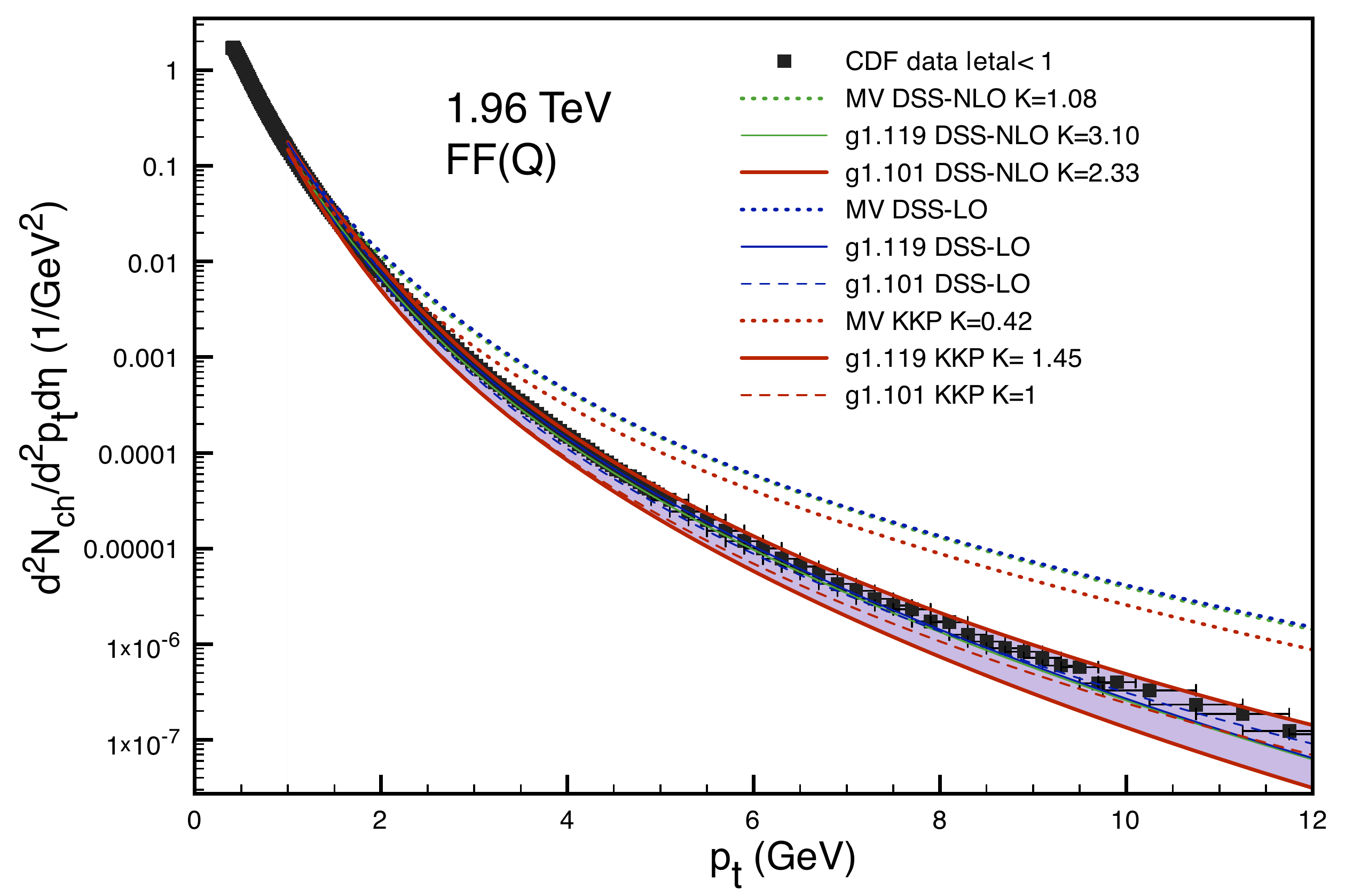}
\includegraphics[width=0.48\textwidth]{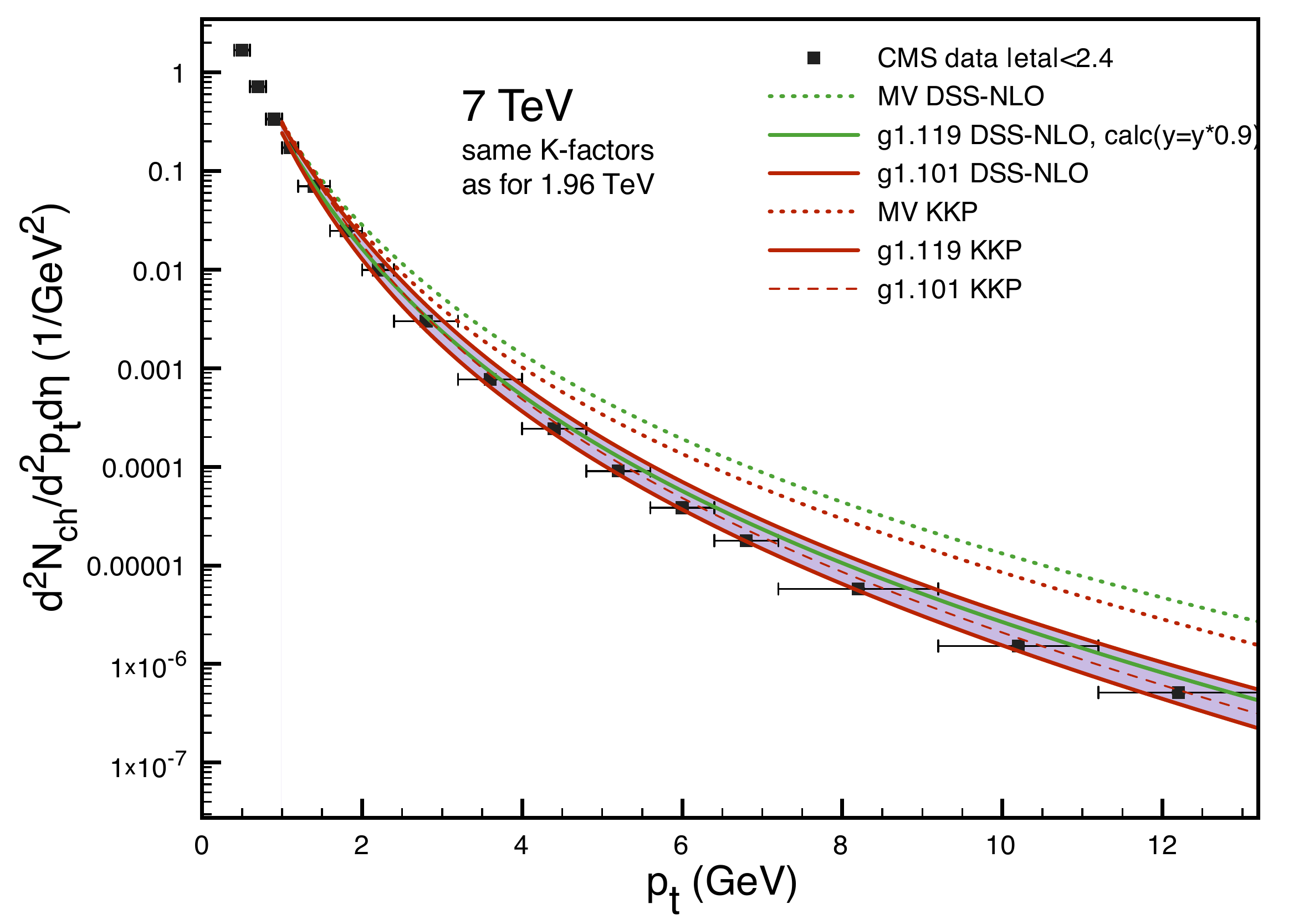}
\end{center}
\vspace*{-0.15cm}
\caption{Transverse momentum distribution of charged particles in the central rapidity region of p+p collisions at $\sqrt{s} = 1.96$ TeV (left), and p+p collisions at $\sqrt{s} = 7$ TeV (right). CDF and CMS data from refs.~{\protect\cite{Aaltonen:2009ne,Khachatryan:2010us}}. }
\label{ppspectra}
\end{figure}
\vspace*{-0.05cm}

{\it Inelastic} corrections and full NLO corrections have been implemented in refs.~\cite{Albacete:2012xq,Stasto:2013cha} respectively. In ref.~\cite{Albacete:2012xq} the inelastic corrections were evaluated with either a fixed or running coupling for the {\it inelastic} term. In this work it was observed that the {\it inelastic} corrections exhibit a harder $p_t$-dependence than the LO contribution, and at some intermediate transverse momentum it overwhelms the LO contribution, the crossing point depending on the particular ugd set employed. This critical $p_t$-value increases with increasing rapidity of the produced hadron, thus thus justifying the validity of the calculation for the very forward region.
Also, the relative weight of the inelastic term depends on the collision system or, equivalently, on the target saturation scale: it is stronger for p+p than for d+Au collisions. These conclusions on the instability of higher order corrections at moderate rapidities were confirmed through the implementation of full NLO corrections performed in ref.~\cite{Stasto:2013cha}. There, again, the best description of data was achieved with the rcBK ugd sets. The bands in \fig{dAuforward} correspond to varying factorization scales $\mu^2= 10$ to 50 GeV$^2$.
\begin{figure}[htbp]
\begin{center}
\hspace*{0.8cm}
\includegraphics[width=0.43\textwidth]{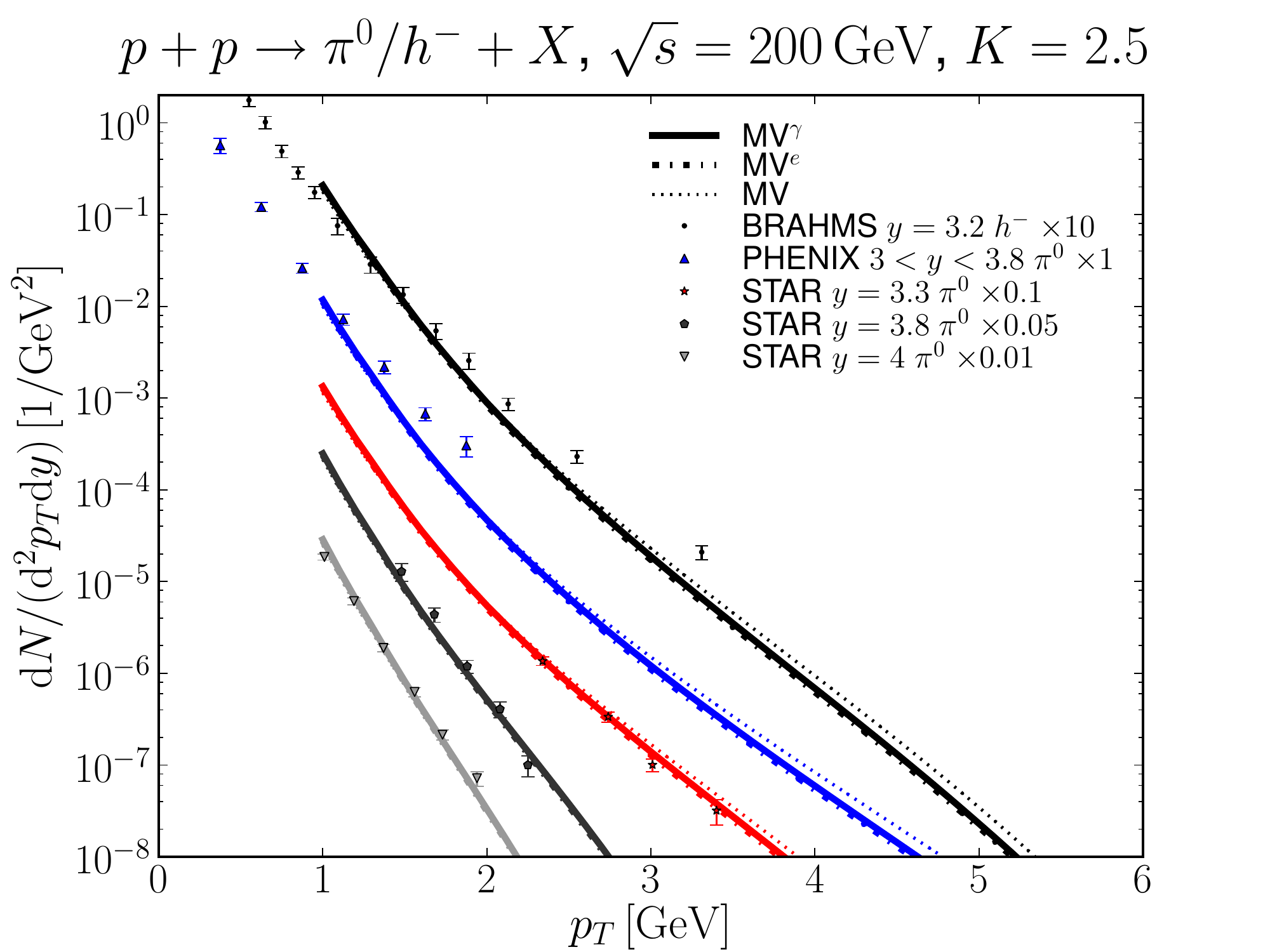}
\hspace*{-0.2cm}
\includegraphics[width=0.41\textwidth]{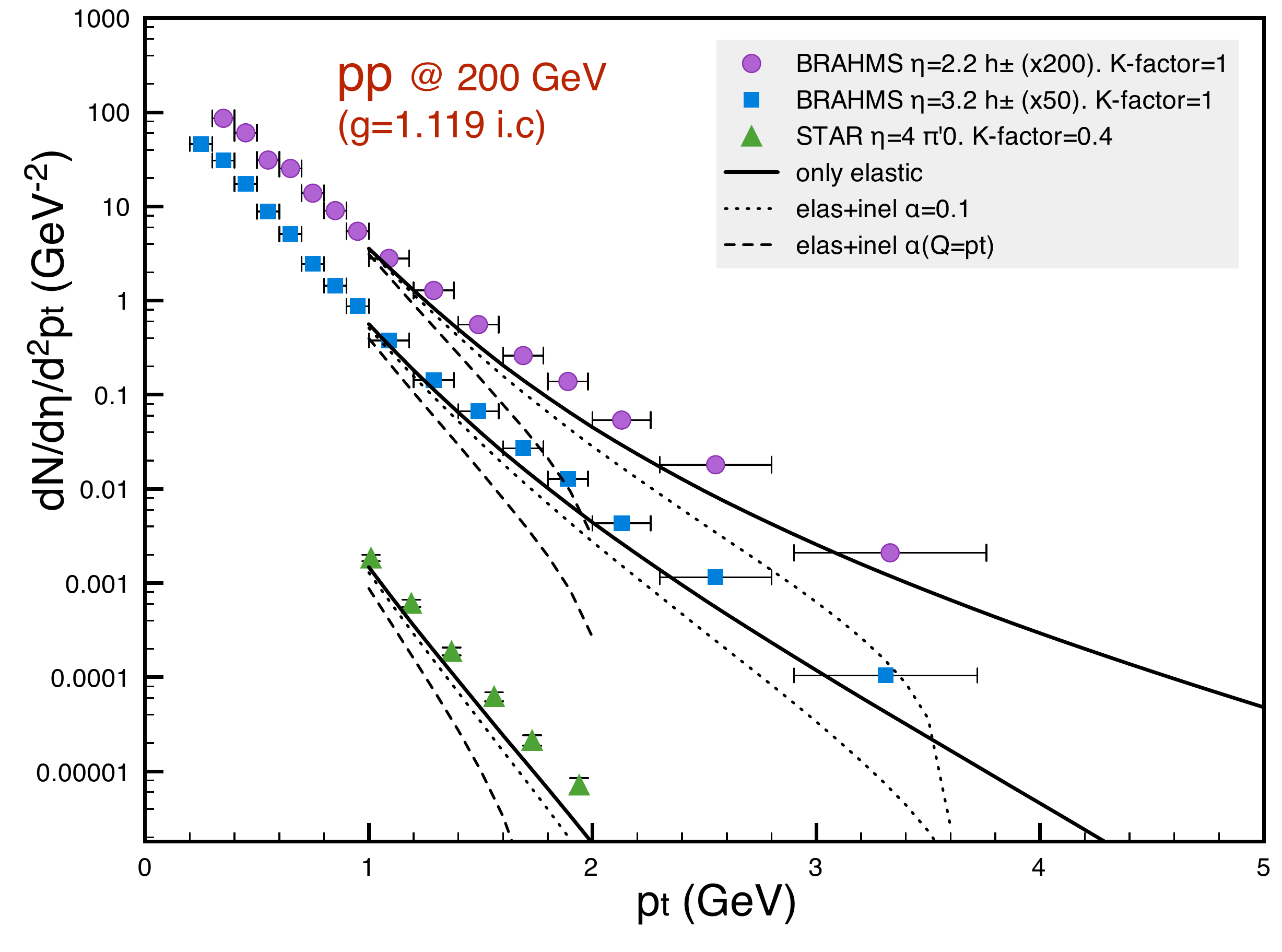}
\includegraphics[width=0.43\textwidth]{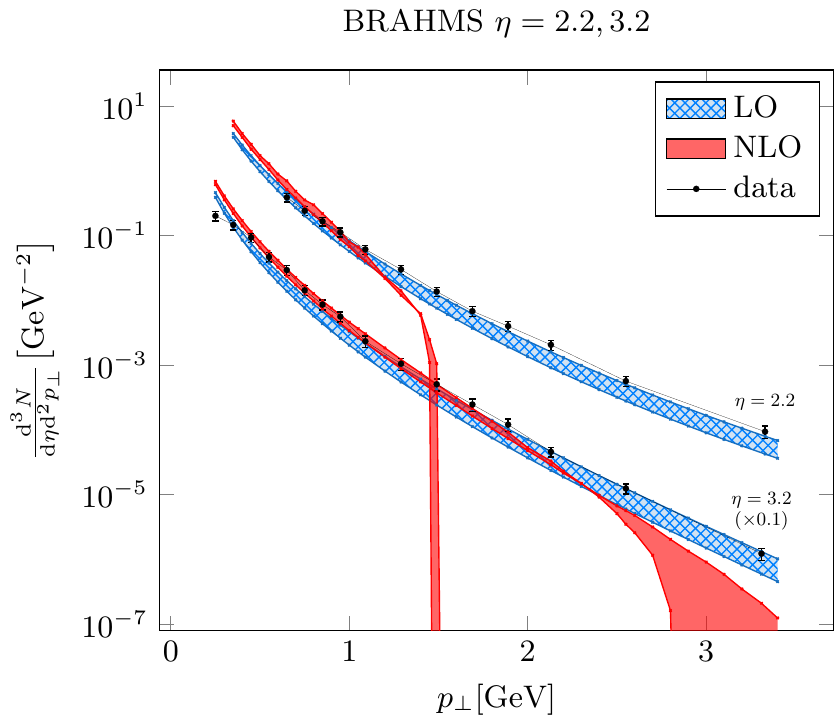}
\includegraphics[width=0.43\textwidth]{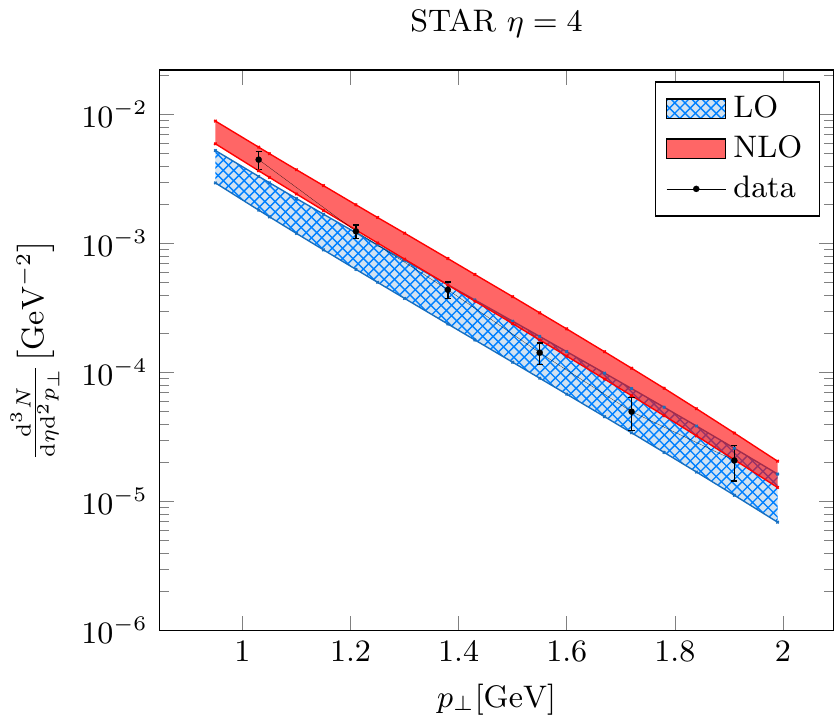}
\end{center}
\vspace*{-0.15cm}
\caption{Negative charged hadrons and neutral pion distribution in RHIC p+p collisions (top panel) and d+Au collisions (bottom panel) for different pseudorapidities (from $\eta =2.2$ to $\eta=4$) compared to the results from the hybrid formalism at LO (top-left{\protect \cite{Lappi:2013zma}}), LO+inelastic corrections (top-right{\protect \cite{Albacete:2012xq}}) and LO and full NLO (bottom pannel{\protect \cite{Stasto:2013cha}}). BRAHMS data from ref.~{\protect \cite{Arsene:2004ux}}, STAR data from ref.~{\protect \cite{Adams:2006uz}} and PHENIX data from ref.~{\protect \cite{Adare:2011sc}}}.
\label{dAuforward}
\end{figure}

\subsubsection*{Nuclear Modification factors}
In relativistic heavy-ion collisions, nuclear effects on single particle production are typically evaluated in terms of ratios of particle yields called nuclear modification factors:
\begin{equation}
R^h_{pA}=\frac{dN^{pA\to hX}/dyd^2p_\perp}{N_{coll}\ dN^{pp\to hX}/dyd^2p_\perp}\ ,
\end{equation} 
where $N_{coll}$ is the number of nucleon-nucleon collisions in the p+A collision. If high-energy nuclear reactions were a mere incoherent superposition of nucleon-nucleon collisions, then the observed $R_{pA}$ would be equal to unity. Before forward RHIC data was available, studies based on the numerical solutions of the BK equation\cite{Albacete:2003iq} indicated that the enhancement of non-linear QCD corrections in the evolution of nuclear ugd's relative to those of proton ugd's should cause an suppression of nuclear modification factors in the forward rapidity region. This expectation was then confirmed by data, thereby supplying a first clear indication of the relevance of saturation effects in nuclear reactions. Indeed, RHIC data for $R_{dAu}$ exhibit two opposite regimes: at mid rapidities the nuclear
modification factors show an enhancement in particle production at intermediate
momenta $|p_t|\sim 2 \div 4$ GeV --commonly referred to as {\it Cronin} peak-- and approach unity smoothly at higher transverse momenta. In turn, a suppression at smaller
momenta is observed. Mid-rapidity data has been successfully described through different formalisms and techniques as leading-twist perturbation theory\cite{Eskola:2009uj}, Glauber-like resummation of multiple scatterings\cite{Accardi:2003jh} or CGC approaches\cite{Kharzeev:2004yx}. Therefore, it is difficult to extract any clean conclusion about the physical origin of the Cronin enhancement. This is probably due to the kinematic region probed in these measurements. For a hadron momentum of $p_t \sim 2$ GeV, one is sensitive to $x_A \sim 0.01 \div 0.1$. In this region different physical mechanisms concur, so neither of the physical assumptions underlying the approaches above is completely fulfilled.

However, at forward rapidities the Cronin enhancement
disappears, turning into an almost homogeneous suppression at high transverse momenta, as predicted in refs.~\cite{Albacete:2003iq, Kharzeev:2002pc,Kharzeev:2003wz}. Indeed, the accurate description description of both p+p and d+Au forward yields described above allows for a good quantitative description of $R_{dAu}$ within the CGC framework.  
However, it has been argued that the observed suppression of particle production at forward rapidities is not an effect associated to the small values of $x_A$ probed in the nuclear wave function, but rather to energy-momentum conservation corrections relevant in the proximity of the kinematic limit $x_F\to 1$\cite{Kopeliovich:2005ym,Frankfurt:2007rn}. Such corrections are not present in the CGC, built upon the eikonal approximation (this may explain why a K-factor is needed to describe the suppression of very forward pions). The energy degradation of the projectile parton, neglected in the CGC, through either elastic scattering or induced gluon brehmstralung would be larger in a nucleus than in proton on account of the stronger color fields of the former, resulting in the relative suppression observed in data. A successful description of forward ratios based on the energy loss calculation was provided in ref.~\cite{Kopeliovich:2005ym}.

Presently the attention focuses in the analyses of data from the LHC p+Pb run at collision energy $\sqrt{s_{NN}}=5.02 $ TeV. Simple kinematical arguments suggest that the degree of suppression expected in mid-rapidity at the LHC due to CGC effects should be roughly equal to the one observed at forward rapidity at RHIC.  However, after a closer look at the kinematics it is clear that single inclusive
hadron production at mid-rapidity at the LHC does not probe the same values of
$x_A$ in the nucleus as forward hadron production at RHIC. The minimal values of
$x_A$ are indeed similar, $x_A\sim 10^{-3}\div 10^{-4}$. However, the mean value is roughly given by $x_A/\langle z\rangle $, where
$\langle z\rangle$ is the average momentum fraction carried by a final state hadron relative to its
mother parton, due to the fragmentation process. At forward rapidity and RHIC
energy, $\langle z \rangle$ is close to 1 due to the proximity of the kinematic limit. In turn, at mid-rapidity at the LHC where no strong kinematical constraints apply, one has $\langle z \rangle \sim 0.3$, meaning
that --for equal momentum of the produced hadron--  the mid-rapidity single-hadron production at the LHC probes the nuclear wave function at larger values of $x_A$. Therefore the suppression of $R_{pA}$ should be less important at mid-rapidity at the LHC as compared to what has been measured at forward rapidities at RHIC.

Indeed, the most recent CGC phenomenological works for $R_{pPb}$ at the LHC agreed to predict a moderate suppression at very small hadron $p_t$ followed by a rapid increase towards unity with increasing $p_t$.  Additionally, the rcBK-MC predictions bands were compatible with a moderate increase above unity at intermediate transverse momenta. The possibility of such enhancement results from the combination of a proper treatment of initial density fluctuations plus the anomalous dimension $\gamma>1$ from the rcBK initial conditions. Available ALICE data for minimum-bias collisions at mid-rapidity are compatible with both NLO nPDF predictions and with CGC-based expectations, see \fig{R_p+Pb}.
Other than the moderate suppression at small momenta $p_t\sim 1\div 3$ GeV, no other cold nuclear matter effects are present in data. In particular there is no Cronin enhancement at low $p_t$ and no suppression at high $p_t$. For $p_t >4$
GeV, $R_{pPb}$ is compatible with unity --but also with a moderate enhancement-- and well described by the leading-twist nuclear
PDF approximation. Therefore more data at forward rapidities is needed to distinguish between nPDF or CGC approaches. 
For fixed transverse momentum of the produced hadron rcBK approaches predict a stronger increase of the suppression with increasing rapidity than nPDF approaches, as shown in \fig{R_p+Pb} (top-right plot).

One important open theoretical question on the CGC side is how to match with standard leading twist perturbation theory at high transverse momentum $p_t\gg Q_s(x)$. Indeed at high transverse momenta the gluon densities in the nuclear wave function cease to be large and saturation effects should be absent. Translated to the observable under discussion, $R_{pPb}$, one would expect that it should approach unity at sufficiently high $p_t$, in agreement with nPDF calculations. However, CGC works based on the LO hybrid formalism predict a flat suppression up to the highest transverse momenta explored in the different works. First phenomenological applications of NLO corrections to the hybrid formalism indicate that they actually go in the right direction of pushing up the nuclear modification factors at moderate $p_t$ (see \fig{R_p+Pb} bottom-left), although, as already discussed, at large enough $p_t$ they become unstable. This instability blurs their physical interpretation and calls for further studies.
\vspace{+0.2cm}

\begin{figure}[htb]
\begin{center}
\begin{minipage}[t]{0.45\textwidth}
\includegraphics[width=\textwidth]{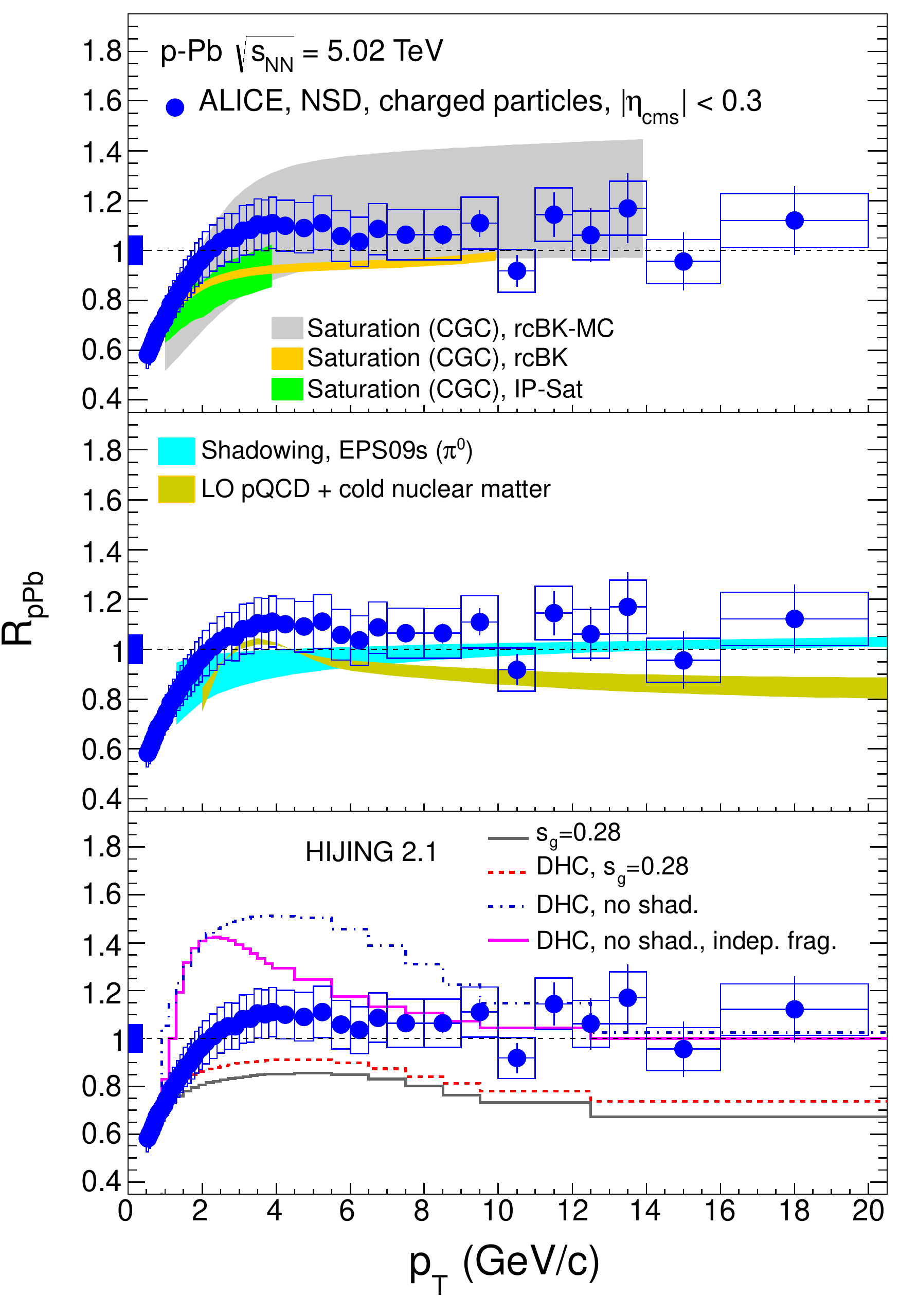}
\end{minipage}
\begin{minipage}[t]{0.5\textwidth}
\vspace{-11.4cm}\includegraphics[width=0.89\textwidth]{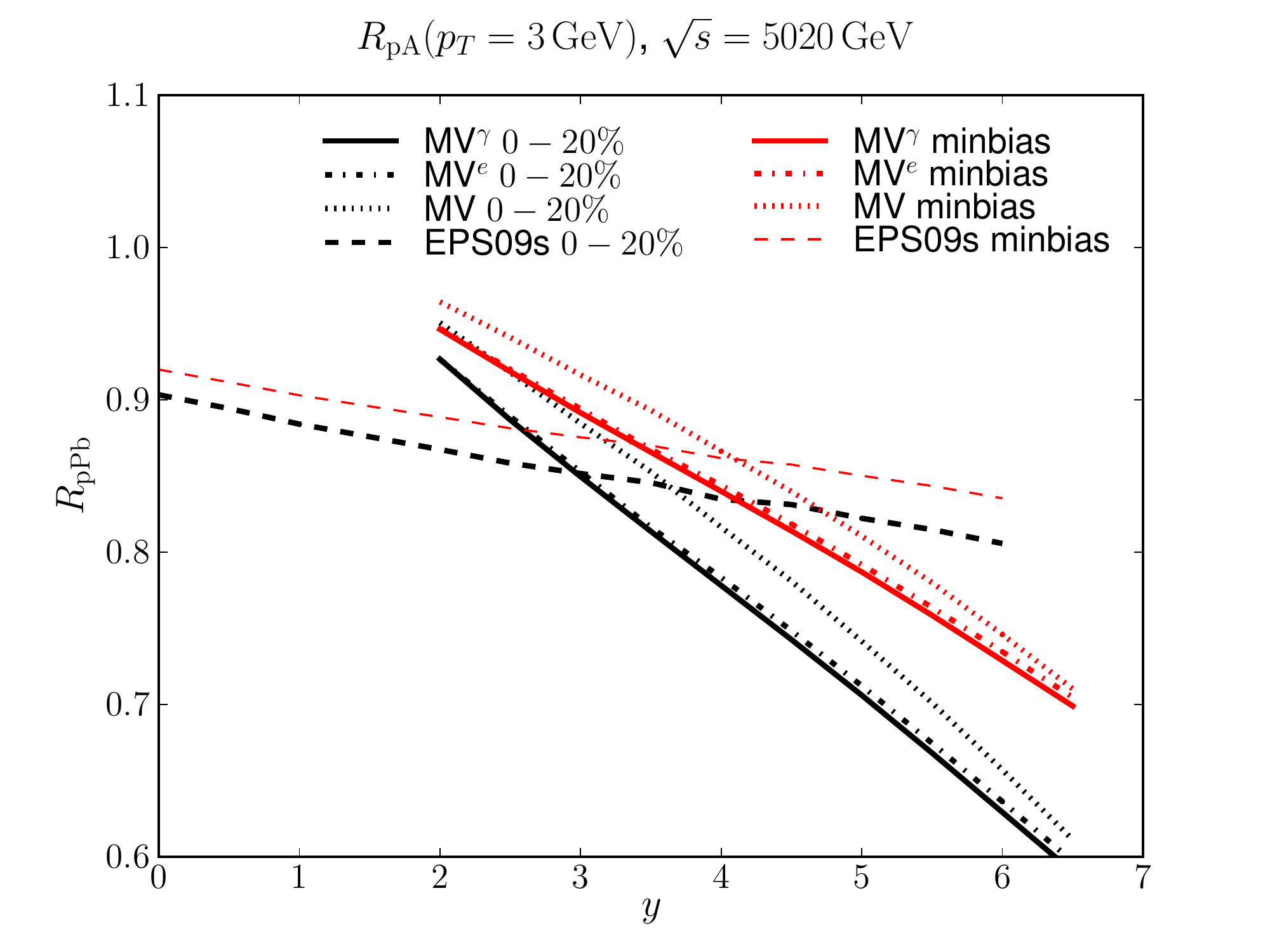}
\vfill\includegraphics[width=0.89\textwidth]{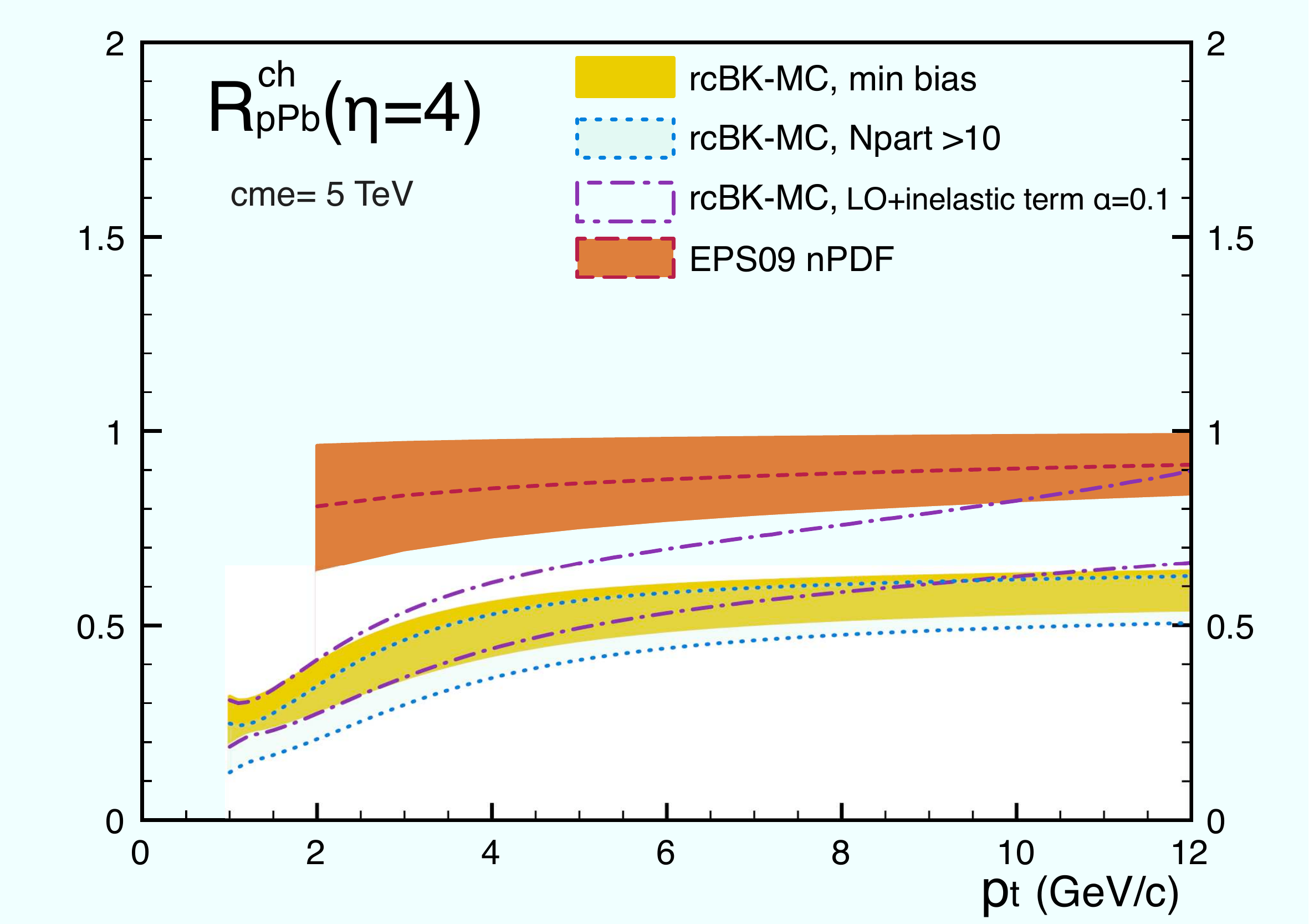}
\end{minipage}
\end{center}
\vspace*{-0.15cm}
\caption[a]{Left: ALICE data {\protect\cite{ALICE:2012mj}} for the
 charged hadron nuclear modification factor at mid-rapidity as a function of
  $p_t$. The theoretical results correspond to the CGC calculations of
 {\protect \cite{Tribedy:2011aa,Albacete:2012xq,Rezaeian:2012ye}}, the nuclear
  PDF approach EPS09\protect{\cite{Helenius:2012wd}}, the cold-nuclear matter
  predictions of \protect{\cite{Kang:2012kc}}, and the HIJING Monte Carlo
 \protect{ \cite{Xu:2012au}}. Right: rapidity (top) and transverse momentum dependence (bottom) of $R_{pPb}$ within rcBK approaches refs.~{\protect \cite{Albacete:2012xq,Lappi:2013zma}} vs LO{\protect\cite{Helenius:2012wd}} (bottom) and NLO {\protect\cite{Helenius}} (top) EPS09 predictions.  }
\label{R_p+Pb}
\end{figure}

\subsection{Double Inclusive}

It is clear from the discussion above that presently available data on single inclusive hadron distributions and nuclear modification factors, although suggestive of the presence of saturation effects, do not allow for a clear discrimination between CGC-based analyses and other approaches. This inconclusive situation triggered the study of other observables sensitive to more exclusive dynamical features like di-hadron azimuthal correlations discussed in this section.

We turn to the discussion of
forward di-hadron correlations measured at RHIC. In the case of
double-inclusive hadron production $pA\!\to\!h_1h_2X$, denoting
$p_{1\perp},$ $p_{2\perp}$ and $y_1,$ $y_2$ the transverse momenta and
rapidities of the final-state particles, the Feynman variables are
$x_i=|p_{i\perp}|e^{y_i}/\sqrt{s_{NN}}$ and $x_p$ and $x_A$ read
\begin{equation}
x_p=x_1+x_2\ ,\hspace{0.5cm}
x_A=x_1\ e^{-2y_1}+x_2\ e^{-2y_2}\ .
\label{kin2}
\end{equation}
We shall first consider the production of two forward particles, this is the only case which is sensitive to values of $x$ as small as in the single-inclusive case: $x_p\!\lesssim\!1$ and $x_A\!\ll\!1$. The central-forward measurement, that we shall discuss after, does not probe such kinematics: moving one particle forward increases significantly the value of $x_p$ compared to the central-central case (for which $x_p=x_A=|p_{\perp}|/\sqrt{s_{NN}}$), but decreases $x_A$ only marginally. In addition, we will focus on the $\Delta\phi$ dependence of the double-inclusive hadron spectrum, where $\Delta\phi$ is the difference between the azimuthal angles of the measured particles $h_1$ and $h_2$.

\subsection*{Forward-forward correlations}

The kinematic range for forward particle detection at RHIC is such that $x_p\!\sim\!0.4$ and $x_A\!\sim\!10^{-3}.$ Therefore the dominant partonic subprocess is initiated by valence quarks in the proton and, at lowest order in $\alpha_s,$ the $pA\!\to\!h_1h_2X$ double-inclusive cross-section is obtained from the $qA\to qgX$ cross-section. Calling $f_{q/p}$ the valence quark density in the proton and  $D_{h/q}$ and $D_{h/g}$ the appropriate hadron fragmentation functions one gets:
\begin{eqnarray}
dN^{pA\to h_1 h_2 X}(P,p_1,p_2)&=&\int_{x_1}^1 dz_1 \int_{x_2}^1 dz_2 \int_{\frac{x_1}{z_1}+\frac{x_2}{z_2}}^1 dx\
f_{q/p}(x,\mu^2)\nonumber\\&&\times\left[dN^{qA\to qgX}\left(xP,\frac{p_1}{z_1},\frac{p_2}{z_2}\right)
D_{h_1/q}(z_1,\mu^2)D_{h_2/g}(z_2,\mu^2)+\right.\nonumber\\&&\left.
dN^{qA\to qgX}\left(xP,\frac{p_2}{z_2},\frac{p_1}{z_1}\right)D_{h_1/g}(z_1,\mu^2)D_{h_2/q}(z_2,\mu^2)\right]\ .
\label{collfact}
\end{eqnarray}
\eq{collfact} assumes independent fragmentation of the two final-state hadrons, therefore it cannot be used to compute the near-side peak around $\Delta\Phi=0$. Doing so would require the use of poorly-known di-hadron fragmentation functions. Rather we will focus on the away-side peak around $\Delta\Phi=\pi$, where saturation effects are important. Note also that if $x_p$ would be smaller (this will be the case at the LHC), the gluon initiated processes $gA\to q\bar{q}X$ and $gA\to ggX$ should also be included in \eq{collfact}. Such gluon initiated processes  yield a more complex color structure of the composite projectile and involve higher $n$-point correlators of Wilson lines that quark initiated processes. This can be understood by recalling the equivalence of gluon lines to $q\bar{q}$ lines implied by the large-$N_c$ limit of the Fierz identity of $SU(3)$. Due to its bigger complexity, the gluon contribution to \eq{collfact} has not yet been calculated, which prevents complete predictions for LHC kinematics where --for moderate transverse momenta of the detected hadrons and intermediate rapidities-- $x_p$ is also small. 

On the nucleus side, \eq{kin2} implies that the gluon longitudinal momentum fraction varies between $x_A$ and $e^{-2y_1}+e^{-2y_2}$. Therefore with large enough rapidities, only the small-$x$ part of the nuclear wave function is relevant when calculating the $qA\to qgX$ cross section (there is no contamination from large-$x$ components). When probing the saturation regime, one expects $dN^{qA\to qgX}$ to be a non-linear function of the nuclear gluon distribution, i.e. that this cross section cannot be factorized further: $dN^{qA\to qgX}\neq f_{g/A}\otimes dN^{qg\to qgX}$. Using the CGC approach to describe the small$-x$ part of the nucleus wave function, the $qA\to qgX$ cross section was calculated in \cite{JalilianMarian:2004da,Nikolaev:2005dd,Baier:2005dv,Marquet:2007vb}, and indeed it was found that, due to the fact that small-$x$ gluons in the nuclear wave function behave coherently and not individually, the nucleus cannot be described by only a single-gluon distribution.

The $qA\!\to\!qgX$ cross section is instead expressed in terms of correlators of Wilson lines (which account for multiple scatterings), with up to a six-point correlator averaged over the CGC wave function of the form
\begin{eqnarray}
 S^{(6)}(x_\perp, x'_\perp,y_\perp, y'_\perp) = \left\langle  -\frac{1}{N_c(N_c^2-1)}\,\text{tr}\left\{ U(x_\perp) U^{\dagger}(x'_\perp)\right\} \right. \nonumber \\ 
\left. + \frac{1}{N_c^2-1}\,\text{tr}\left\{ U(y_\perp) U^{\dagger}(y'_\perp)\right\}\,\text{tr}\left\{ U(x_\perp)U^{\dagger}(x'_\perp) U(y_\perp) U^{\dagger}(y'_\perp) \right\} \right\rangle_Y \,, 
\label{quad}
\end{eqnarray}
where $(x_\perp, x'_\perp)$ and $(y_\perp, y'_\perp)$ refer to the transverse position of the quark and gluon after the multiple scatterings with the nucleus in the amplitude and complex conjugate amplitude respectively.

At the moment, it is practically difficult to evaluate the six-point function in \eq{quad}. In the large-$N_c$ limit it factorizes into a dipole scattering amplitude times a trace of four Wilson lines or quadrupole, corresponding to the last factor in \eq{quad}. The latter can in principle be obtained by solving an evolution equation written down in \cite{JalilianMarian:2004da}. However, this implies a significant amount of numerical work, and also requires to introduce initial conditions for the quadrupole evolution which, unlike the initial conditions for dipole evolution that are well constrained from analyses of HERA DIS data, are presently unknown. In practice, further approximations have been used, and this is where different phenomenological studies of forward di-hadron correlations in p+A collisions differ:
\begin{itemize}
\item in Ref.~\cite{Tuchin:2009nf}, the quadrupole contribution is simply disregarded, the cross-section is obtained solely from the dipole amplitude, using the so-called $k_t$-factorized formula recovered in the limit $Q_s/|p_{\perp1,2}|\ll1$. Saturation effects are nevertheless included in the UDGs, and it is done using the GBW parametrization \cite{Golec-Biernat:1998js}.
\item in Ref.~\cite{Albacete:2010pg}, the quadrupole is evaluated using the so-called Gaussian approximation of B-JIMWLK evolution, however only the {\it elastic} contribution is kept. The non-linear evolution is obtained using the rcBK equation.
\item in Ref.~\cite{Stasto:2011ru}, the complete Gaussian expression of the quadrupole is used, however only the so-called correlation limit $Q_s\sim|p_{\perp1}+p_{\perp2}|\ll|p_{\perp1,2}|$ is considered. Saturation effects are included using the GBW parametrization. The gluon-initiated processes calculated in \cite{Dominguez:2011wm} are included for the first time.
\item in Ref.~\cite{Lappi:2012nh}, the complete Gaussian expression of the quadrupole is used, and the non-linear evolution is obtained using the rcBK equation. Gluon-initiated processes are not included.
\end{itemize}

\begin{figure}[t]
\begin{center}
\includegraphics[width=0.8\textwidth]{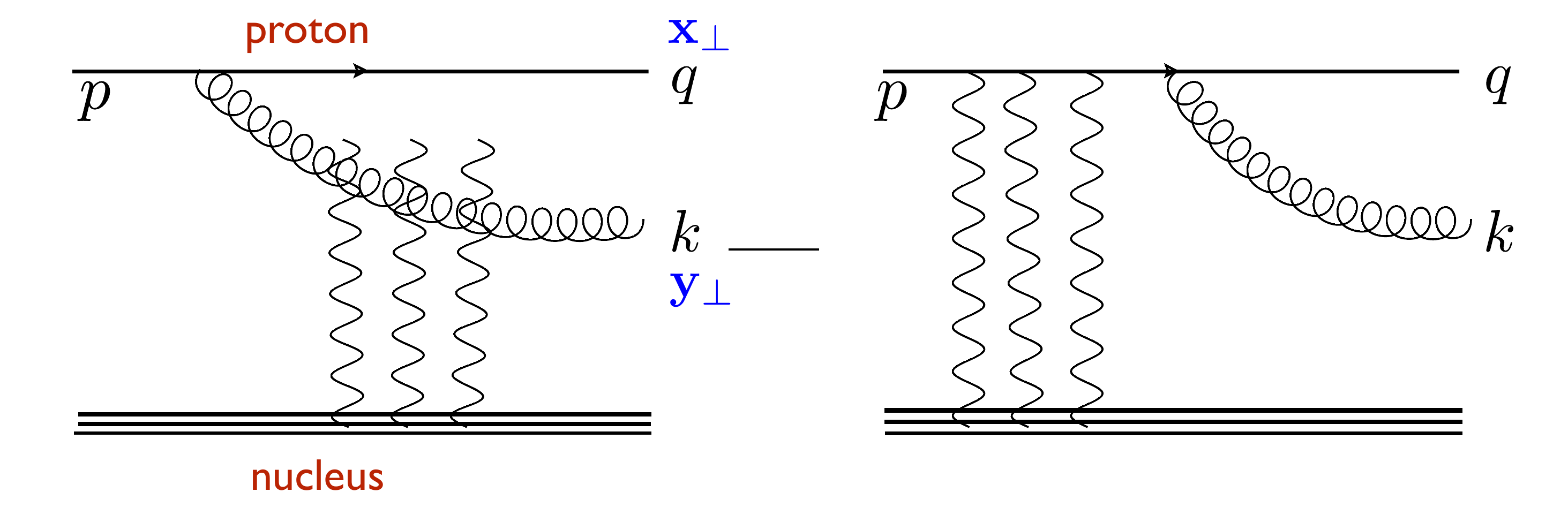}
\end{center}
\vspace*{-0.5cm}
\caption[a]{Diagrams contributing to the double inclusive cross-section in proton-nucleus collisions.}
\label{fig:double}
\end{figure}

We now come to the comparison with data. Nuclear effects on di-hadron correlations are typically evaluated in terms of the coincidence probability to, given a trigger particle in a certain momentum range, produce an associated particle in another momentum range. The coincidence probability is given by $CP(\Delta\phi)=N_{pair}(\Delta\phi)/N_{trig}$ with
\begin{equation}
N_{pair}(\Delta\phi)=\int\limits_{y_i,|p_{i\perp}|}\frac{dN^{pA\to h_1 h_2 X}}{d^3p_1 d^3p_2}\ ,\quad
N_{trig}=\int\limits_{y,\ p_\perp}\frac{dN^{pA\to hX}}{d^3p}\ .
\label{kinint}
\end{equation}

\begin{figure}[t]
\begin{center}
\includegraphics[width=0.6\textwidth]{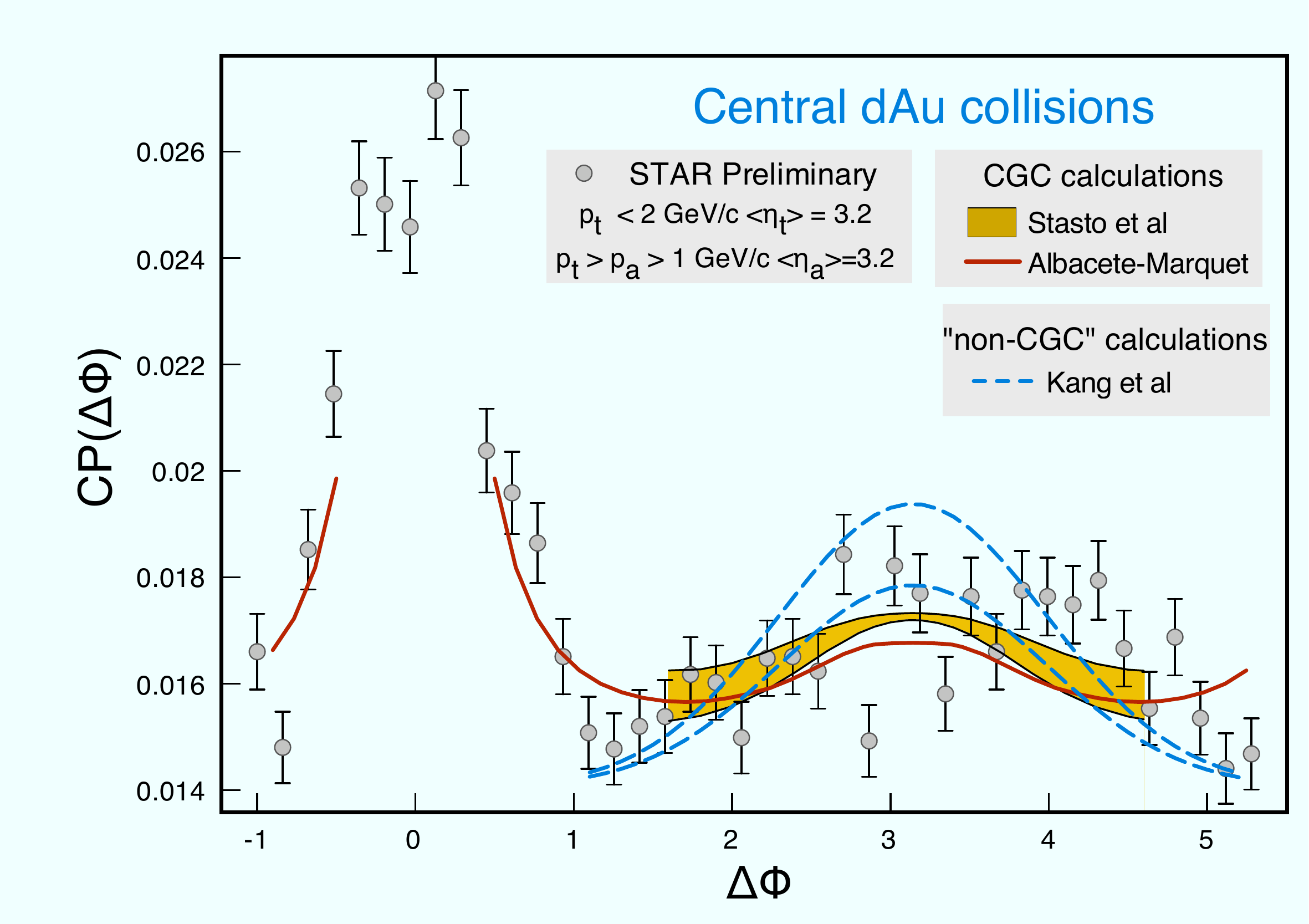}
\end{center}
\vspace*{-0.5cm}
\caption[a]{STAR preliminary data~{\protect\cite{Braidot:2010zh}} for the Coincident Probability between pairs of hadrons as a function of the relative azimuthal angle in d+Au collisions at RHIC. The theoretical results correspond to two CGC-based calculations~{\protect\cite{Albacete:2010pg}} and {\protect\cite{Stasto:2011ru}} and a {\it higher-twist} one~{\protect\cite{Kang:2011bp}}. Figure from{\protect~\cite{Albacete:2012td}}.}
\label{fig:corr}
\end{figure}

The STAR data for the coincidence probability obtained with two neutral pions are displayed in \fig{fig:corr} for central d+Au collisions. The nuclear modification of the di-pion azimuthal correlation is quite impressive, considering that the prominent away-side peak seen in p+p collisions is absent in central d+Au collisions, in agreement with the behavior first predicted in \cite{Marquet:2007vb}. The data are compared with the CGC calculations of \cite{Albacete:2010pg} and \cite{Stasto:2011ru}, as well as with a non-CGC approach discussed below. As mentioned before, the complete near-side peak is not computed, as \eq{collfact} does not apply around $\Delta\phi=0$. Note also that, since uncorrelated background has not been extracted from the data, the overall normalization of the theory points has been adjusted by adding a constant shift.

The physics of the disappearance of the away-side peak is the
following. The two measured hadrons predominantly come from a quark
and a gluon, which were back-to-back while part of the initiating
valence quark wave function. During the interaction, if they are put
on shell by a single parton from the target carrying zero transverse
momentum, as is the case when non-linear effects are not important,
then the hadrons are emitted back-to-back (up to a possible transverse
momentum broadening during the fragmentation process). By contrast, in
the saturation regime, the quark and antiquark independently receive a coherent
transverse momentum kick whose magnitude is of order of the nuclear saturation scale $Q_{sA}$. This breaks the initial back-to-back correlation in transverse momentum space  and depletes the correlation function around $\Delta\phi=\pi$ for hadron
momenta not much higher than $Q_{sA}$.

The sizable width of the away-side peak in \fig{fig:corr} cannot be
described within the leading-twist collinear factorization framework,
which completely neglects non-linear effects or, equivalently, multiple scatterings. While it may not be
obvious from single-inclusive measurements, this indicates that power
corrections are important when $|p_\perp|\sim 2$ GeV. At such low
transverse momenta, collinear factorization does not provide a good
description of particle production in p+A collisions. In contrast,
CGC calculations which do incorporate non-linear effects by resuming power
corrections can reproduce the suppression phenomenon. Ref.~\cite{Kang:2011bp} uses a non-CGC framework where only the first power corrections to leading-twist collinear factorization are considered. Additionally, the model of Ref.~\cite{Kang:2011bp} also includes nuclear energy loss corrections.

\subsection*{Forward-central correlations}

\begin{figure}[t]
\begin{center}
\includegraphics[width=0.52\textwidth]
                {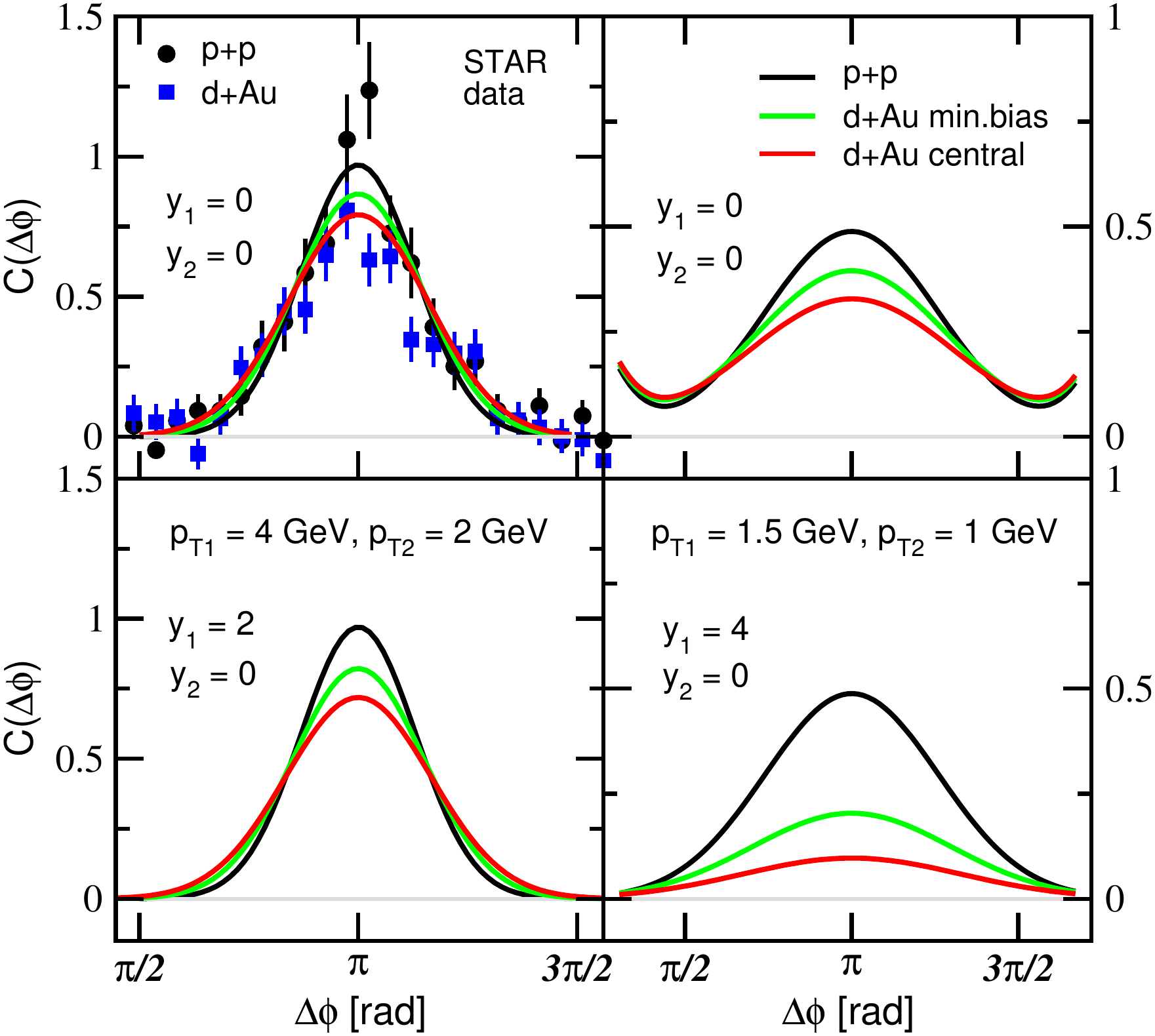}
\includegraphics[width=0.47\textwidth]
                {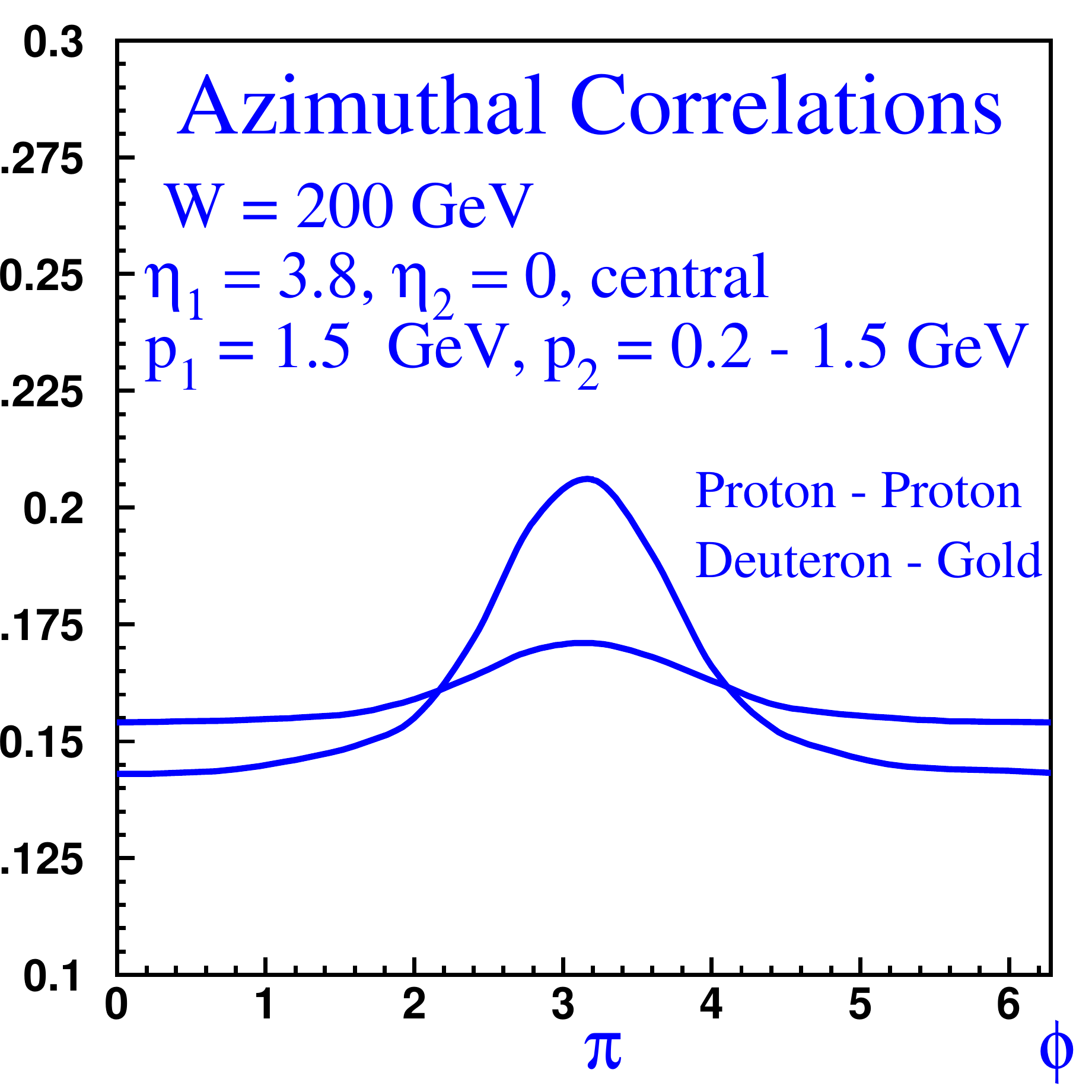}
\end{center}
\caption[a]{The coincidence probability at mid-rapidity as a function of $\Delta\phi$. RHIC data show that in the
central-central case the away-side peak is similar in d+Au and p+p collisions. In the central-forward case, the Glauber-eikonal (left plot from {\protect\cite{Qiu:2004da}}) and CGC (right plot from {\protect\cite{Kharzeev:2004bw}}) calculations predict that the away-side peak is suppressed in d+Au compared to p+p collisions, in agreement with later data.}
\label{fig:fwd-cent}
\end{figure}

The first measurements performed at RHIC by the PHENIX and STAR collaborations \cite{Adams:2006uz,Adler:2006hi} compared forward-central correlations to central-central ones. In the central-central case, the coincidence probability features a near-side peak around $\Delta\phi=0,$ when both measured particles belong to the same mini-jet, and an away-side peak around $\Delta\phi=\pi,$ corresponding to hadrons produced back-to-back. In the central-forward case, there is naturally no near-side peak. Both the Glauber multiple scattering \cite{Qiu:2004da} and CGC \cite{Kharzeev:2004bw} approaches can qualitatively describe the data, including the depletion of the away-side peak in d+Au collisions when going from central-central to central-forward production. This is illustrated in \fig{fig:fwd-cent}. Such a depletion does not occur in p+p collisions, since it is due to nuclear-enhanced power corrections, and therefore the p+A to p+p ratio of the integrated coincidence probabilities
\begin{equation}
I_{pA}=\frac{\int d\phi\ CP_{pA}(\Delta\phi)}{\int d\phi\ CP_{pp}(\Delta\phi)}
\end{equation}
is below unity. In fig.~\ref{fig:IandJdA} (left), recent PHENIX data on $I_{dAu}$ are displayed as a function of centrality. At the moment, since $x_A$ is not that small, it is not clear whether the mechanism for this suppression is due to saturation effects rather than incoherent multiple scatterings.

\begin{figure}[t]
\begin{center}
\includegraphics[width=0.4\textwidth]
                {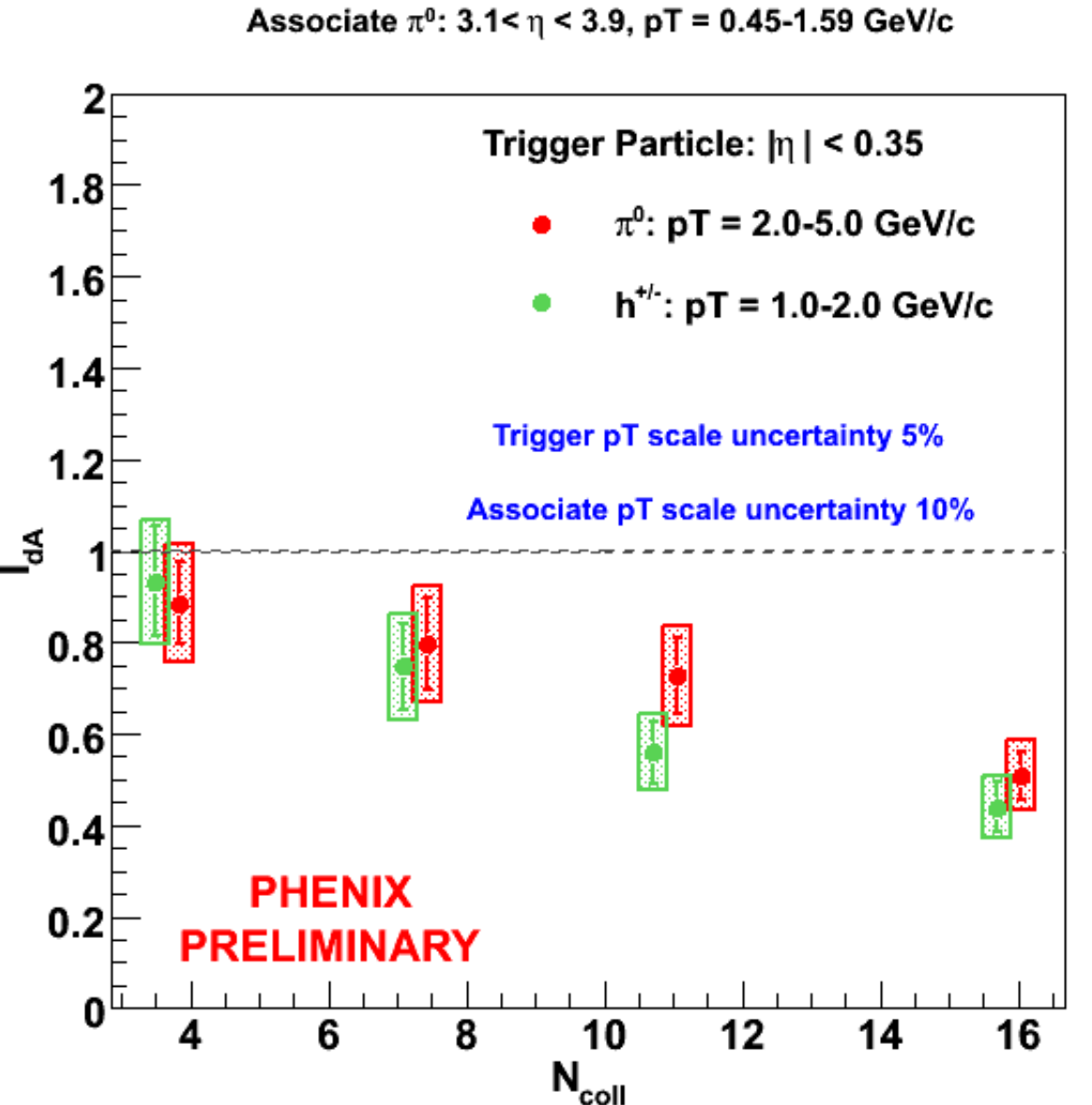}
\includegraphics[width=0.59\textwidth]
                {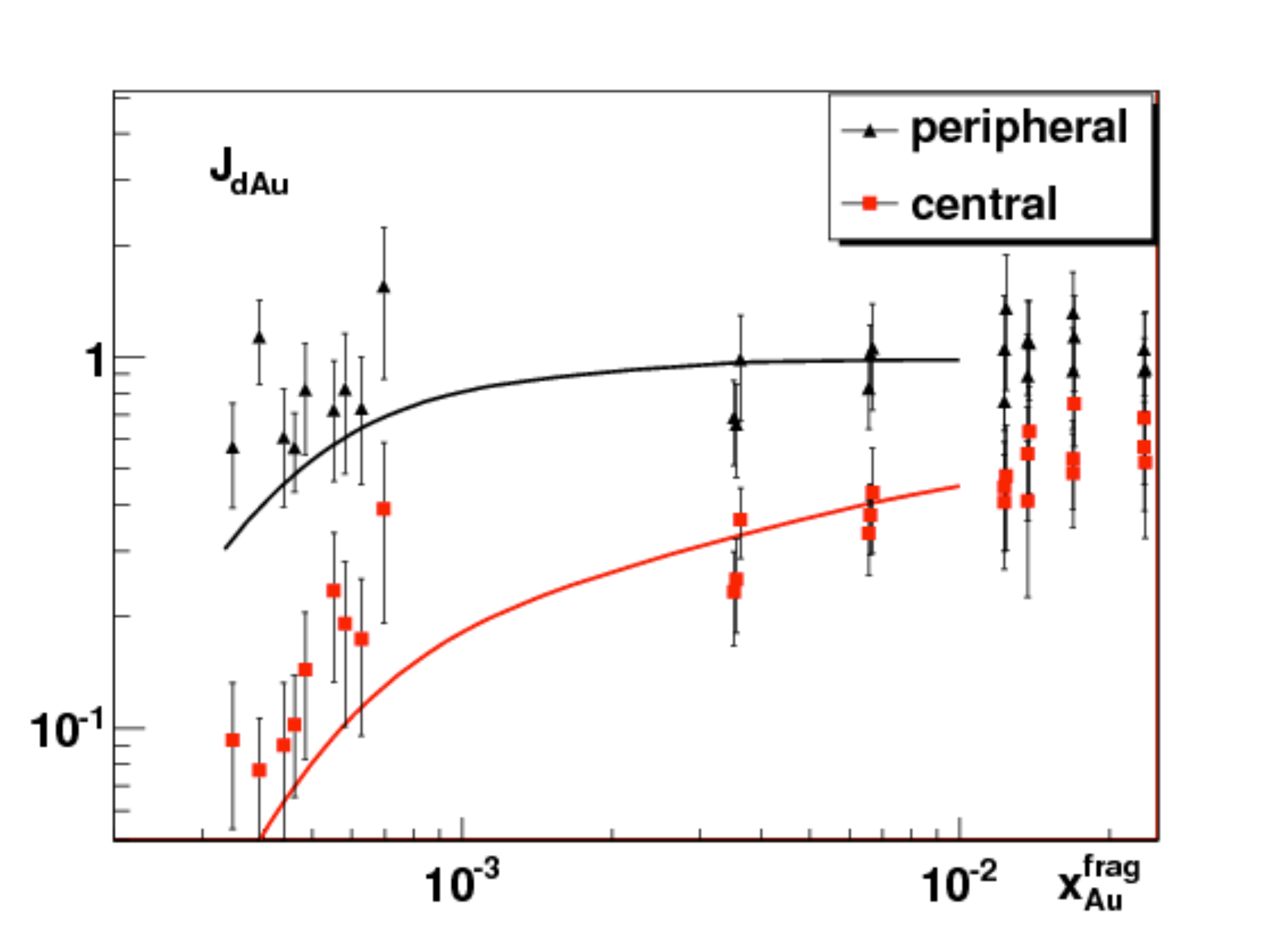}
\end{center}
\caption[a]{Left: central-forward preliminary $I_{dAu}$ data as a function of centrality~{\protect\cite{Meredith:2009fp}}. In central collisions, the integral of the coincidence probabilty is about half that in p+p collisions, reflecting the depletion of the away-side peak. Right: the nuclear suppression factor
$J_{dA}$ in dAu collisions as function of $x_A$. The experimental data are from PHENIX~{\protect\cite{Adare:2011sc}} and the theory calculations are from ref.~{\protect\cite{Stasto:2011ru}}, with pedestal contributions subtracted.}
\label{fig:IandJdA}
\end{figure}

The most recent measurements compare the forward-central case to the forward-forward one \cite{Adare:2011sc}. In particular, in order to quantify the stronger suppression going from the former case to the latter, a new observable was considerered:
\begin{equation}
J_{pA}=\frac{\int d\phi\ N_{pair}^{pA}(\Delta\phi)}{N_{coll}\ \int d\phi\ N_{pair}^{pp}(\Delta\phi)}\ .
\end{equation}
This allows to remove from $I_{pA}$ the effects of trigger particle suppression at forward rapidity, and to focus on the magnitude of the di-hadron suppression. The results are displayed in \fig{fig:IandJdA} (right), along with a successful CGC description.

To conclude this section, let us briefly discuss the recent LHC data on di-hadron correlations in p+Pb collisions. So far, the measurements have only been done at mid-rapidity, i.e. for central-central correlations. Nevertheless, 
after subtracting of a double-ridge to make the near-side peak centrality-independent, a sensible comparison with the RHIC data can be made. The origin of this double ridge and the physics associated to it will be discussed in the next section. The subtracted ALICE data \cite{Abelev:2012ola} are shown in fig.~\ref{fig:alicemidrap}, they show that, at mid-rapidity at the LHC, there may be an onset of suppression of the away-side peak with increasing centrality.

\begin{figure}[h!]
\begin{center}
\includegraphics[width=0.59\textwidth]{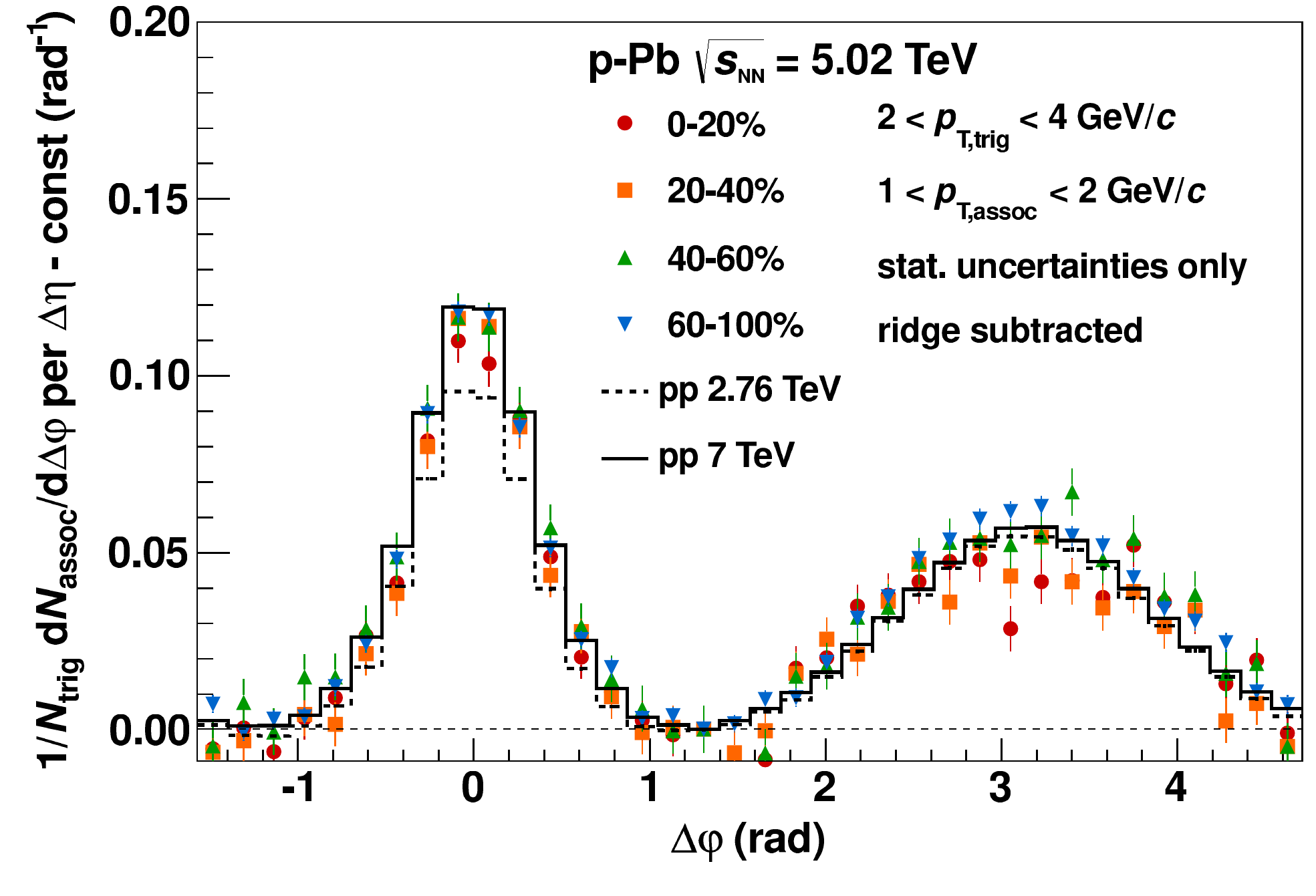}
\end{center}
\caption[a]{Associated yield per trigger particle as a function of $\Delta\phi$ averaged over $|\Delta\eta| < 1.8$ for pairs of charged particles
with $2<p_{T,trig}<$~4 GeV and $1<p_{T,asso}<$ 2 GeV in p+Pb collisions at $\sqrt{s} =5.02$ TeV for
different event classes, compared to p+p collisions at $\sqrt{s} = 2.76$ and 7 TeV.
For the event classes 0--20\%, 20--40\% and 40--60\% the long-range contribution on the near-side
$1.2<|\Delta\eta|<1.8$ and $|\Delta\phi|<\pi/2$ has been subtracted from both the near side and the away side.}
\label{fig:alicemidrap}
\end{figure}

\begin{figure}[htb]
\begin{center}
\includegraphics[width=0.44\textwidth]
                {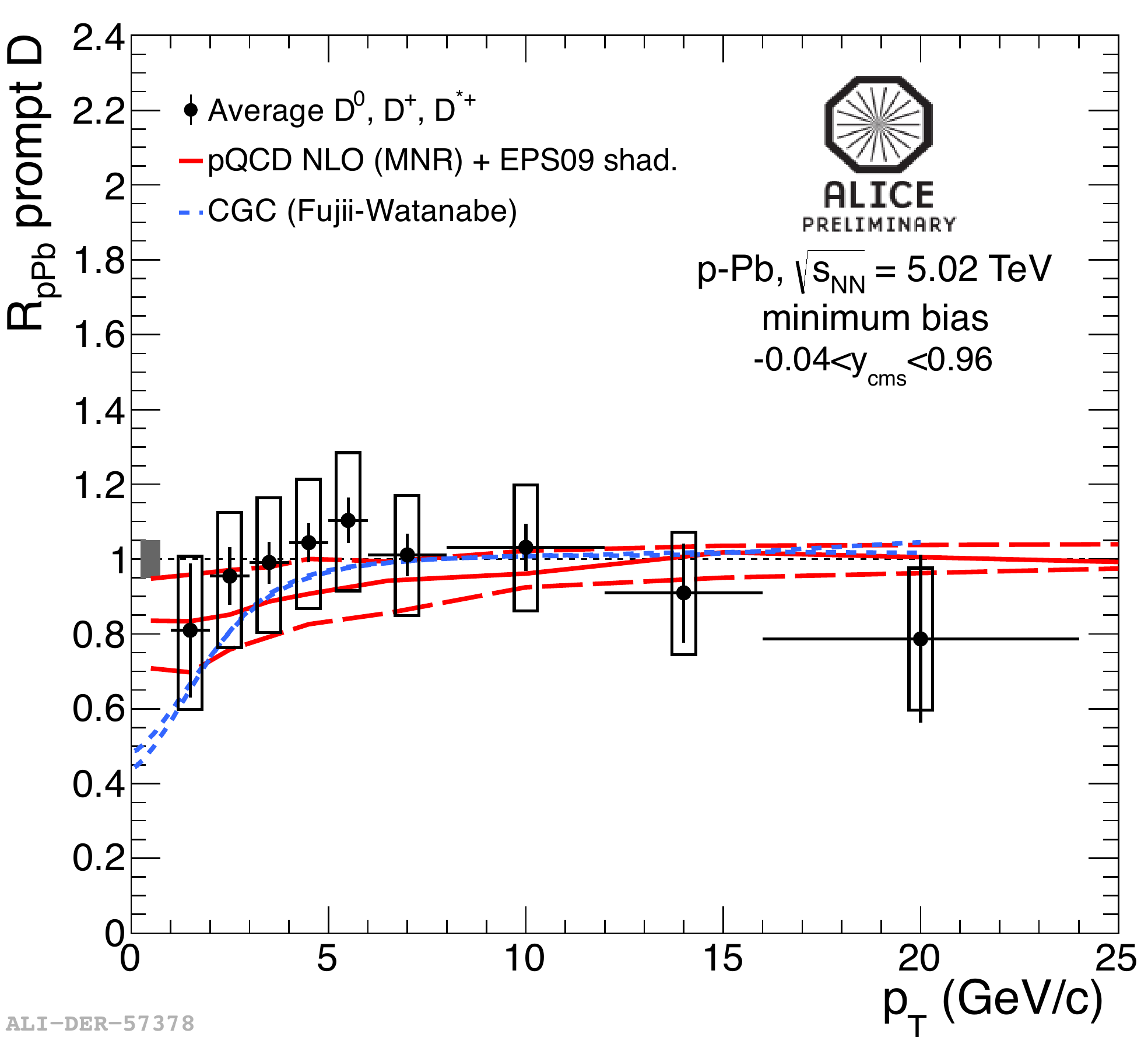}
\includegraphics[width=0.55\textwidth]
                {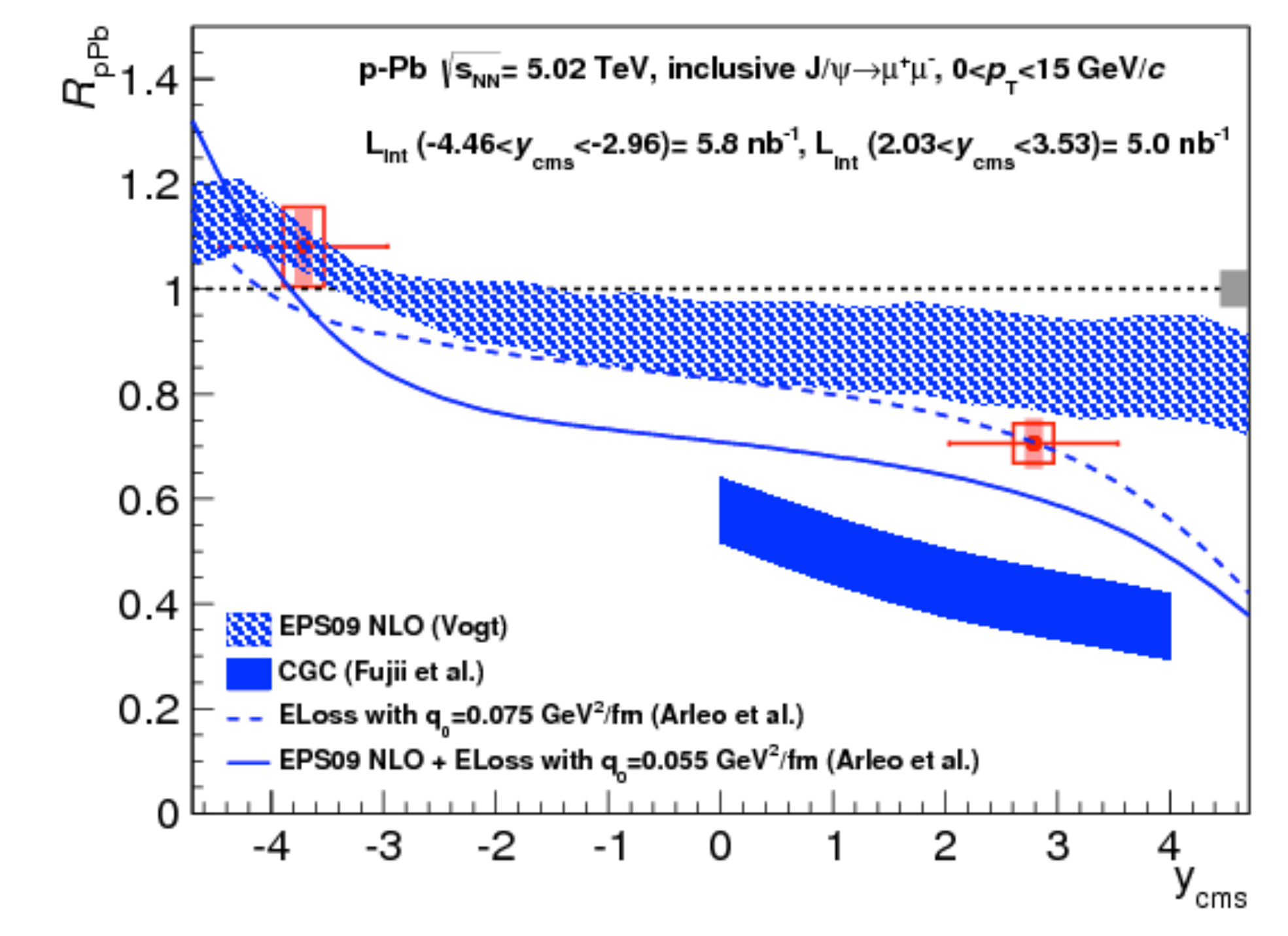}
\end{center}
\caption[a]{Left: ALICE preliminary data {\protect\cite{fortheALICE:2013ica}} for the D meson nuclear modification factor at mid-rapidity as a function of $p_T$. Right: ALICE data {\protect\cite{Abelev:2013yxa}} for the nuclear modification factor for  $J/\Psi$ $p_t$-integrated yields as a function of rapidity.}
\label{fig:heavy-mesons}
\end{figure}

\subsection{Other observables}

We conclude this section with a brief discussion of a few other observables considered in the literature, in the context of p+Pb collisions at the LHC. We shall first discuss heavy flavor mesons, measured so far only at mid-rapidity like light hadrons, and quarkonia, the only forward measurement available up to now. We finish with forward photons and photon-hadron correlations, which could become crucial measurements, should suitable detector upgrades be installed in the future.

\subsubsection*{Open heavy flavors and quarkonia}

Fig.~\ref{fig:heavy-mesons} (left) shows a comparison of experimental data for the $R_{pPb}$ of D mesons as a function of transverse momentum, with calculations performed in the collinear-factorization and CGC approaches \cite{Fujii:2013yja}. The conclusions are the same than for light hadrons (see Fig.~\ref{R_p+Pb}): both calculations are consistent with the data, the amount of non-linear effects is small, as expected at mid-rapidity. 

Quarkonia are special, since already in p+p collisions, the production mechanism is not fully understood. Therefore is it difficult to precisely quantify how much their production is affected by non-linear effects. Nevertheless, they are particularly interesting since there exists forward-rapidity measurements, for which one expects such effects to be appreciable. In practice, saturation effects have a different impact depending on the production mechanism assumed, therefore at this point, no strong conclusions about non linear effects can be drawn from quarkonia data comparisons with CGC calculations.

As an example, Fig.~\ref{fig:heavy-mesons} (right) shows a comparison of experimental data for the $J/\Psi$ $R_{pPb}$ as a function of rapidity, with a CGC calculation in the color evaporation model (CEM) \cite{Fujii:2013gxa}. The data are also compared with a CEM calculation that does not include saturation effects, and with calculations which take into account cold-matter-induced energy loss \cite{Arleo:2012hn,Arleo:2012rs}. This effect could indeed be very relevant, and so far is absent from CGC calculations, including the most recent ones performed in the context of the color-singlet and color-octet models \cite{Kang:2013hta,Qiu:2013qka}.

\begin{figure}[htb]
\begin{center}
\includegraphics[width=0.57\textwidth]
                {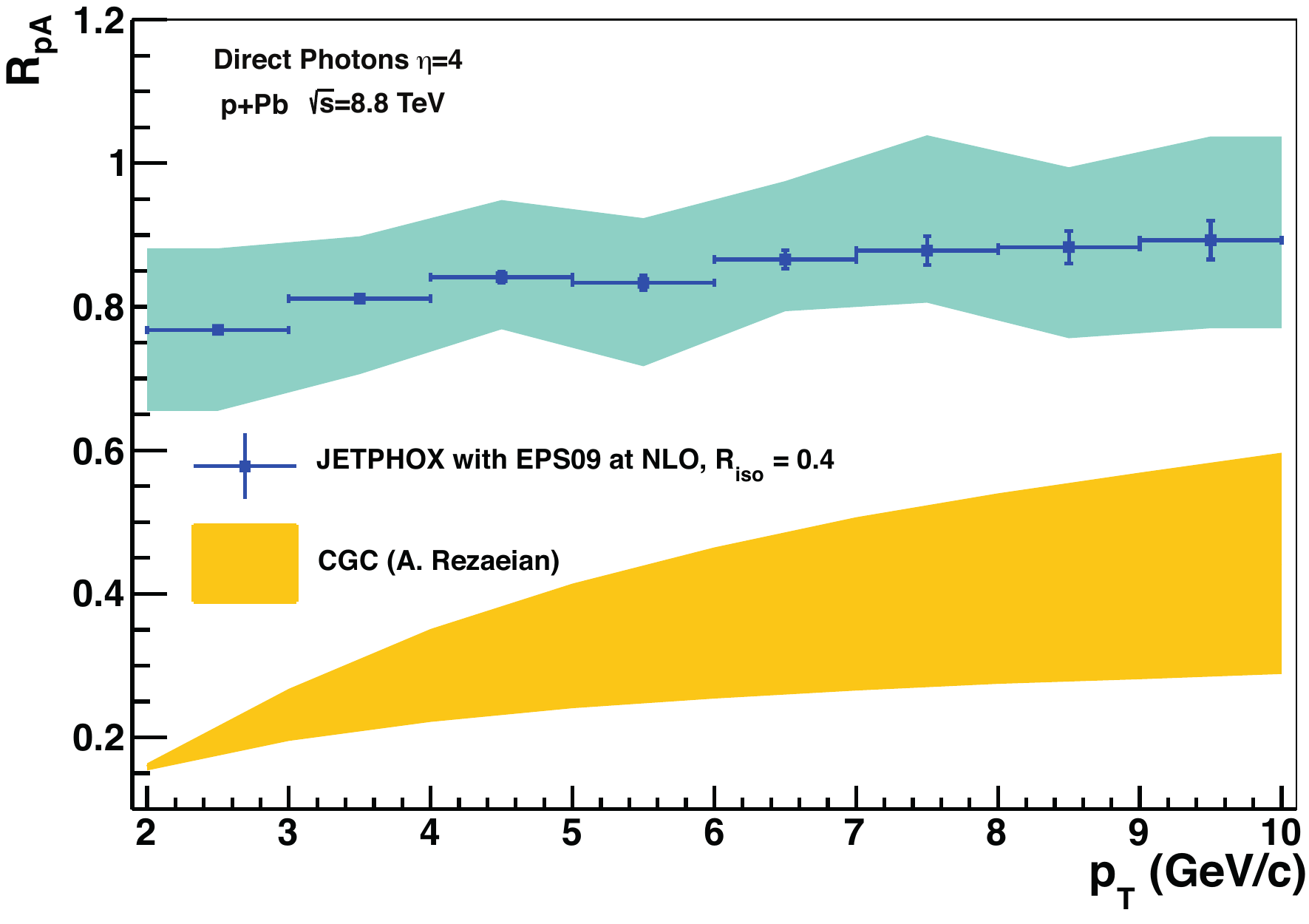}
\includegraphics[width=0.42\textwidth]
                {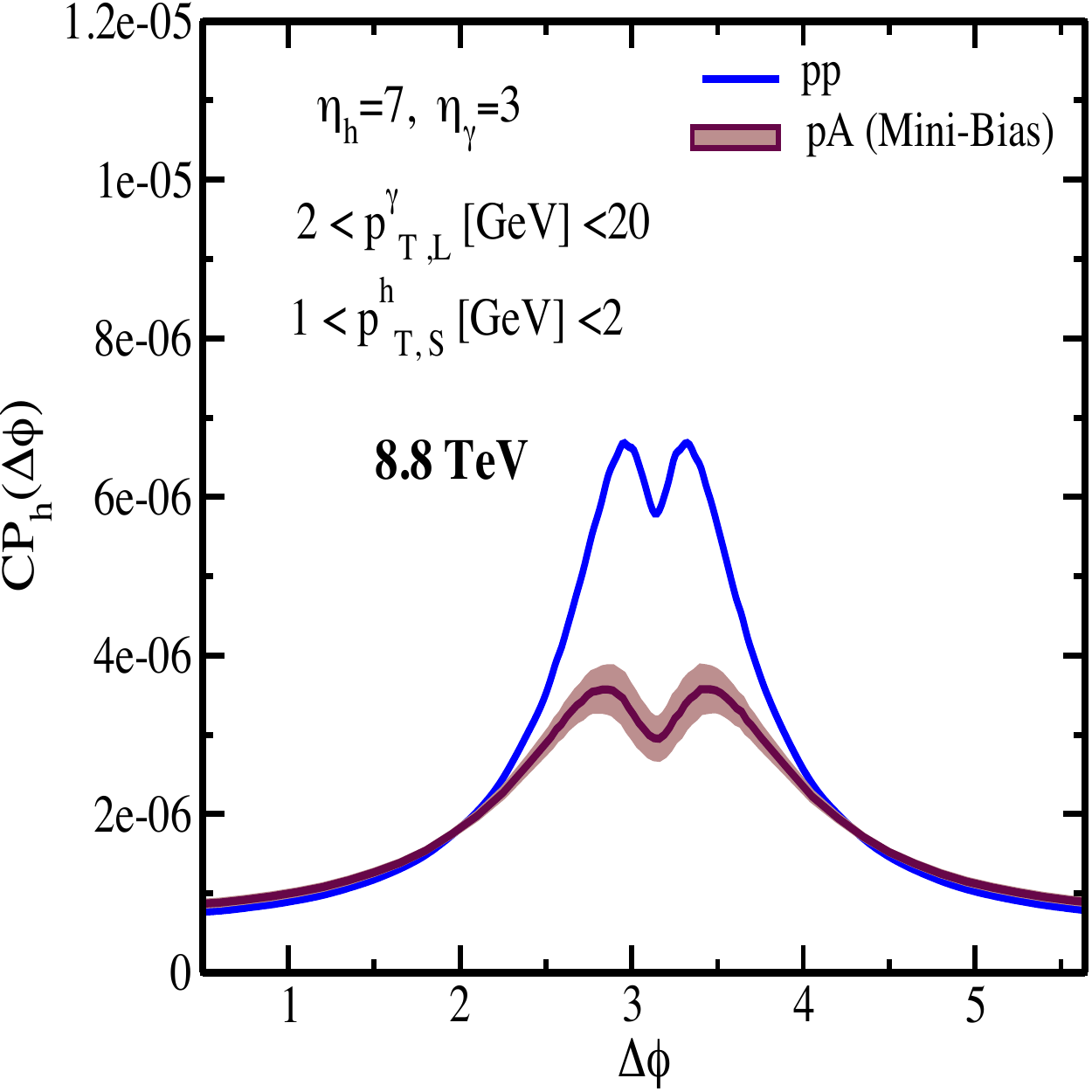}
\end{center}
\caption[a]{Left: predictions for the forward direct photon $R_pPb$ at the LHC, from \cite{T.PeitzmannfortheALICEFoCal:2013fja}. Right: forward photon-hadron azimuthal correlations in p+p and p+A collisions, showing the photon-triggered coincidence probability, from {\protect\cite{Rezaeian:2012wa}}.}
\label{fig:photons}
\end{figure}

\subsubsection*{Photons and photon-hadron correlations}

From the theory perspective, photons are the best probe of non-linear effects: the calculations are relatively simple \cite{Gelis:2002ki,Baier:2004tj}, fragmentation photons and direct photons can be separated \cite{JalilianMarian:2012bd}, and they do not suffer from cold matter final-state effects such as energy loss. In addition they are massless and therefore can probe smaller values of $x$, even more so with direct photons which do not come from hadron fragmentation. Finally at the LHC (p+Pb vs p+p), isospin effects will not alter the interpretation of the data, as was the case at RHIC (d+Au vs p+p).

Experimentally though, photons are not easy to measure. Fig.~\ref{fig:photons} (left) shows the large differences expected between calculations in the collinear factorization and CGC approaches, for the nuclear modification factor $R_{pPb}$ in the case of forward direct photons. Such a measurement could be carried out with the future FoCal upgrade to ALICE, and could provide the cleanest evidence for parton saturation at the LHC.

Photon-hadron correlations are also very interesting, since their theoretical formulation involves only dipole amplitudes, as opposed to di-hadrons which also involve quadrupoles. An example of how saturation effects modify photon-hadron azimuthal correlations in p+A collisions compared to p+p collisions at the LHC, is displayed in Fig.~\ref{fig:photons} (right). A similar pattern is predicted in the case of Drell Yan lepton-pair-hadron correlations \cite{Stasto:2012ru}.

\section{Bulk properties of multi-particle production}
\subsection{Total Multiplicities}
Two main features of RHIC and LHC data on multiplicities in nucleus-nucleus collisions can be highlighted (see \fig{energynpart}): 
First, the energy dependence of mid-rapidity multiplicities in A+A collisions is well reproduced by a power law, $dN_{ch}/d\eta(\eta = 0) \sim s^{0.15}$. This observation seems to rule out the logarithmic trend observed for lower energies data and is in generic agreement with pQCD based approaches. Second, the centrality dependence of LHC multiplicities is, up to an overall scale factor, very similar to the one measured at RHIC. These two features suggest a factorization of the energy and centrality dependence of multiplicities which, in turn, admit a natural explanation in the CGC formalism. There, mid-rapidity multiplicities densities per unit transverse area rise proportional to the saturation scale squared of the colliding nuclei. This is so regardless of the precise formalism used to calculate them --$k_t$-factorization or Classical Yang Mills methods as we shall see-- since the saturation scale is the only dimensionful dynamical scale in the problem. In turn, the saturation scale is, in a first approximation, proportional to the local nuclear density or, equivalently, to the number of participants:
\be
\frac{dN^{A+A}}{d\eta\,d^2b}(\eta=0)\sim Q_{sA}^2(x,b)\quad \Rightarrow\quad \frac{dN^{A+A}}{d\eta}(\eta=0)\sim  \sqrt{s_{NN}}^{\lambda} \, N_{part}\,.
\label{multsat}
\ee
This is a highly non-trivial result that indicates that the main features of inclusive particle production may be understood within weakly-coupled, albeit non-linear, QCD. As we argue below, and notwithstanding the relevance of non-perturbative effects, this idea is realized to a rather good approximations in LHC and RHIC data.

\begin{figure}
\hskip 1cm 
\includegraphics[width=70mm]{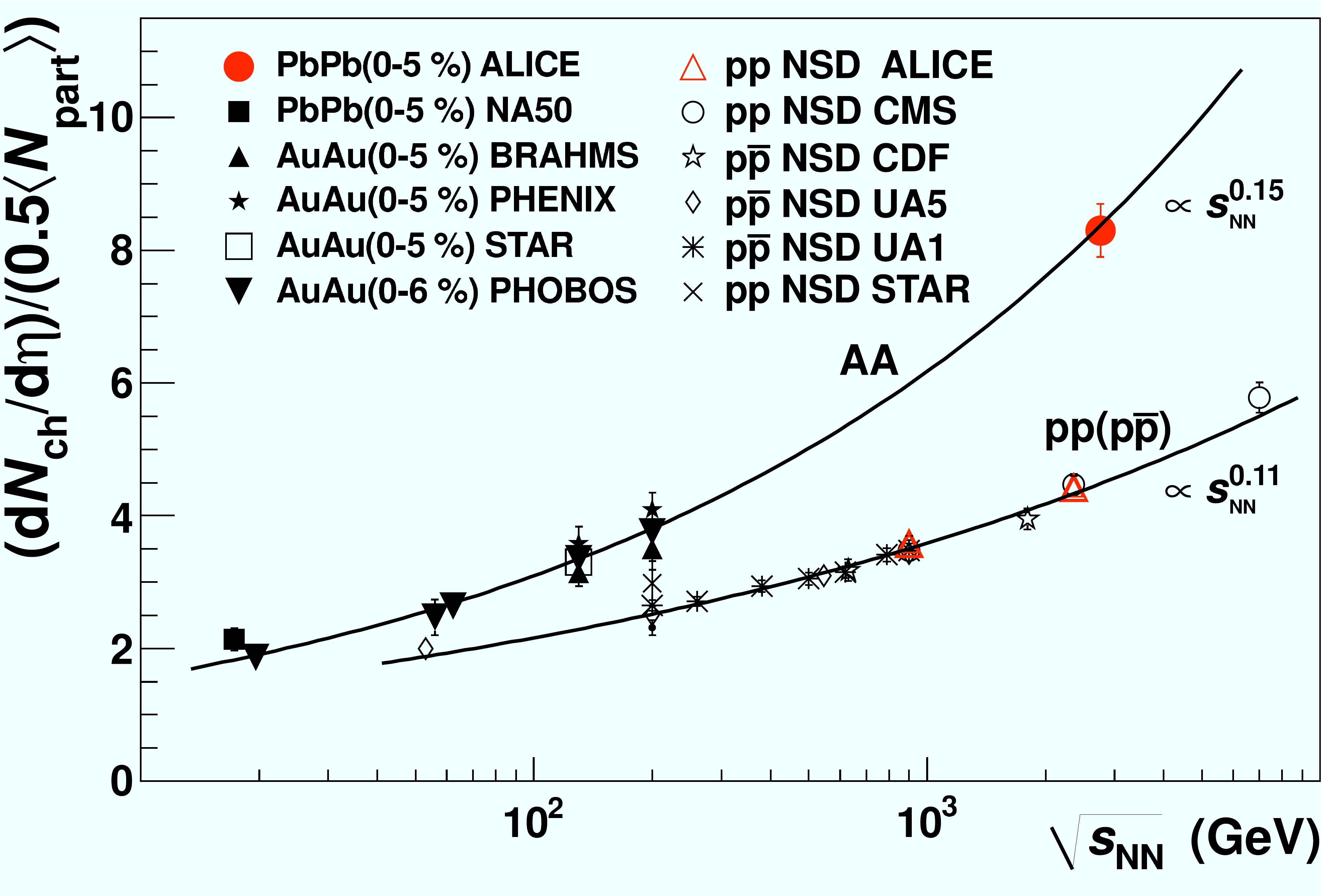}
\hskip 1cm
\includegraphics[width=67mm]{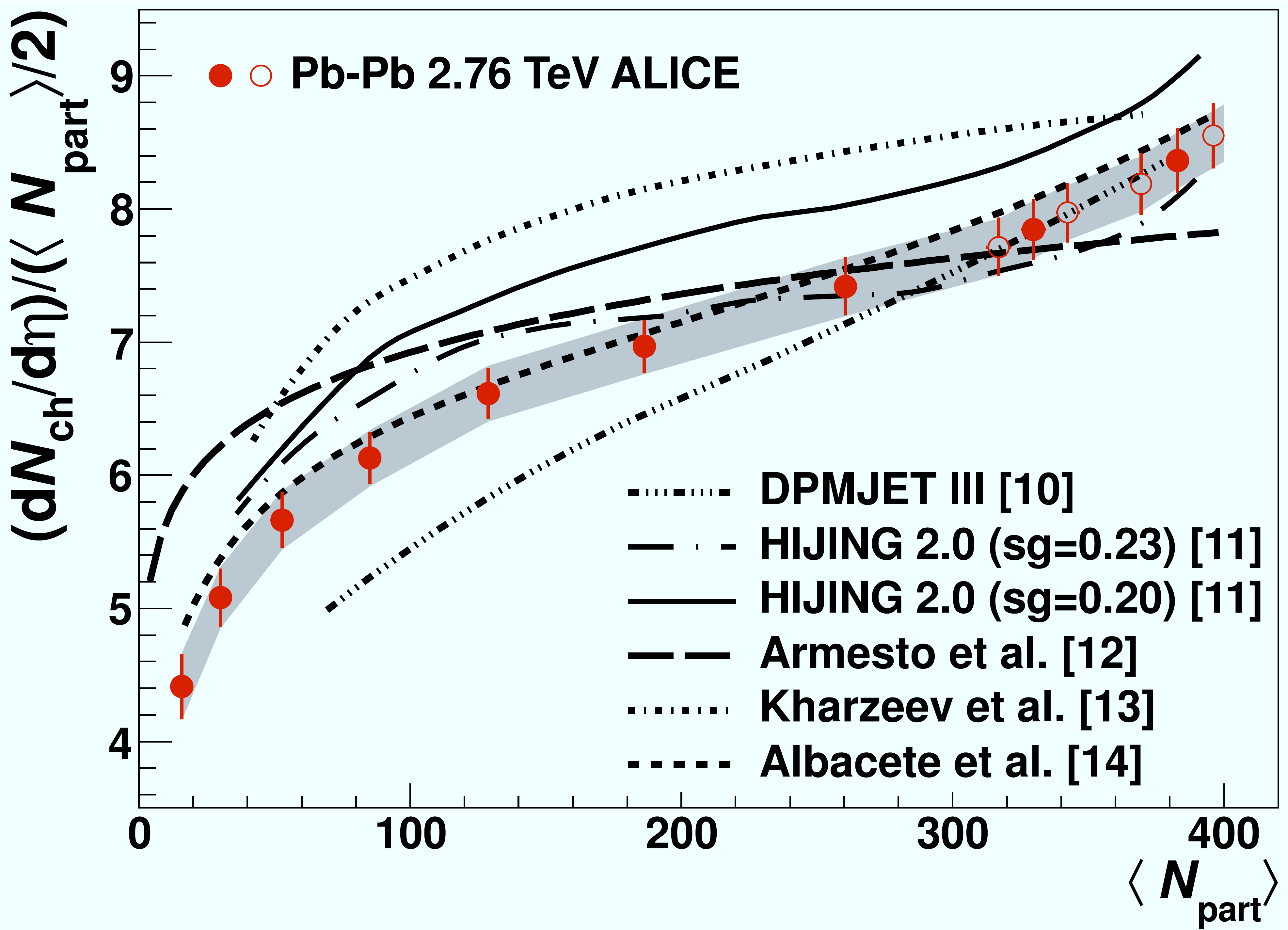}
\caption{Energy (left) and centrality (right) dependence of mid-rapidity multiplicities in A+A and p+p collisions. Figures and LHC data by the ALICE collaboration {\protect\cite{Aamodt:2010pb,Aamodt:2010cz}}.}
\label{energynpart}
\end{figure}

The first attempt to calculate total multiplicities in the CGC framework following the ideas above is due to 
Kharzeev, Levin and Nardi~\cite{Kharzeev:2001yq,Kharzeev:2004if,Kharzeev:2007zt} who
obtained a very good description of the geometry dependence of hadron production in nucleus-nucleus collisions at
RHIC energy. The KLN model relies in the use of $k_t$-factorization \eq{kt2} to calculate initial gluon production. This approach has then been improved to incorporate more realistic ugd's compatible with HERA data in refs\cite{Rezaeian:2011ia,Rezaeian:2012ye} and with known features of small-$x$ evolution refs~\cite{Albacete:2007sm} and and to improve the description of the collision geometry through the use of Monte Carlo simulations, refs.~\cite{Drescher:2006ca, Drescher:2007ax, Albacete:2010ad,Albacete:2012xq}.

The above referred works present some uncertainties. First of all, the very use of $k_t$-factorization is not justified for nucleus-nucleus collisions, where it fails to account for some relevant diagrams\cite{Blaizot:2010kh,Baier:2011ap}. This problem is absent in the case of {\it dilute-dense} scattering, i.e proton-nucleus collisions. 
Next, and due precisely to the absence of such multiple scatterings, gluon production diverges for small transverse momenta $k_t\to 0$. The infrared divergence is typically fixed through the introduction of either an energy-independent cut-off or by limiting the transverse momentum integration in \eq{kt2} to the momentum of the produced hadron (gluon), $p_t$. This difficulty is avoided by the use of exact numerical solutions of the classical Yang-Mills (CYM) calculations in the presence of two sources (nuclei). This procedure is theoretically more solid, as it accounts for multiple scattering both on the projectile and target side, yielding an IR-finite gluon yield. Such is the basis of the IP-Glasma model~\cite{Schenke:2012hg}, which assumes
that the effective action far from the valence sources remains a local
Gaussian and that the nuclear wave function is given by an MV model action with an energy dependent saturation momentum built from the IP-Sat fits to HERA data (see section 2). 
We expect, however, that these calculations will be improved
in the foreseeable future since a complete CGC calculation should combine both the use of CYM to describe particle production (as done in the IP-Glasma model) with the use of an effective action that solves the B-JIMWLK equations (as done in $k_t$-factorization models with rcBK evolved ugd's). 

Other sources of uncertainties common to all CGC-based models are the choice for the scale of the running coupling  and, most importantly, the assumption that the final-state hadron yields are directly proportional to the initial gluon production. This last assumption, sometimes referred to as parton-hadron duality, permits to completely bypass the thermalization and hadronization stages of the collision. However, the number of hadrons per produced gluon  could depend on energy~\cite{Levin:2011hr}. Or, phrased in a thermal language, entropy production in the Glasma and QGP phases could be energy dependent\cite{Muller:2011ra}. Furthermore, it is clear that the
distribution of hadrons in rapidity need not be {\em identical} to
that of gluons, especially when their density per unit of rapidity is
high. Specifically, pseudo-rapidity distributions $dN/d\eta$ of
unidentified charged hadrons involve a jacobian transformation from rapidity to pseudo-rapidity , $\partial y/\partial \eta$, 
which depends on the mass and transverse momentum of the observed hadron. However, it is known that $\langle
p_t\rangle$ increases with energy, with the mass numbers of projectile
and target, and is affected by soft final-state interactions (at least
in case of A+A collisions). The degree of modeling involved in different works for dealing with the non-perturbative aspects of the $y \leftrightarrow \eta$ transformation explain to a large extent the spread in the effective value of $\lambda\sim 0.2\div0.3$ in \eq{multsat} and, therefore, the in the CGC predictions for LHC multiplicities. 

The common hope is that the free parameters in these models --IR-cutoff, scale for the strong coupling, {\it fudge} factor that relates gluon and hadron multiplicity etc-- are not much dependent on energy such that, once they have been fixed RHIC energies the extrapolation of models to LHC energies is indeed controlled by the small-$x$ evolution of the nuclear saturation scale as described by the B-JIMWLK or BK equations. Indeed, and despite these caveats, CGC models provide a rather good and consistent description of the bulk features of multi particle production in p+p, p+Pb and Pb+Pb data at the LHC. 

\subsubsection*{Multiplicities in proton-nucleus collisions}
We first start by discussing some of the most recent CGC calculations for p+Pb multiplicities at the LHC, all based in the use of $k_t$-factorization. We stress the fact all these calculations were made public before any p+Pb data were available\footnote{Albeit, in most cases the initial predictions assumed a
  collision energy of $\sqrt{s}=4.4$~TeV which was later updated to
  5~TeV; also, initial predictions did not account for the rapidity
  shift by $\Delta y = \frac{1}{2}\log (82/208)\simeq - 0.465$ of the
  experimental detectors towards the Pb beam and were updated later.}. 
Fig.~{\protect \ref{fig:pPb5000_dNdeta}} shows the prediction of the KLN
model~{\protect \cite{Dumitru:2011wq}} for 5~TeV (left plot) and of the  rcBK-MC~{\protect\cite{Albacete:2012xq}} model for two different UGD sets (centre plot). 
In the KLN prediction fluctuations in the number of particles produced in each collision are accounted for through a negative binomial distribution, whereas fluctuations in the transverse positions of nucleons are neglected. In turn, the MC-rcBK does not account for fluctuations at the sub-nucleon level but does a detailed stochastic description of the collision geometry. 

In both cases the original predictions
(depicted by open or closed squares in both plots) employed the exact same $y\to\eta$
transformation, specifically the same rapidity independent ratio of
hadron mass to transverse momentum $m/p_t=0.24$, as for p+p
collisions. At $\eta=0$ the predictions essentially coincide with the
data, in fact much better than its true level of accuracy. Away from
mid-rapidity they exhibit a slightly steeper dependence on $\eta$ than
the data. At such level of detail, however, one can not discard
hadronization effects entirely; this is illustrated in the figures by
the ``updated curves'' (stars in the MC-KLN plot and filled triangles in the MC-rcBK one) which introduce a slight rapidity and atomic number dependence of $m/\langle p_t\rangle$, with a sign that follows from the
rapidity dependence of $\langle p_t\rangle\sim Q_{s,{\rm Pb}}$. The better agreement of data after these small modifications illustrates the sensitivity to detailed properties of the hadrons in
the final state.

\begin{figure}[htbp]
\begin{center}
\includegraphics[width=0.33\textwidth]{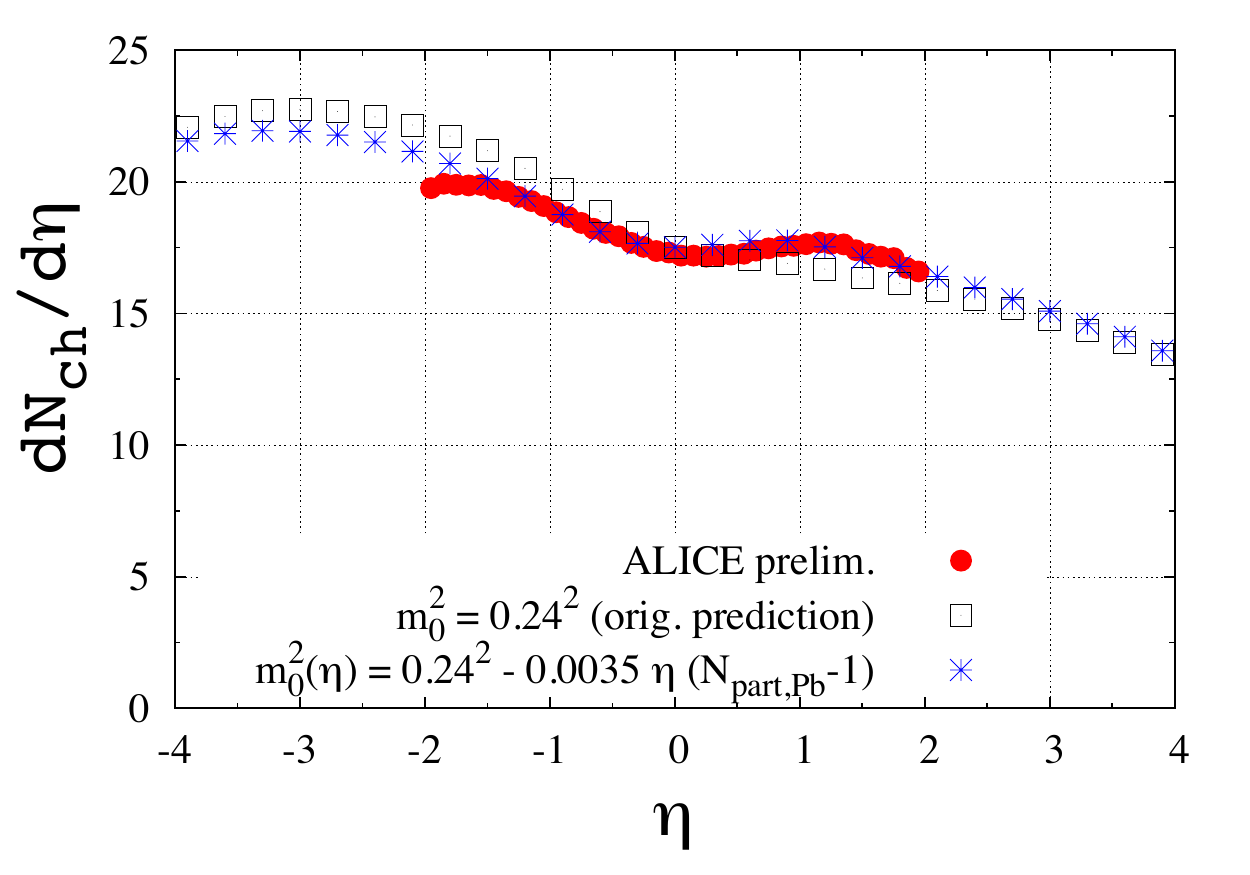}
\includegraphics[width=0.33\textwidth]{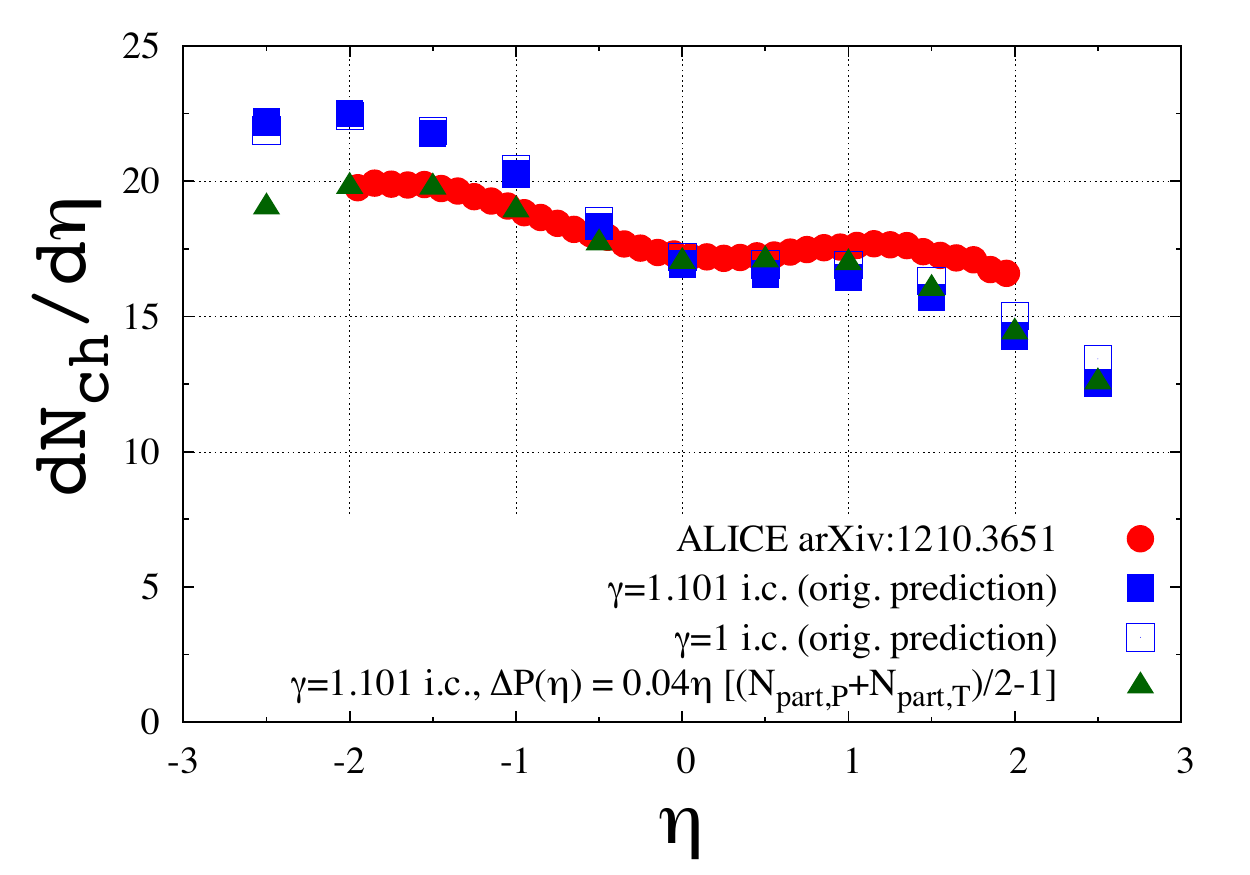}
\includegraphics[width=0.3\textwidth]{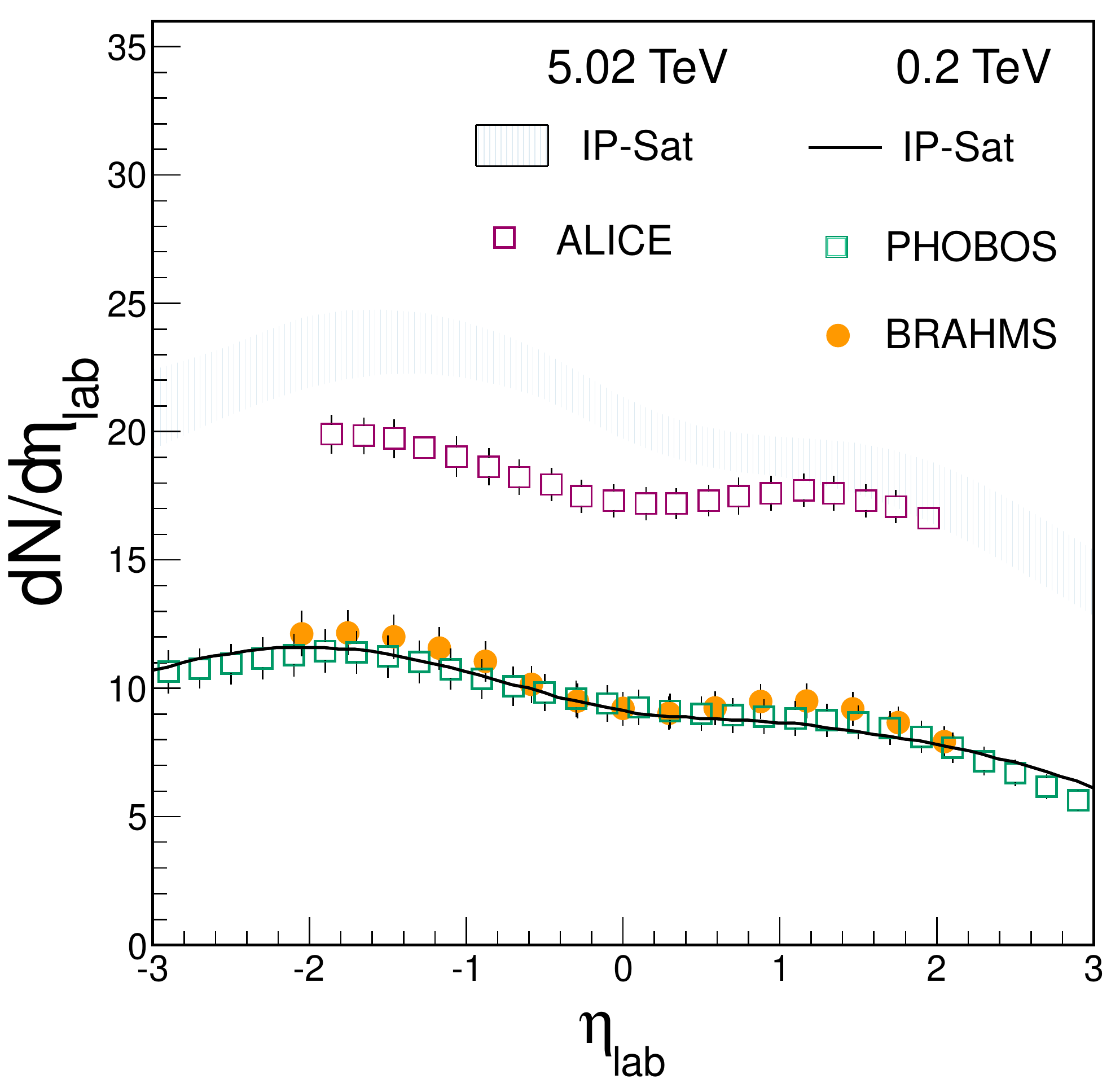}
\end{center}
\vspace*{-0.15cm}
\caption{Left: charged particle rapidity distribution for minimum-bias
  p+Pb collisions at 5~TeV from the KLN gluon saturation model~{\protect\cite{Dumitru:2011wq}}. Center: rcBK
  UGD~{\protect\cite{Albacete:2012xq}}.  Right: IP-Sat UGD~{\protect\cite{Tribedy:2011aa}}. 
 ALICE data from ref.~{\protect\cite{ALICE:2012xs}}.}
\label{fig:pPb5000_dNdeta}
\end{figure}

Yet another prediction for $dN/d\eta$ in p+Pb collisions employs
$k_t$-factorization with UGDs from the IP-sat
model~\cite{Tribedy:2011aa}. An updated result for $\sqrt{s}=5$~TeV,
and including the above-mentioned rapidity shift, is shown again in
\fig{fig:pPb5000_dNdeta} (right plot). These authors also provide
uncertainty estimates shown by the bands. This prediction is also
close to the data although it over-predicts the multiplicity at
$\eta\lesssim1$ somewhat. In part, this may be due to neglecting target
thickness fluctuations which effectively suppress the multiplicity in
the dense regime at $p_t<Q_s$ (see fig.~7 and related discussion in
ref.~\cite{Albacete:2012xq}). Finally another set of predictions for
$\sqrt{s}=4.4$~TeV based on the b-CGC model for the nuclear ugd's was presented in ref.~\cite{Rezaeian:2011ia} which is slightly higher than the above (again, without accounting for target thickness
fluctuations). 

All in all, it is quite remarkable that all of these
implementations which differ in their hadronization prescription or in
the treatment of the nuclear geometry and of Glauber fluctuations lead
to similar predictions at $\eta=0$ (deviations from the central value
are $\pm15\%$ or less); they all involved an extrapolation from d+Au
collisions at RHIC by more than an order of magnitude (factor of 25) in
the center-of-mass energy. This is consistent with expectations that the energy
dependence of particle production is determined mainly by the growth
of the gluon saturation scale rather than by genuinely
non-perturbative soft physics.

\subsubsection*{KNO scaling in p+p and p+A collisions}
In recent years calculations of single-inclusive multiplicities within
the CGC framework have been extended to {\em multiplicity
  distributions} and to their evolution with energy and system
size. Much of this initial progress on multi-particle production in the
non-linear regime is based on the framework of classical fields. Some
basic insight into quantum evolution effects on various moments of the
multiplicity distributions has also been developed\cite{Kovner:2006wr}
but an explicit discussion of their properties is still lacking.

Assuming dominance of classical fields, the probability to produce $q$
particles is given by\cite{Gelis:2009wh}
\be
\left < \frac{d^q N}{dy_1\cdots dy_q} \right>_{\rm conn.} = C_q\,
\left < \frac{d N}{dy_1} \right> \cdots
\left < \frac{d N}{dy_q} \right>
\ee
with the reduced moments
\be \label{eq:C_q_MV}
C_q = \frac{(q-1)!}{k^{q-1}}~.
\ee
This expression is valid with logarithmic accuracy and was derived
under the assumption that all transverse momentum integrals over
$p_{T,1} \cdots p_{T,q}$ are effectively cut off in the infrared at
the non-linear scale.

The fluctuation parameter $k$ in eq.~(\ref{eq:C_q_MV}) is of order
\be
k \sim (N_c^2-1)\, Q_s^2\, S_\perp~.
\ee
The numerical pre-factor (in the classical approximation) has been
determined by a numerical computation to all orders in the valence
charge density in ref.~\cite{Lappi:2009xa}.

The multiplicity distribution is therefore approximately a negative
binomial distribution (NBD)~\cite{Gelis:2009wh},
\be \label{eq:NBD}
P(n) = \frac{\Gamma(k+n)}{\Gamma(k) \, \Gamma(n+1)}
\frac{\bar n^n k^k}{(\bar n + k)^{n+k}}~.
\ee
Indeed, multiplicity distributions observed in high-energy p+p
collisions (in the central region) can be described reasonably well by
a NBD, see for example
refs.~\cite{Dumitru:2012yr,Tribedy:2011aa}. The
parameter $k^{-1}$ determines the variance of the
distribution\footnote{The width is given by $\bar{n}\,
  \sqrt{k^{-1}+\bar{n}^{-1}}\sim \bar{n}/\surd k$ where the latter
  approximation applies in the limit $\bar{n}/k\gg1$.} and
can be obtained from the (inclusive) double-gluon multiplicity:
\be \label{eq:dN2_k}
\left < \frac{d^2 N}{dy_1 dy_2} \right>_{\rm conn.} = \frac{1}{k}\;
\left < \frac{d N}{dy_1} \right> \;
\left < \frac{d N}{dy_2} \right>~.
\ee
This expression shows that the perturbative expansion of $k^{-1}$
starts at ${\cal O}(\alpha_s^0)$ since the connected diagrams on the
l.h.s. of \eq{eq:dN2_k} involve the same number of sources and
vertices as the disconnected diagrams on the r.h.s.\ of that
equation~\cite{Dumitru:2012tw}. Since the {\em mean} multiplicity in
the classical limit is of order $\bar{n}\equiv \langle dN/dy\rangle
\sim 1/\alpha_s$ it follows that $\bar{n}/k\sim 1/\alpha_s\gg
1$. This, in fact, corresponds to the limit where the NBD multiplicity
distributions exhibit KNO scaling\cite{Koba:1972ng}. The
semi-classical strong-field limit of a Gaussian effective theory
relates the emergence of KNO scaling in $p_\perp$-integrated
multiplicity distributions to properties of small-$x$ gluons around the
saturation scale~\cite{Dumitru:2012tw}.

\begin{figure}[htb]
\begin{center}
\includegraphics[width=0.5\textwidth]{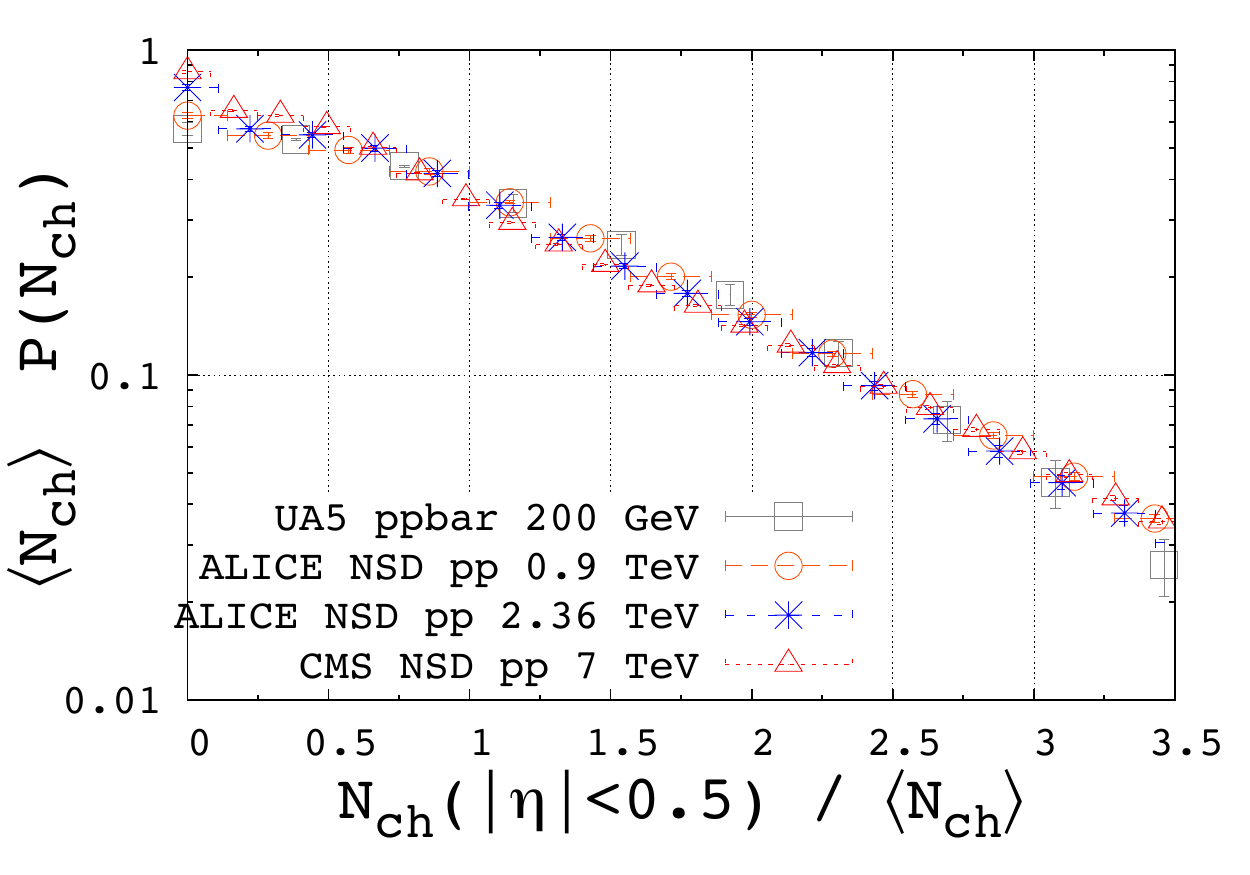}
\end{center}
\vspace*{-0.15cm}
\caption[a]{KNO scaling plot of charged particle multiplicity
  distributions in non-single diffractive $pp\,/\,p\overline{p}$
  collisions at various energies as measured by the
  UA5~{\protect\cite{Ansorge:1988kn}}, ALICE~{\protect\cite{Aamodt:2010ft}} and
  CMS~{\protect\cite{Khachatryan:2010nk}} collaborations, respectively. Note
  that we restrict to the central rapidity region $|\eta|<0.5$ and to
  the bulk of the distributions up to 3.5 times the mean
  multiplicity.}
\label{fig:KNO_LHCdata}
\end{figure}
KNO scaling conjectures that the particle multiplicity distribution in
high-energy hadronic collisions is {\em universal} (i.e., energy
independent) if expressed in terms of the fractional multiplicity
$z\equiv n/\bar n$. This is satisfied in first approximation in the
central (pseudo-) rapidity region at energies $\ge200$~GeV as shown in
\fig{fig:KNO_LHCdata}. For a detailed review of pre-LHC data
see ref.~{\protect\cite{GrosseOetringhaus:2009kz}}.

However, more precise NBD fits to the multiplicity distributions
reveal that $\bar{n}/k$ increases somewhat with energy, and so KNO
scaling is not exact. An increase of $\bar{n}/k\sim1/\alpha_s$ with
energy may be partially explained by running of the coupling with
$Q_s(x)$ but this goes beyond the classical approximation. Corrections
beyond the classical limit are also indicated by the fact that at
lower energies higher-order factorial moments $G_q$ of the
distribution are significantly different from the reduced moments
$C_q$~\cite{Zajc:1986pn}:
\be
G_q \equiv \frac{\langle n\, (n-1)\cdots (n-q+1)\rangle}{\bar{n}^q} ~~~,~~~
C_q \equiv \frac{\langle n^q\rangle}{\bar{n}^q}~~.
\ee
Note that the difference of $G_q$ and $C_q$ is subleading in the
density of valence charges $\rho$. In fact, while
eq.~(\ref{eq:C_q_MV}) has been derived for a quadratic MV-model action
which applies near infinite density, at finite density the action
contains additional terms
$\sim\rho^4$\cite{Dumitru:2011zz,Dumitru:2011ax}. These operators
provide corrections to the moments of the multiplicity distribution~\cite{Dumitru:2012tw}.

Finally we also point out that approximate KNO scaling has been
predicted to persist even for min-bias p+Pb collisions at LHC
energies~\cite{Dumitru:2012yr}, with deviations from the KNO curve
from p+p collisions at a level of ``only'' $\sim25\%$. This is in
spite of Glauber fluctuations of the number of target participants
which of course exist for a heavy-ion target. The multiplicity
distribution in p+A collisions thus tests the expectation that
intrinsic particle production fluctuations are dominated by the dilute
source, i.e.\ that $k\sim {\rm min}(T_A,T_B)$ is proportional to the
thickness of the more dilute collision partner. Furthermore, it can
provide valuable constraints for models of initial-state fluctuations
in A+A collisions which are currently of great interest (see section \ref{sec:IC}).

\subsubsection*{Nucleus-nucleus collisions}

CGC models also provide a good description of the energy, pseudo-rapidity and centrality dependence on multiplicities in nucleus-nucleus RHIC and LHC collisions. Concerning the collision energy dependence of multiplicities, and somewhat unexpectedly, LHC data seem to indicate a stronger energy dependence in mid-rapidity multiplicities in p+p collisions than in nucleus-nucleus collisions. While no clear explanation of this is observation is yet available, several possibilities have been recently proposed based on the ideas of additional entropy production in the pre-equilibrium phase\cite{Baier:2011ap}, enhanced parton showers in A+A collisions due to the larger average transverse momentum of the initially produced minijets compared to p+p collisions\cite{Levin:2011hr} or to non-trivial high-$Q^2$ effects intertwined with impact parameter dependence\cite{Lappi:2011gu}.

We now focus in the centrality dependence of the multiplicity and
transverse energy in heavy-ion collisions at LHC energies as obtained
from $k_t$-factorization with rcBK UGDs. We recall that
$k_t$-factorization is not expected to provide accurate results for
$p_t$ integrated observables in A+A
collisions. Furthermore,
we restrict to the initial ``gluon liberation'' process and do not
account for subsequent gluon multiplication towards
thermalization~\cite{Baier:2000sb,Blaizot:2010kh,Baier:2011ap}.
Nevertheless, it is of course important to check that basic trends
such as the centrality dependence of $dN/dy$ are consistent with
theoretical expectations regarding the dependence of the saturation
momentum on the thickness of a
nucleus~\cite{Kharzeev:2000ph,Kharzeev:2007zt}.

\begin{figure}[htb]
\begin{center}
\includegraphics[width=0.49\textwidth]{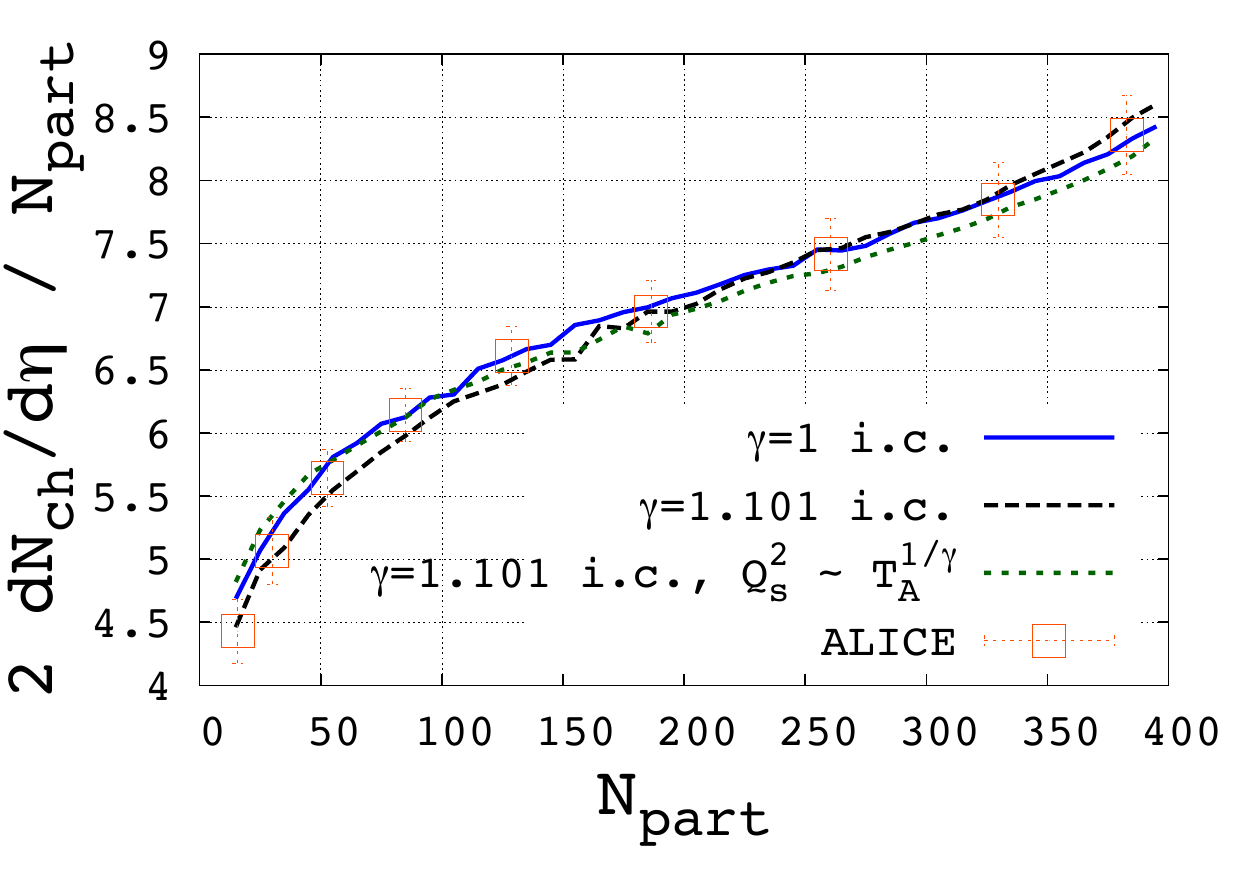}
\includegraphics[width=0.49\textwidth]{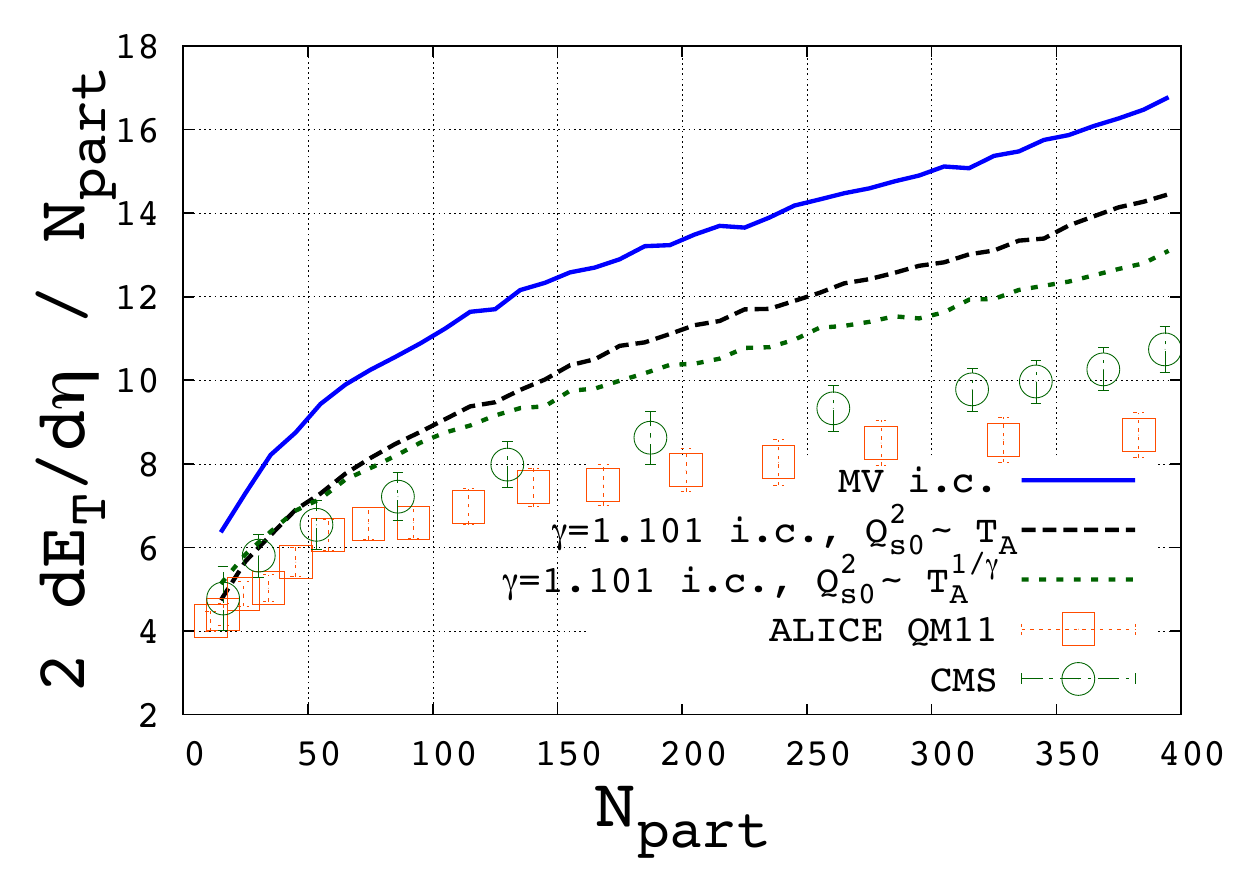}
\end{center}
\vspace*{-0.14cm}
\caption[a]{Left: Centrality dependence of the charged particle multiplicity
  at midrapidity, $\eta=0$; Pb+Pb collisions at 2760~GeV. We compare
  our calculation for two UGDs to data by the ALICE collaboration.
  Right: Centrality dependence of the transverse energy at $\eta=0$.}
\label{fig:PbPb_Centrality}
\end{figure}
We shall focus on the centrality dependence of the charged particle
multiplicity at central rapidity, $\eta=0$, which we determine along
the lines described at the beginning of section~\ref{sec:3}. The result is shown
in Fig.~\ref{fig:PbPb_Centrality} for the UGDs with MV-model
($\gamma=1$) and AAMQS ($\gamma=1.101$ and $Q_s^2(x_0)\sim T_A$ or
$Q_s^2(x_0)\sim T_A^{1/\gamma}$) initial conditions.  We use $K=1.43$
for the former and $K=2.0$, 2.3 for the latter. The number of
final hadrons per gluon is $\kappa_g=5$ in all three cases.

Aside from normalization factors, all UGD sets give a rather similar
centrality dependence of the multiplicity, and are in good agreement
with ALICE data~\cite{Aamodt:2010pb,Aamodt:2010cz}. On the other hand,
they differ somewhat in their prediction for the transverse
energy. This is of course due to the fact that the $\gamma>1$
initial condition suppresses the high-$k_t$ tail of the
UGD\footnote{One should keep in mind though that our estimate of the
  initial transverse energy carries a significant uncertainty of at
  least $\pm15\%$ related to our choice of $K$-factor; it is {\em not}
  determined accurately by the multiplicity since the latter involves
  only the product of $K$-factor and gluon $\to$ hadron multiplication
  factor $\kappa_g$.}. With the $K$-factors mentioned above these UGDs
match the measured $E_t$ in peripheral collisions. This is a sensible
result since one does not expect large final-state effects in very
peripheral collisions.  For the most central collisions the energy
deposited initially at central rapidity is about 0.5\% of the energy
of the beams and exceeds the preliminary measurements by
ALICE~\cite{Collaboration:2011rta} and CMS~\cite{Chatrchyan:2012mb} by
roughly 50\%.  This leaves room for $-p\, \Delta V$ work performed by
longitudinal hydrodynamic
expansion~\cite{Gyulassy:1997ib,Eskola:1999fc,Dumitru:2000up}.

\subsection{The ridge}

In this section, we discuss the ``ridge'' structure observed in
heavy-ion collisions, and in high-multiplicity p+p and p+Pb
collisions at the LHC. A review dedicated entirely to this phenomenon
has appeared recently in ref.~\cite{Kovner:2012jm}. The ``ridge'' refers to two-particle correlations at {\em small} relative azimuth, $\Delta\phi\sim0$, which extends over at least several units of relative (pseudo-)rapidity $\Delta\eta$, perhaps even all the way to the sources. These do not correspond to particles
produced in a binary collision since those would appear in opposite
azimuthal hemispheres, due to transverse momentum conservation.

\begin{figure}[t]
\begin{center}
\includegraphics[width=0.59\textwidth]
                {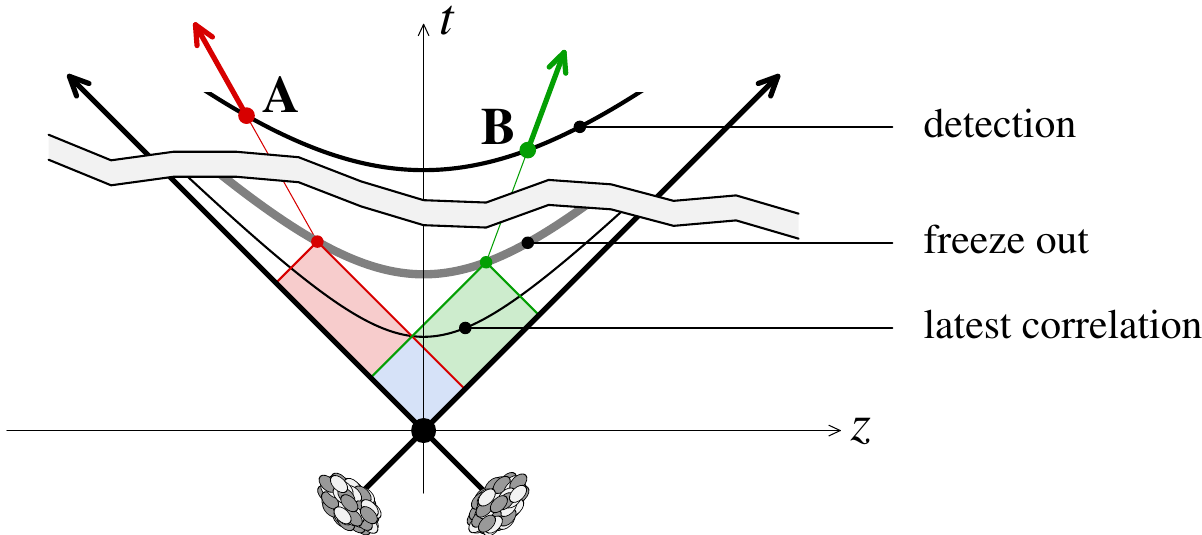}
\includegraphics[width=0.4\textwidth]
                {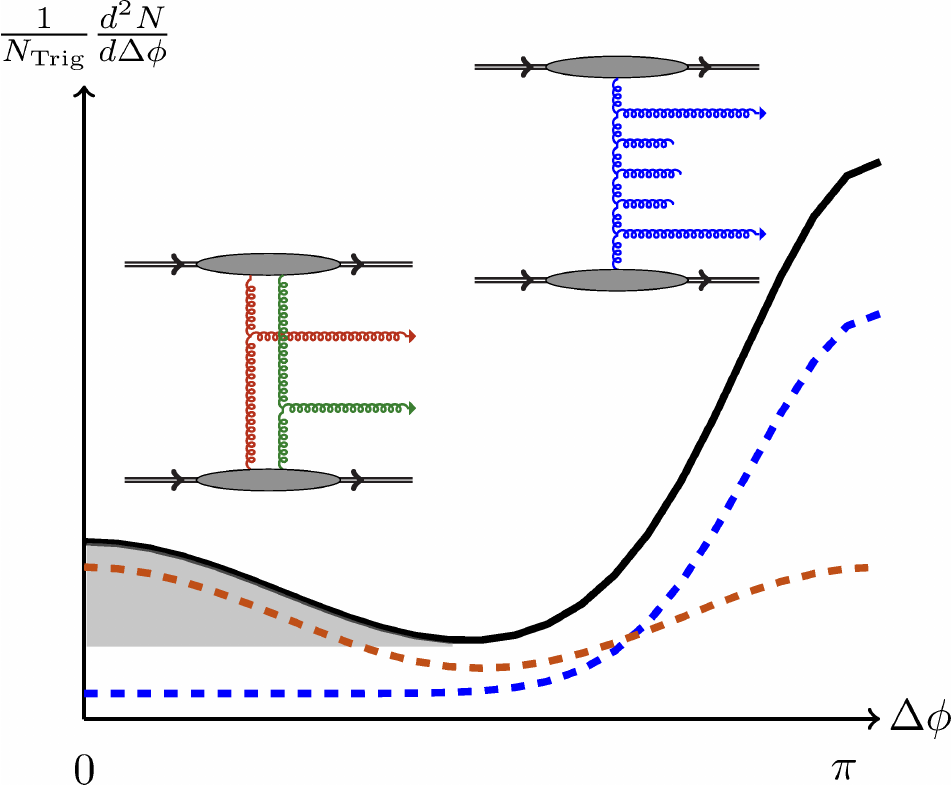}
\end{center}
\caption[a]{Left: the space-time location of events that may correlate two particles is the intersection of their past light-cones. Correlations between particles widely separated in rapidity are due to early-time dynamics~\cite{Dumitru:2008wn}. Right: Illustration of the diagrams that dominate at
different $\Delta\phi$~{\protect\cite{Dusling:2012wy}}.}
\label{fig:ridge_dijet}
\end{figure}

What makes the ridge particularly interesting is that particles
separated by a large pseudo-rapidity interval $\Delta\eta$ are causally
disconnected and hence cannot be correlated, unless they were
produced early on at a time on the order of $\tau\sim \tau_f\, \exp
(-\Delta\eta/2)$ or less~\cite{Dumitru:2008wn}, see fig.~\ref{fig:ridge_dijet} (left). In the absence of final-state collective flow, the ridge reflects the non-trivial momentum correlations built up in the very early stages of the the scattering process, as illustrated for instance in fig.~\ref{fig:ridge_dijet} (right).

By contrast, if the system later becomes ``a perfect fluid'', those initial QCD momentum correlations are altered. Some may be washed away while others may be enhanced. Actually, the ridge was first observed in Au+Au collisions at RHIC~\cite{Adams:2005ph,Abelev:2009af,Alver:2009id}, and therefore was difficult to interpret directly from CGC calculations. Contrary to the case of the total multiplicities discussed before, the post-Glasma phases cannot be ignored: both the amplitude as well as the azimuthal collimation of the initial momentum correlations are strongly modified by the fluid behavior of the QCD matter in the later stages. Nevertheless, the momentum correlations created by the fluid behavior still largely reflect properties of the initial state, albeit in another way: it is the initial spatial distribution of the QCD matter which matters in this case, instead of the momentum distribution, as we shall explain below.

\subsubsection*{The ridge in high-multiplicity proton-proton collisions}

Let us start with high-multiplicity p+p collisions, for which, contrary to the case of minimum-bias collisions, the strong-field Glasma approach may be applicable. In addition, no final-state collective flow is expected. Indeed, the duration of any final-state interactions for such small systems (in the transverse direction) clearly must be much shorter than in heavy-ion collisions: hydrodynamics predicts that (energy) density gradients decay on a time scale $\sim R/c_S$ where $c_s\simeq 1/\sqrt{3}$ denotes the relativistic speed of
sound\footnote{Here, $R$ denotes the scale of gradients which may even
  be substantially smaller than the overall radial extent of the
  collision zone. However, the radial expansion of very small density
  spikes is damped by viscosity.}. Within the CGC framework it is only the local density of gluons that matters but not the size and life-time of the system.

\begin{figure}[t]
\begin{center}
\includegraphics[width=0.47\textwidth]
                {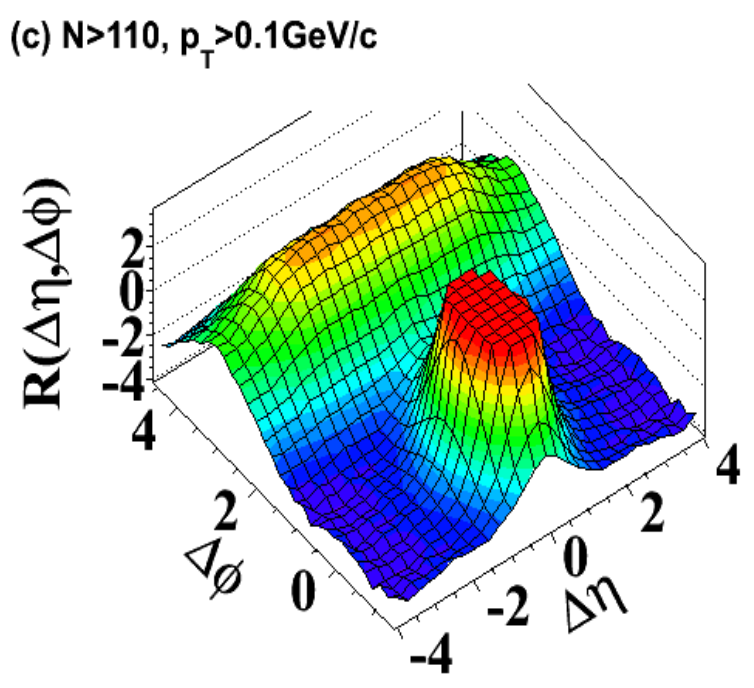}
\includegraphics[width=0.52\textwidth]
                {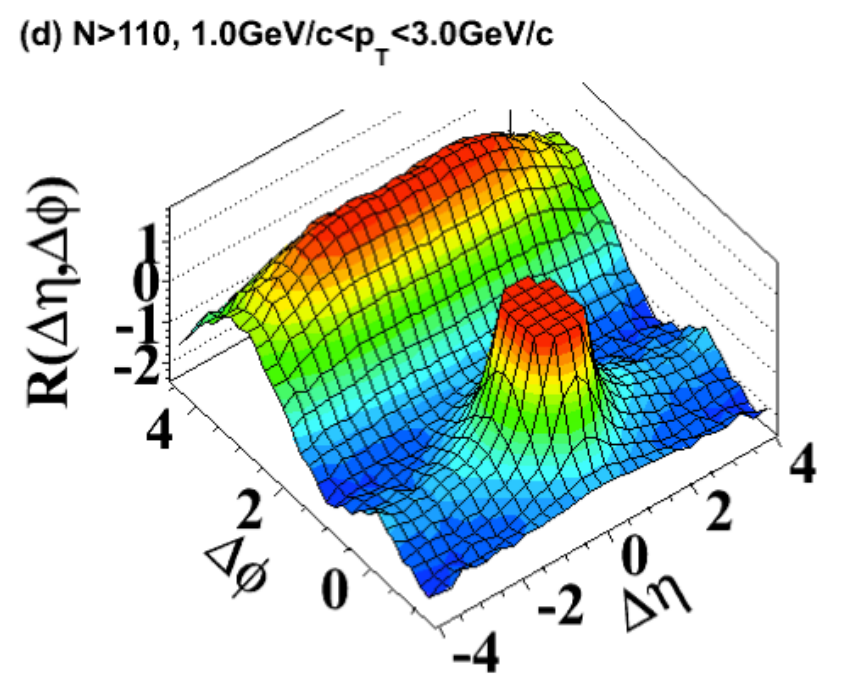}
\end{center}
\caption[a]{CMS measurement of di-hadron correlations in high-multiplicity p+p collisions {\protect\cite{Khachatryan:2010gv}}, as a function of $\Delta\phi$ and $\Delta\eta$. The data is consistent with the CGC picture, as the so-called near-side ridge is only observed when $1<p_t<3$ GeV.}
\label{fig:ridge_pp}
\end{figure}
The discovery by CMS of ridge-like correlations in high
multiplicity p+p collisions at 7~TeV~\cite{Khachatryan:2010gv}, displayed in fig.~\ref{fig:ridge_dijet}, represented a major breakthrough. The $p_t$-integrated correlation function (left plot) looks like that of a standard QCD di-jet event,
which backs up the no-flow hypothesis, while the presence of an
additional``double-ridge'' in the correlation function
for transverse momenta of order of the saturation scale (right plot) is consistent with the strong-color-field picture of \fig{fig:ridge_dijet} (right). Two hard gluons from the same ladder are nearly back to back, by conservation of transverse momentum. On the other hand, those produced in separate ladders can
have any relative angle.

The double-ridge structure appears in the case of correlated production, where the two ladders need to be connected through the sources after squaring the
amplitude. This is achieved by flipping the sources of the two ladders
in the amplitude and its c.c.; the resulting diagrams are enhanced if
the density of sources is non-perturbatively high~\cite{Dumitru:2008wn}.
In fact, this leads to a variety of
contributions (incl.\ HBT-like correlations) as discussed in more
detail in ref.~\cite{Kovchegov:2012nd}. It has been realized early on
that the ``ridge'' correlations are actually not uniform in
$\Delta\phi$ but that the relative angle peaks at $\Delta\phi\sim0$
and $\Delta\phi\sim\pi$~\cite{Dumitru:2010RBRC,Dusling:2009ni,Dumitru:2010iy}.

Another underlying picture for correlated gluon production, also invoking the presence of strong color fields~\cite{Kovner:2010xk,Kovner:2011pe,Kovner:2012jm}, is as follows. Two incoming gluons from the projectile wave function with different
rapidities scatter off the dense target field. If their relative
impact parameter is less than the correlation length $1/Q_{s,T}$ of
the dense gluons in the target, and if their color charges are alike,
they prefer to scatter to a relative azimuthal angle $\Delta\phi\sim0$
and lead to a ``near side'' ridge. Projectile gluons with opposite
color charges, a configuration which is equally likely to occur, on
the other hand prefer to scatter to
$\Delta\sim\pi$. Ref.~\cite{Kovner:2012jm} makes the interesting
prediction that this ``degeneracy'' of the near-side and away-side
ridges is lifted in case of projectile quarks; this could in principle
be tested if one could trigger on hadrons which are more likely to
originate from quark fragmentation.

\subsubsection*{The ridge in nucleus-nucleus collisions}

\begin{figure}[t]
\begin{center}
\includegraphics[width=0.49\textwidth]
                {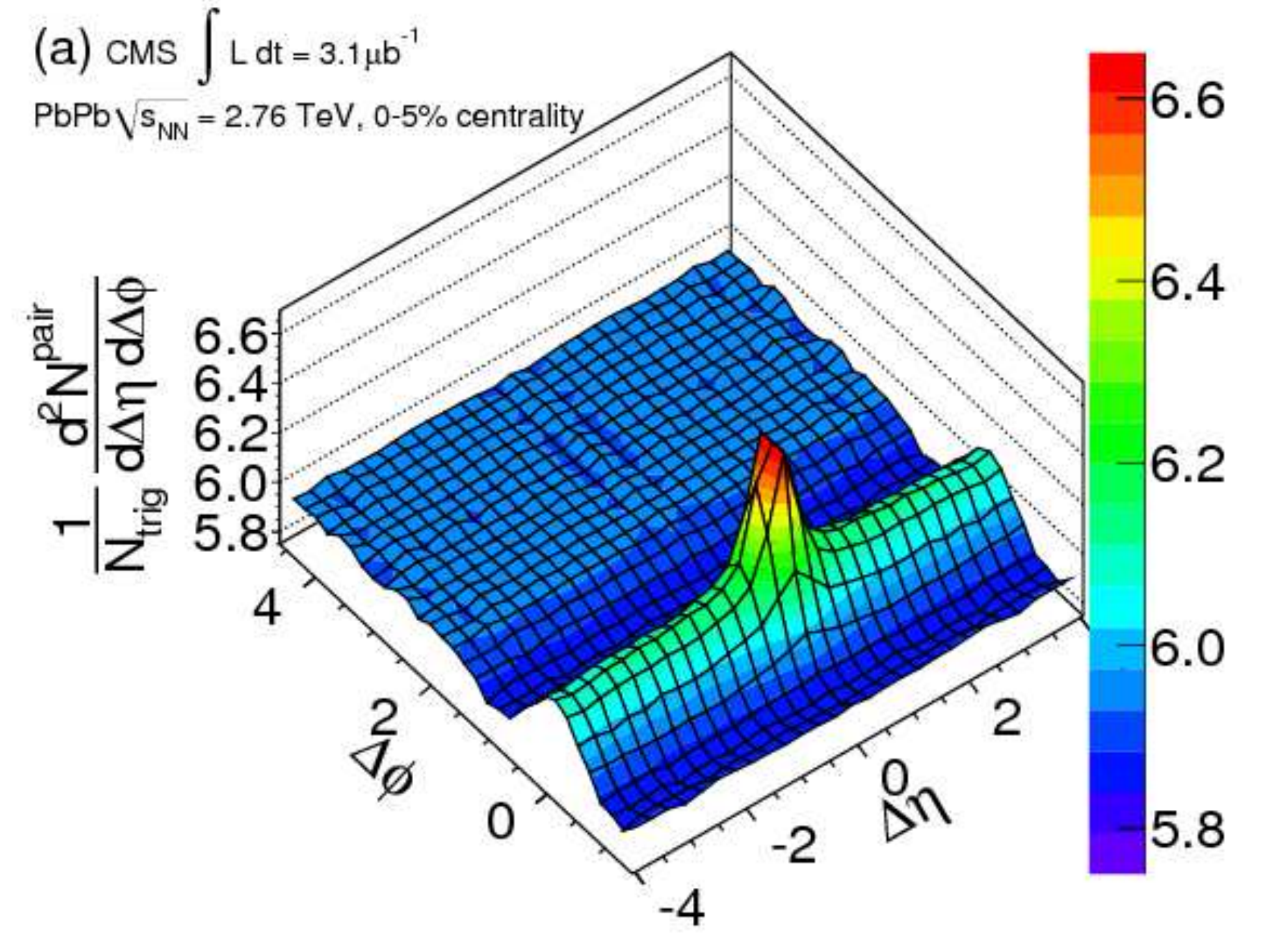}
\includegraphics[width=0.49\textwidth]
                {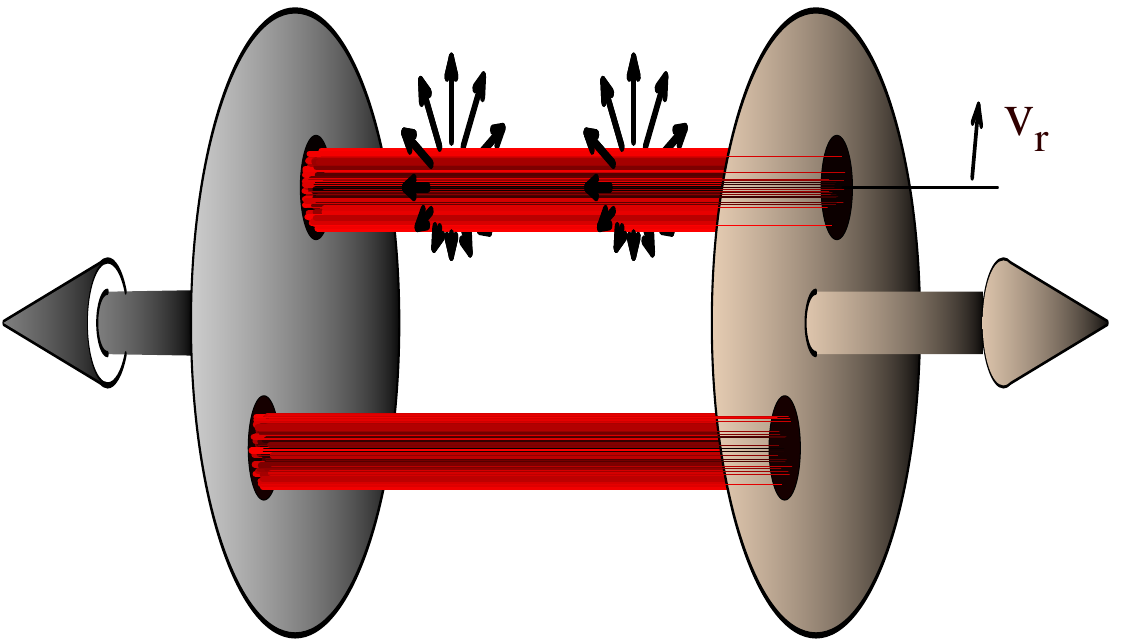}
\end{center}
\caption[a]{Left: CMS measurement of di-hadron correlations in central Pb+Pb collisions~{\protect\cite{Chatrchyan:2011eka}}, featuring long-range rapidity correlations at $\Delta\phi=0$ on both sides of the jet peak. Right: figure showing that particles emitted from the same flux tube are correlated in rapidity in a nearly boost-invariant way, and explaining how radial flow turn this a priori $\Delta\phi$-independent correlation into a ridge structure.}
\label{fig:ridge_AA}
\end{figure}
The CMS measurement of di-hadron correlations in central Pb+Pb collisions~\cite{Chatrchyan:2011eka} is shown in \fig{fig:ridge_AA} (left). In this context, the strong collective flow of the nearly-ideal hydrodynamic phase is expected to dramatically alter the pre-hydro momenta of the particles, both in the partonic and hadronic phases. In particular, it can turn $\Delta\phi$-independent, but spatially-anisotropic, correlations into a ridge-like structure. This is illustrated for instance in \fig{fig:ridge_AA} (right), where the flux-tube Glasma picture is used as an example for the initial spatial distribution of the gluons. In magnitude, such a flow-created ridge overwhelms the one due to the momentum correlations created in the scattering process.

At the same time, without initial spatial correlations, hydrodynamics does not create a ridge. Therefore, what becomes crucial about the pre-hydro QCD matter is its spatial distribution and correlations, instead of its properties in momentum space. In fact, bulk observables in heavy-ion collisions reflect these properties of the initial state as much as those of the quark-gluon plasma (QGP) phase. As initial-state studies have historically been given less importance overall, we have reached a point where the main source of error in the extraction of medium parameters (e.g. $\eta/s$) is our insufficient understanding of the initial colliding nuclei, and more precisely of the spatial fluctuations of the initial small-$x$ Glasma field.

Even though a QCD-based description is conceptually more satisfactory, in practice the Glauber-model initial conditions provide an alternative to the CGC approach to describe the nuclear geometry and it's fluctuations, which is equally successful in the case of heavy-ion collisions. The same cannot be said for p+A collisions, QCD dynamics cannot be ignored in this case, as detailed below and in section \ref{sec:IC}.

\subsubsection*{The ridge in high-multiplicity proton-nucleus collisions}

\begin{figure}[t]
\begin{center}
\includegraphics[width=0.53\textwidth]
                {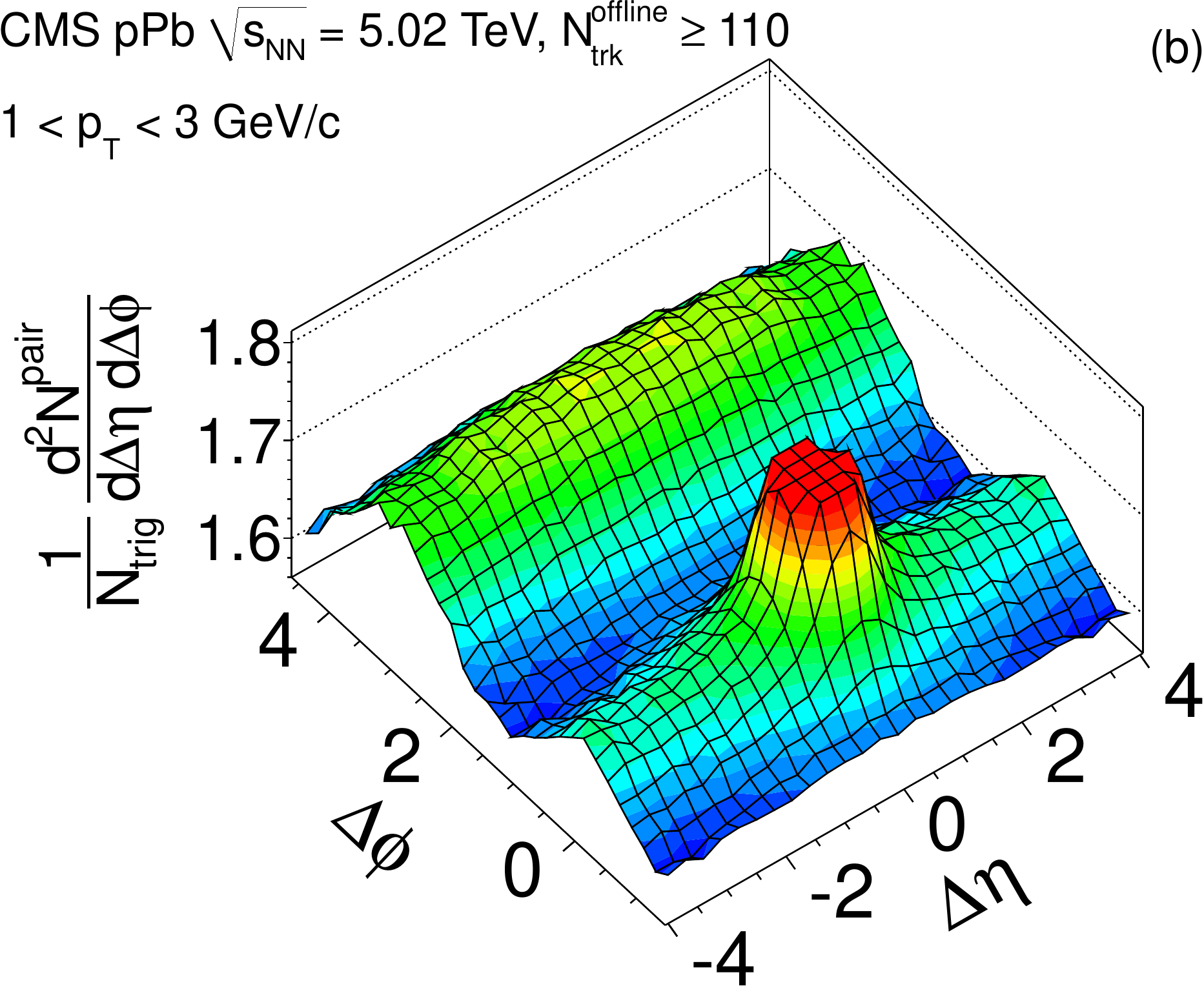}
\includegraphics[width=0.46\textwidth]
                {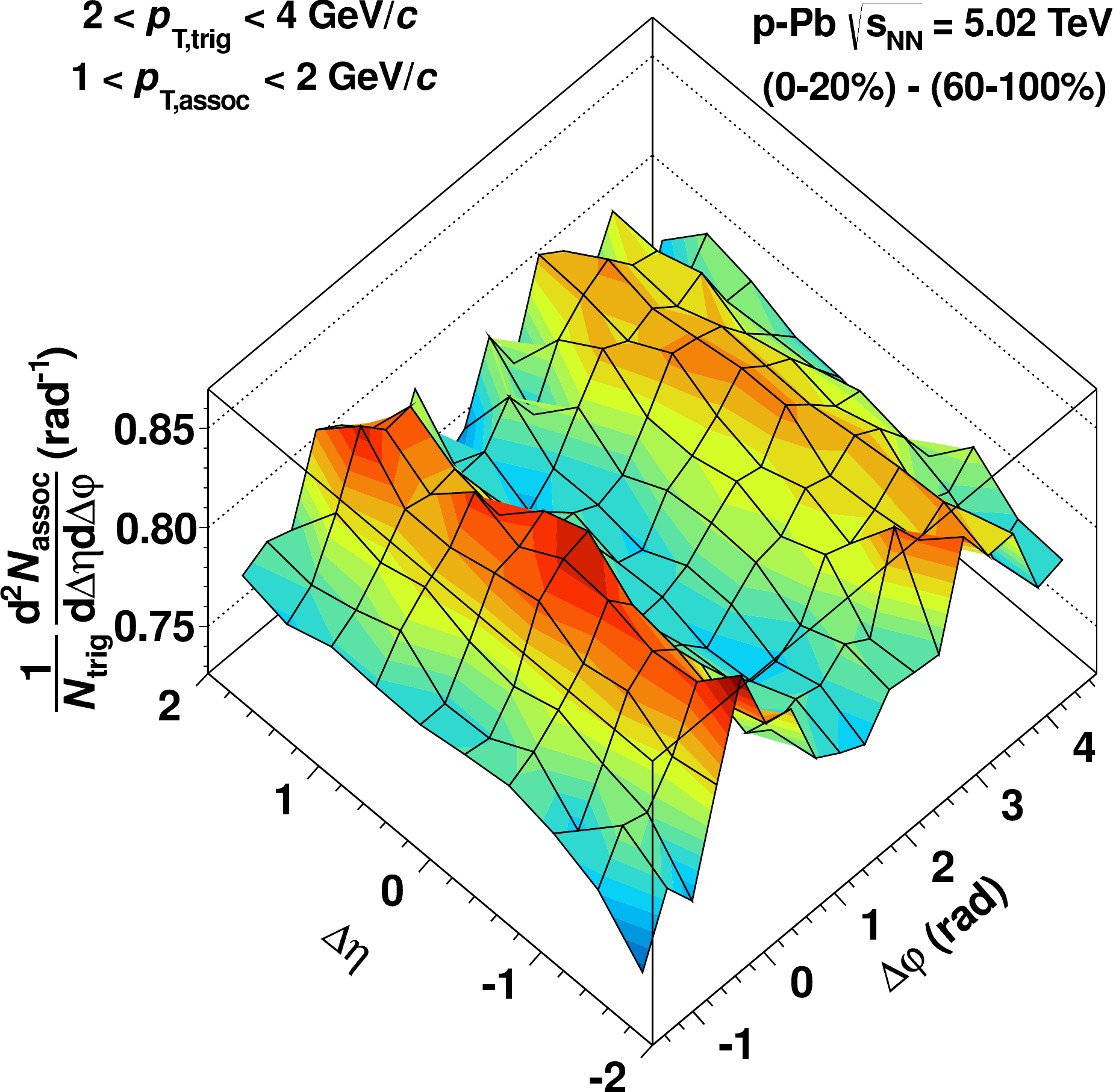}
\end{center}
\caption[a]{Left: the ridge in high-multiplicity p+Pb collisions at 5~TeV as
  seen by CMS~{\protect\cite{CMS:2012qk}}; particles with $1~{\rm GeV}\le p_t\le
  3~{\rm GeV}$. Right: the ALICE ridge with {\em subtracted} jet yield
  reveals the double-ridge structure~{\protect\cite{Abelev:2012cya}}. ATLAS results
  for the p+Pb ridge are presented in ref.~{\protect\cite{Aad:2012ridge}}.}
\label{fig:ridge_pPb}
\end{figure}
CMS has recently confirmed the presence of ridge correlations also for
high-multiplicity p+Pb collisions at 5~TeV as shown in
\fig{fig:ridge_pPb}. While the radial extent of the collision
zone and the particle density is comparable to that from
high-multiplicity p+p collisions the observed amplitude of the ridge
is larger. In the absence of final-state collective flow (we will come back to this hypothesis below), then  the p+A ridge reflects the QCD momentum correlations generated during the scattering process, same as in p+p collisions. In the CGC picture, the increased magnitude of the ridge in naturally explained by the larger saturation scale of the nucleus compared to the proton.

\begin{figure}[t]
\begin{center}
\includegraphics[width=0.4\textwidth]
                {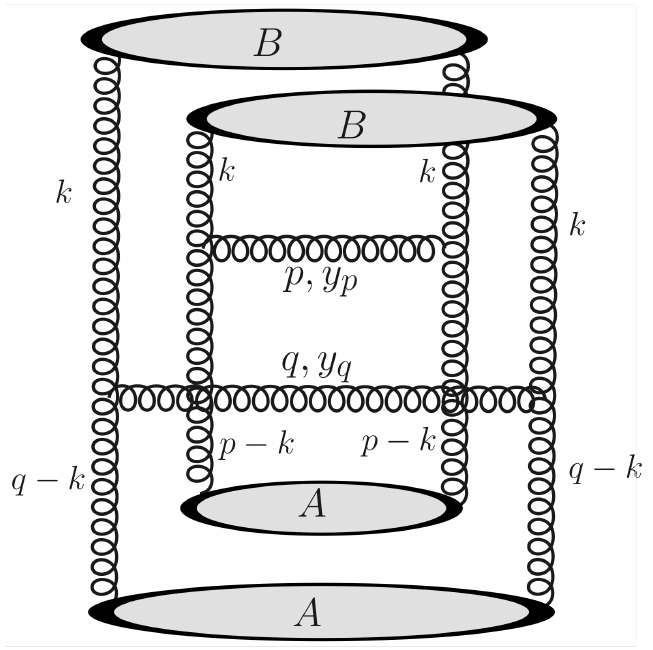}
\includegraphics[width=0.59\textwidth]
                {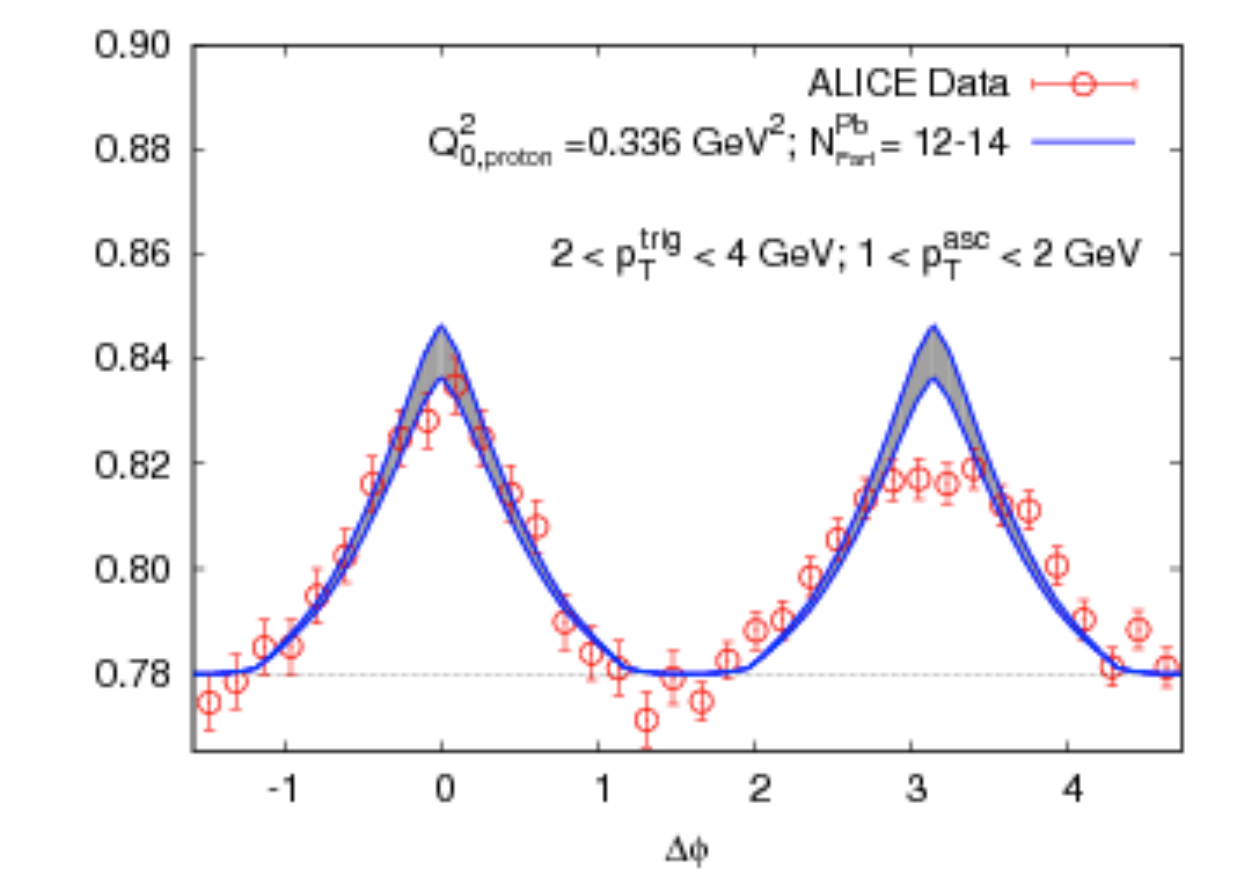}
\end{center}
\caption[a]{Left: a typical diagram that generates an angular collimation around $\Delta\phi=0$ nearly independent of $\Delta\eta$, the rapidity separation between the two hadrons. However, it is non-zero only for hadron transverse momenta comparable to $Q_s$, and it is dominant only at large $\Delta\eta$ away from the jet peak. Right: Comparison of Glasma graphs to ALICE data for the collimated yield per trigger per $\Delta\eta$ for central (0-20\%) events with (60-100\%) peripheral events subtracted versus $\Delta\phi$, from ref.~{\protect\cite{Dusling:2013oia}}.}
\label{fig:cgc-alice}
\end{figure}
The current state-of-the-art quantitative results are due to
Dusling and Venugopalan~\cite{Dusling:2012iga,Dusling:2012wy}. \fig{fig:cgc-alice} shows the diagram responsible for the double-ridge $\Delta\phi$ dependence
in the CGC calculation (left), and the comparison with the ALICE data (the right part of Fig.~\ref{fig:ridge_pPb}) at fixed large $\Delta\eta$ (right). As a matter of fact, the entire ``matrix'' of correlations as a function of the transverse momenta of the trigger and associated particles, and for different bins of multiplicity is well described. This can be seen in fig.~\ref{fig:ridge_matrix} (left).

\begin{figure}[t]
\begin{center}
\includegraphics[width=0.45\textwidth]{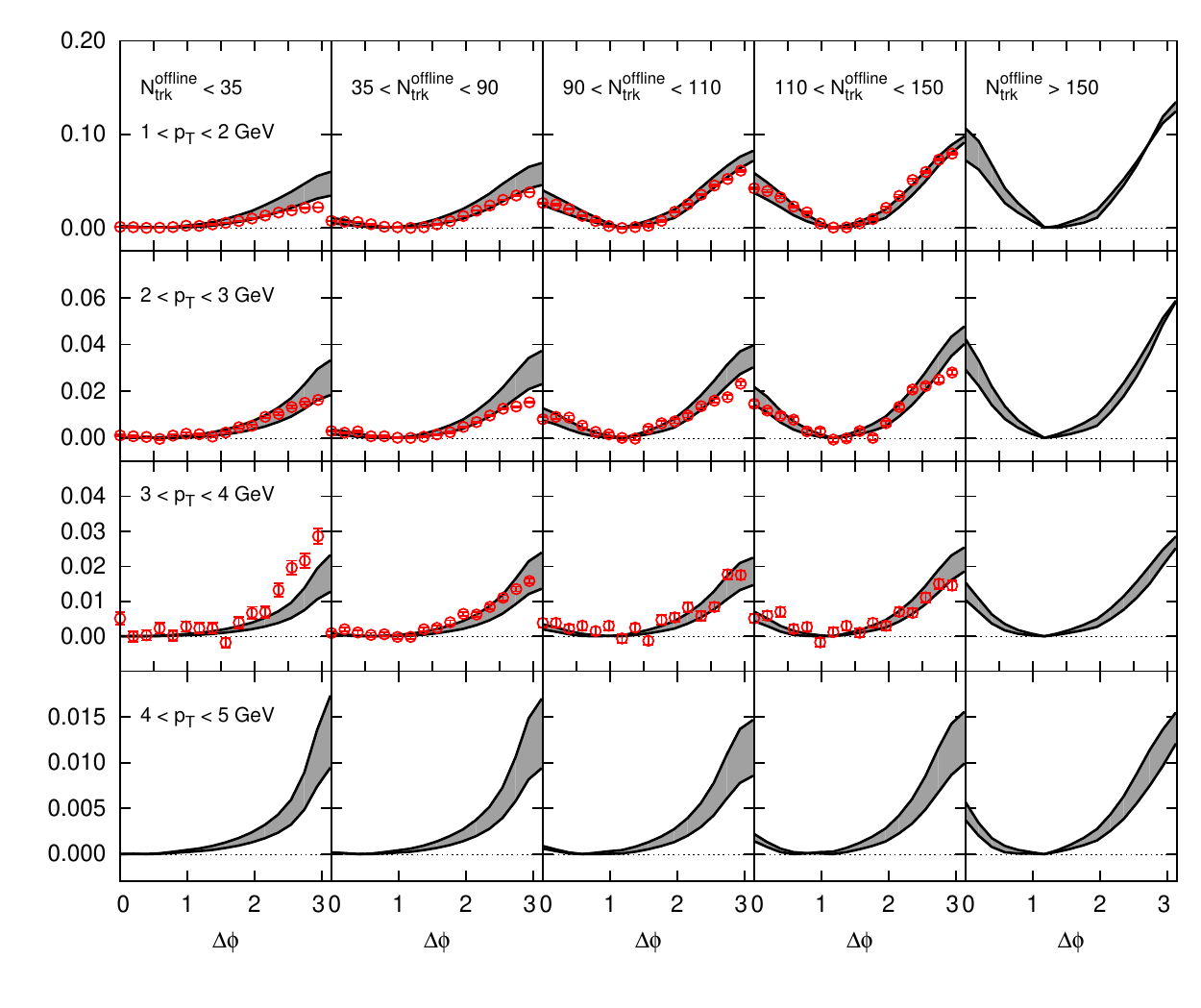}
\includegraphics[width=0.54\textwidth]{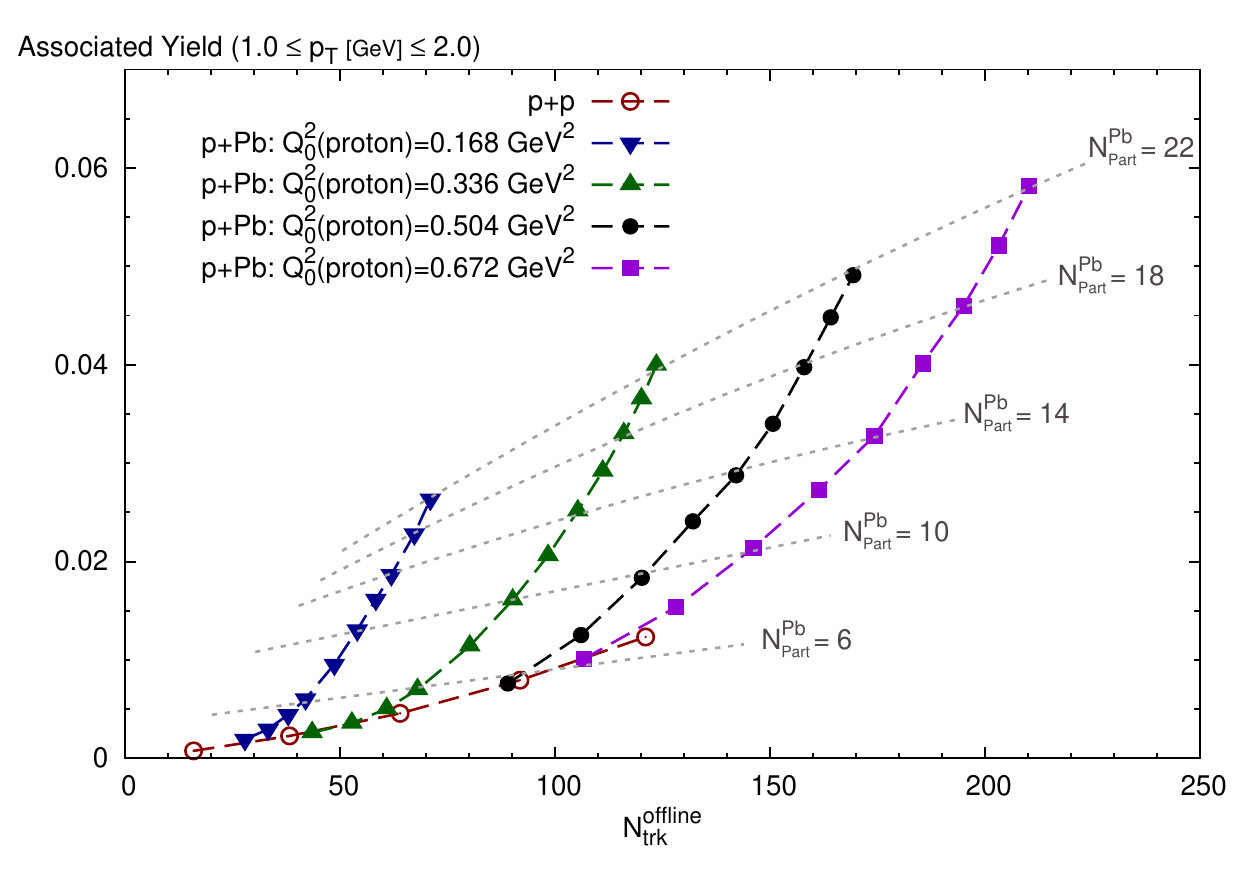}
\end{center}
\caption[a]{Left: correlated particle yield after ZYAM subtraction as
  a function of relative azimuth in different centrality bins. Right:
  associated near-side yield as a function of multiplicity $N_{\rm
    trk}^{{\rm offline}}$: dashed curves / symbols show how the yield
  increases with the number of participants in the Pb target, thin
  dotted lines track its dependence on the valence charge density in
  the proton. Both figures from ref.~{\protect\cite{Dusling:2012wy}}.}
\label{fig:ridge_matrix}
\end{figure}
Fig.~\ref{fig:ridge_matrix} (right) shows how the associated yield increases with the overall multiplicity $N_{\rm trk}^{{\rm offline}}$. For p+p collisions a very high multiplicity is due to rare extreme density fluctuations in either of the nucleons; in p+Pb collisions high multiplicity fluctuations are helped by the presence of additional Glauber fluctuations of the target thickness. Folding
the two, ref.~\cite{Dusling:2012wy} obtains a very natural explanation
for the increased near-side yield in p+Pb as compared to p+p at the
same multiplicity $N_{\rm trk}^{{\rm offline}}$.

If the no-flow hypothesis in p+A collisions is true, then the GCG should also be able to reproduce the large triangular flow observed experimentally. There are some indications that in fact the CGC alone is not able to do that, using the IP-glasma model followed by classical Yang-Mills evolution of the Glasma field \cite{Raju}. This would mean that p+A collisions are more like A+A collisions, and that the p+A ridge is a result of hydrodynamic evolution. In this hypothesis, the problem would be, as in heavy-ion collisions, to describe the nature and the dynamics of the pre-hydro fluctuations. This also bodes well for the CGC approach, since in proton-nucleus collisions one cannot rely on the Glauber model anymore.

Let us conclude this section by pointing out that, if p+A collisions are ``contaminated'' by flow, the A dependence of the intrinsic two-particle correlations could still be measured in e+A collisions.

\subsection{Initial Conditions for transport}
\label{sec:IC}

As outlined in the previous section, the CGC is of utmost importance in the description of high-energy collisions, as it provides a framework to determine the early stages immediately after a hadronic or a heavy-ion collision. This refers to the distribution of produced particles in momentum space, relevant in the absence of a subsequent collective behavior of the system, or by contrast in transverse coordinate space, to characterize the initial condition for the following thermalization and (viscous) hydrodynamic stages.

In this section we deal with the latter case. Hydrodynamics converts initial spatial deformations into momentum anisotropies in the final state through anisotropic pressure gradients. The conversion efficiency essentially depends on the shear viscosity, i.e. on how ideal the hydrodynamic phase is. A proper combination of CGC initial condition with hydrodynamics is necessary in order to interpret the so-called bulk observables, and correctly extract medium parameters. We start by discussing the case of heavy-ion collisions, and then we briefly address proton-nucleus collisions, for which much less has been done, due to the lack of relevant data until very recently.

The detailed dynamics of thermalization is often ignored and the $p_t$-integrated distribution of produced particles in the transverse plane is used directly to initialize hydrodynamic simulations. This appears reasonable if one is interested mainly in the features of $dN/dyd^2b_T$ over length scales larger than the
thermalization time $\tau_{\rm th}$. To actually understand the thermalization process in high-energy heavy-ion collisions is a much more complicated task, it requires to describe how the full Glasma energy momentum tensor is converted into one compatible with hydrodynamics. This topic is currently under intense investigation; we refer the interested reader to a recent review \cite{Gelis:2012ri}.

\begin{figure}[t]
\begin{center}
\includegraphics[width=0.34\textwidth]{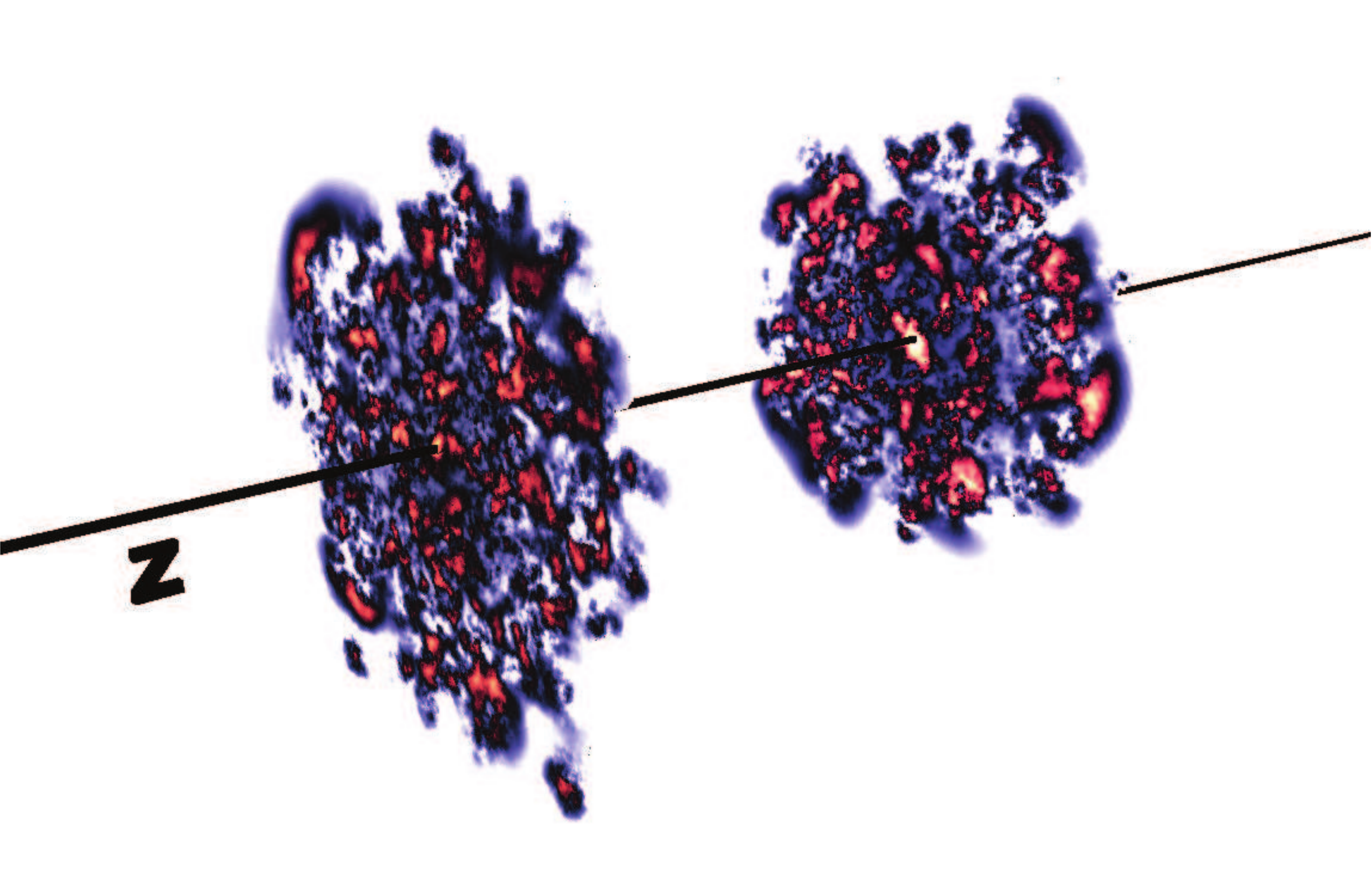}
\hfill
\includegraphics[width=0.3\textwidth]{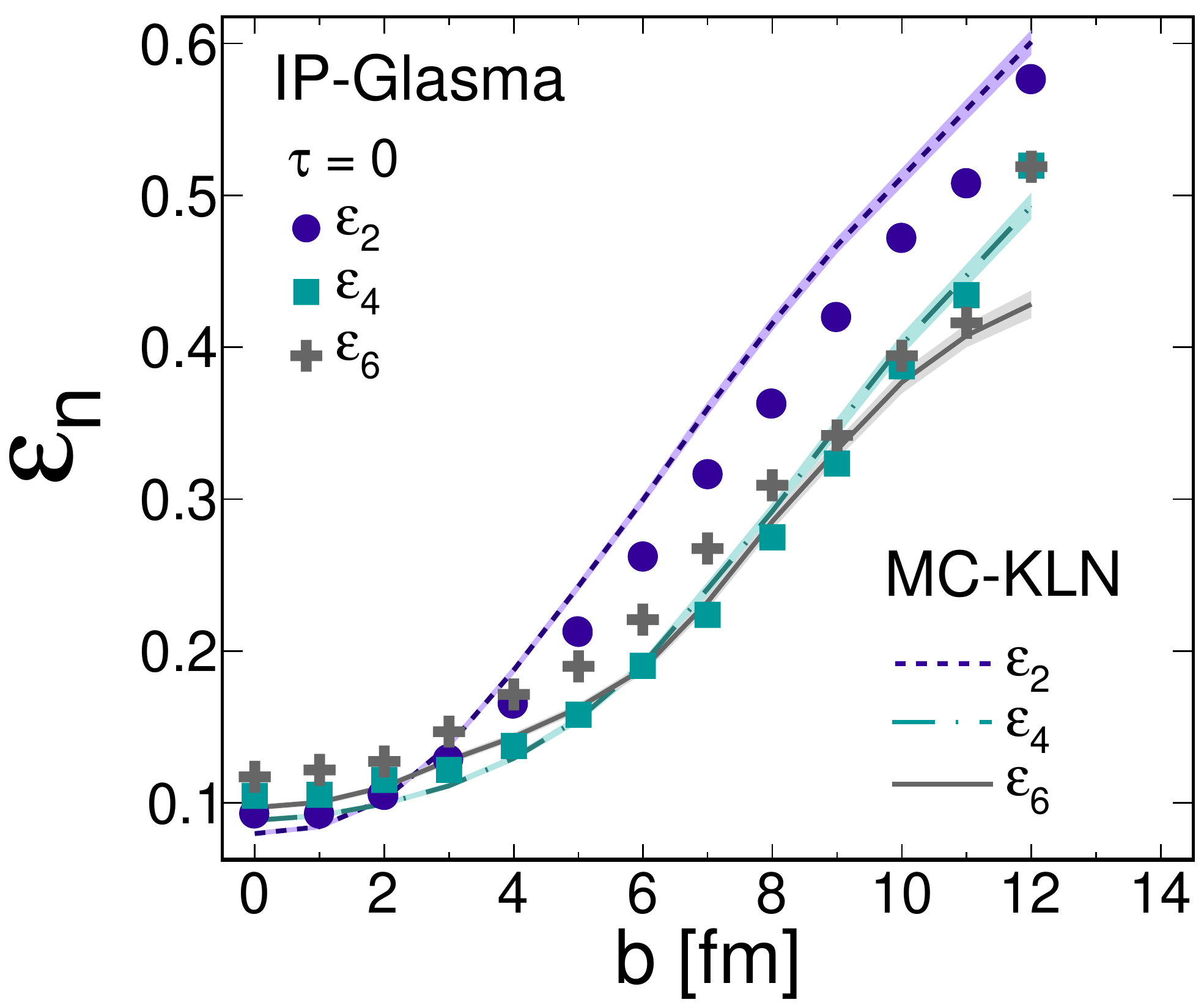}
\hfill
\includegraphics[width=0.31\textwidth]{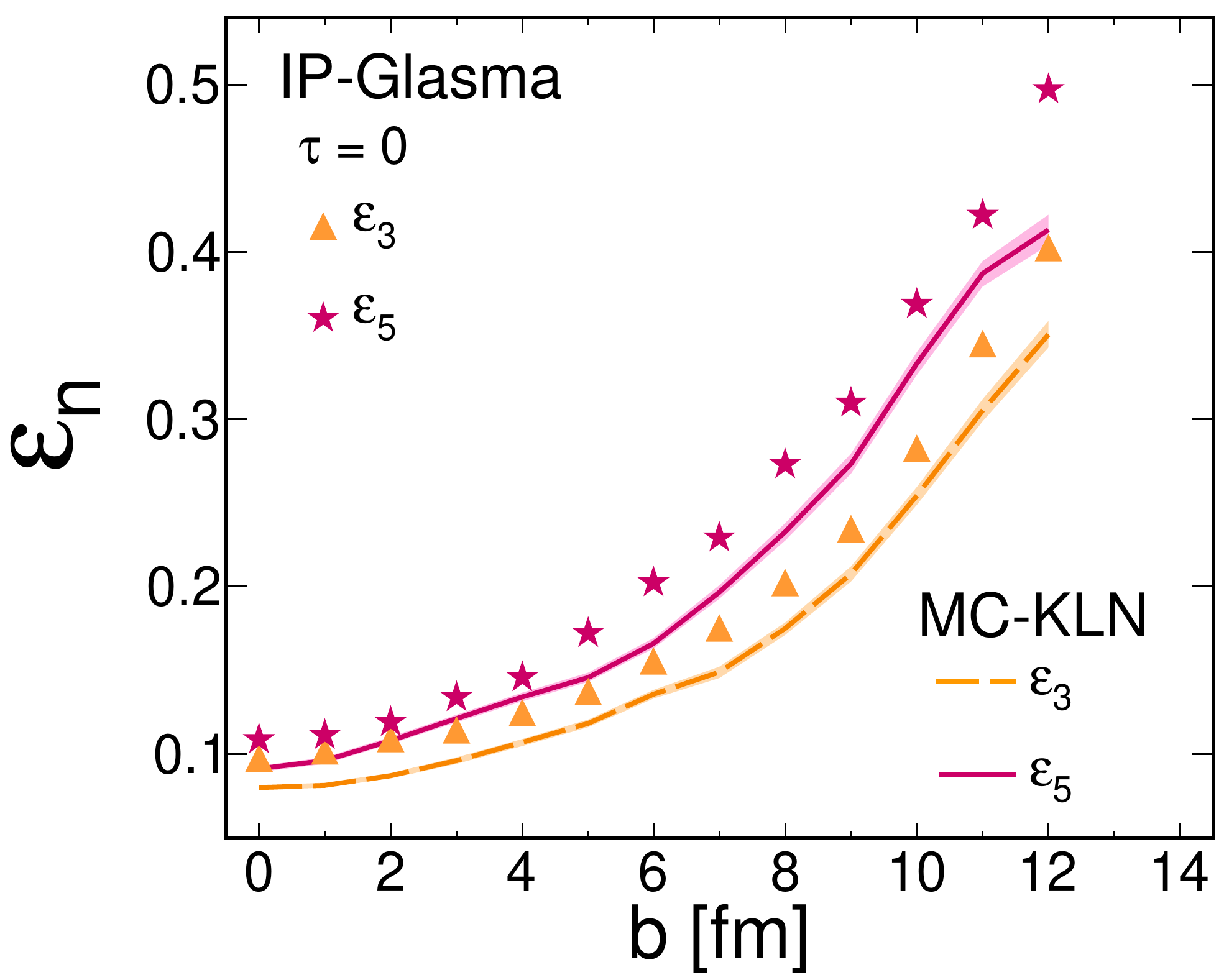}
\end{center}
\caption{Left: picture of two colliding nuclei in heavy-ion collisions, and of their fluctuating small-x gluon fields. Right: from ref.~{\protect\cite{Schenke:2012hg}}, predictions of two CGC-motivated models for the first even and odd eccentricity harmonics, crucial inputs to QGP evolution models.}
\label{fig:eccentric}
\end{figure}

Instead of resorting to the CGC formalism a much simpler shortcut to
the initial $dN/dyd^2r_T$ is to assume that the local density of
particles produced at a point $r_T$ is simply proportional to the
average density of projectile and target participants, $dN/dyd^2r_T
\sim (\rho_{\rm part}^A(r_T) + \rho_{\rm part}^B(r_T))/2$. This
``wounded nucleon'' model, often times also labeled ``Glauber model'',
clearly must capture the rough features of $dN/dyd^2r_T$. On the other
hand, it is evident from the increase of $(1/N_{\rm part})\; dN/d\eta$ or
$(1/N_{\rm part})\; dE_T/d\eta$ from peripheral to central Pb+Pb
collisions by about a factor of two (see fig.\ref{fig:PbPb_Centrality}) that such a model is far from accurate.
There is also no reason to believe that the second harmonic moment of the
density distribution in the transverse plane, the so-called
eccentricity $\varepsilon = \langle y^{2}{-}x^{2}\rangle/ \langle
y^{2}{+}x^{2}\rangle$ should be well described (the same goes for higher
moments). Indeed, a variety of CGC models which do describe the
centrality dependence of $dN/dy$ predict higher eccentricity
$\varepsilon$ than the ``wounded nucleon model''
\cite{Hirano:2005xf,Drescher:2006pi,Drescher:2007cd,Drescher:2006ca,Schenke:2012hg}.
This is illustrated in fig.~\ref{fig:eccentric}, for $\varepsilon$ and high-order eccentricity harmonics.

The ``wounded nucleon model'' for soft particle production can be
improved by adding a semi-hard component. It has to incorporate
impact parameter and $Q^2$ dependent shadowing in order to smoothly
interpolate to p+p collisions in very peripheral collisions and to lead 
$\sim N_{\rm coll}$ scaling at high $p_t$~\cite{Helenius:2012wd};
also, the energy dependence of the low-$p_t$ cutoff required by
leading-twist calculations needs to be fixed carefully to reproduce
measured multiplicities.

The initial spatial particle distribution exhibits large
fluctuations. They manifest in non-zero elliptic flow $v_2$ in central
heavy-ion collisions~\cite{Alver:2006wh} as well as in a large
``triangular flow'' component $v_3$~\cite{Alver:2010gr}. One source
of fluctuations is due to the locations of participant
nucleons~\cite{Alver:2006wh}; these have also been incorporated early
on in Monte-Carlo implementations of the $k_t$-factorization formula
with KLN UGDs~\cite{Drescher:2006ca,Drescher:2007ax}.

\begin{figure}[htb]
\begin{center}
\includegraphics[width=0.45\textwidth]
                {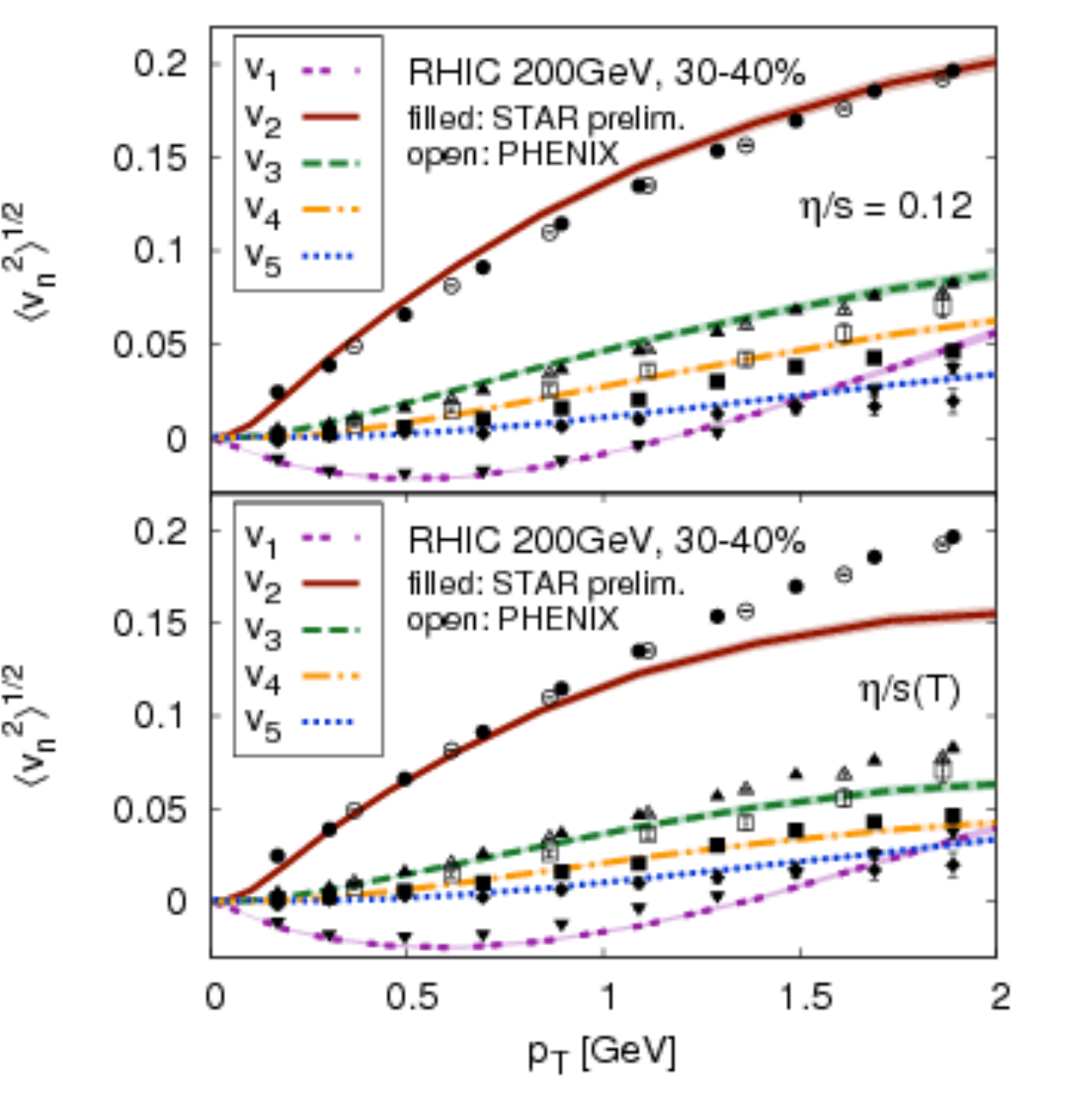}
\includegraphics[width=0.54\textwidth]
                {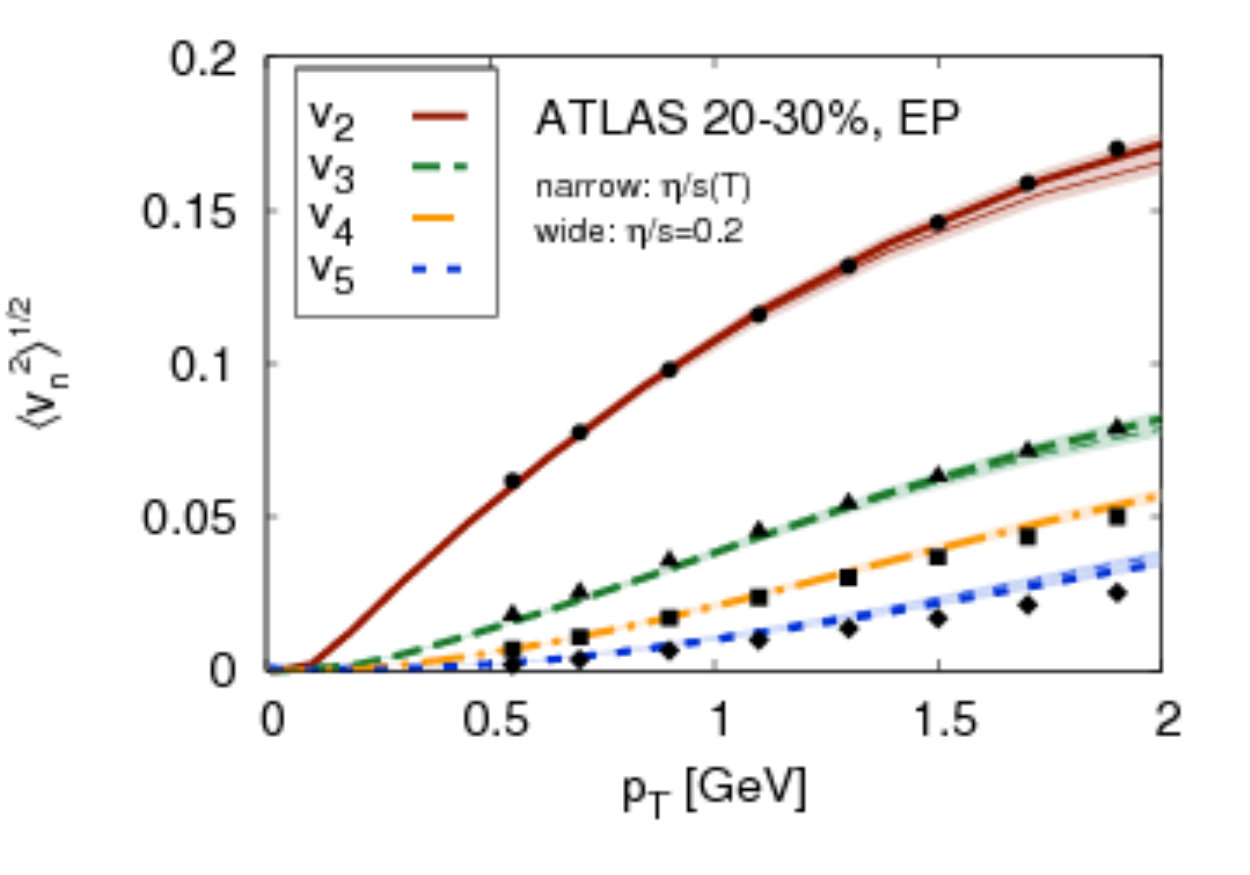}
\end{center}
\caption[a]{Theory/experience comparison for $v_n(p_T)$, using for RHIC (LHC) a constant $\eta/s=0.12$ (0.2) or a temperature dependent $\eta/s(T)$ as parametrized in \cite{Niemi:2011ix}. From ref.~{\protect\cite{Gale:2012rq}}, bands indicate statistical errors. Left: data by the PHENIX~{\protect\cite{Adare:2011tg}} (open symbols) and STAR~{\protect\cite{Pandit:2013vza}} (preliminary, filled symbols) collaborations. Right: data by the ATLAS collaboration using the event-plane (EP) method~{\protect\cite{ATLAS:2012at}} (points).}
\label{fig:eliptic}
\end{figure}

However, even for a fixed (local) number of participants there are
intrinsic particle production fluctuations. This is most evident from
the wide multiplicity distribution in non-single diffractive p+p
collisions. A suitable
extrapolation to A+A collisions has to be included both in ``wounded
nucleon''~\cite{Qin:2010pf} as well as in CGC
based~\cite{Dumitru:2012yr} Monte-Carlo models.  In the CGC approach,
intrinsic particle production fluctuations are expected to occur on
sub-nucleon distance scales on the order of
$\sim1/Q_s$~\cite{Schenke:2012wb}. It is interesting to note that early
CGC initial state models which do {\em not} incorporate intrinsic particle
production fluctuations~\cite{Drescher:2006ca,Drescher:2007ax} appear
to be inconsistent with the distributions of angular flow harmonics
measured by the ATLAS collaboration~\cite{Jia:2012ve} while recent
approaches that include sub-nucleonic quantum fluctuations and pre-equilibrium dynamics of the glasma fields can describe higher eccentricity harmonics very
well~\cite{Gale:2012rq}.

\begin{figure}[htb]
\begin{center}
\includegraphics[width=0.49\textwidth]{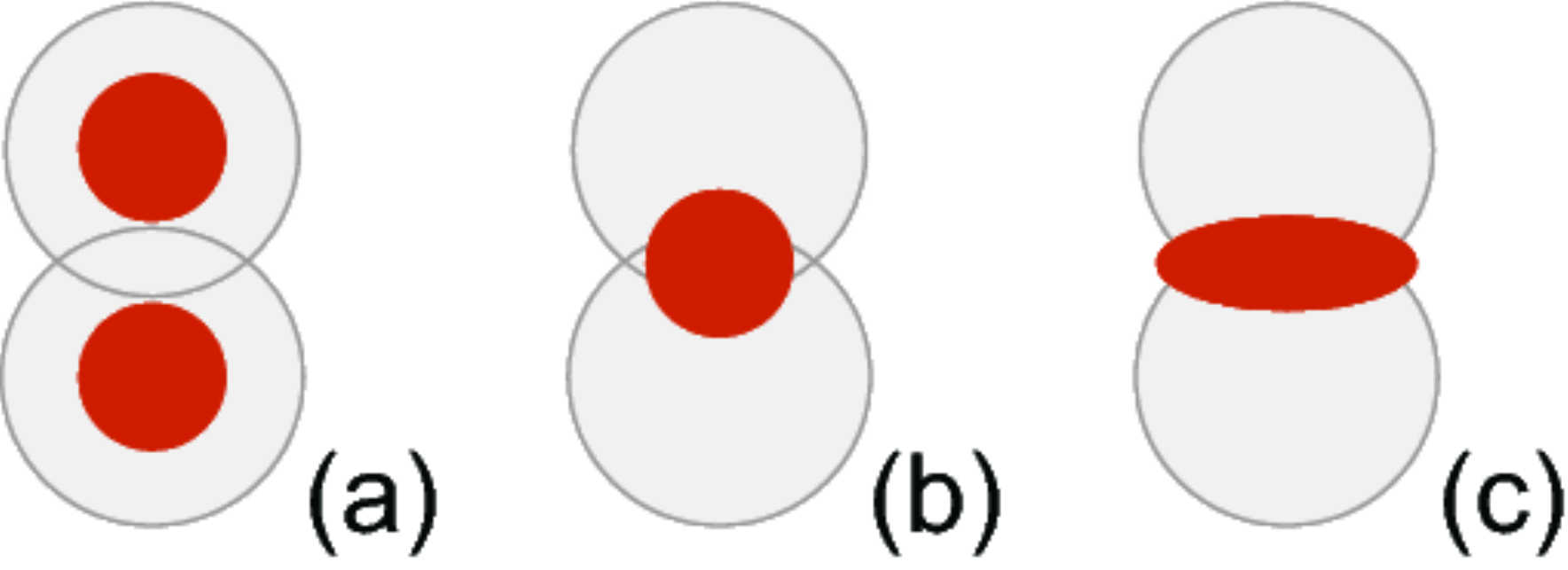}
\vspace*{-0.5cm} 
\includegraphics[width=0.49\textwidth]{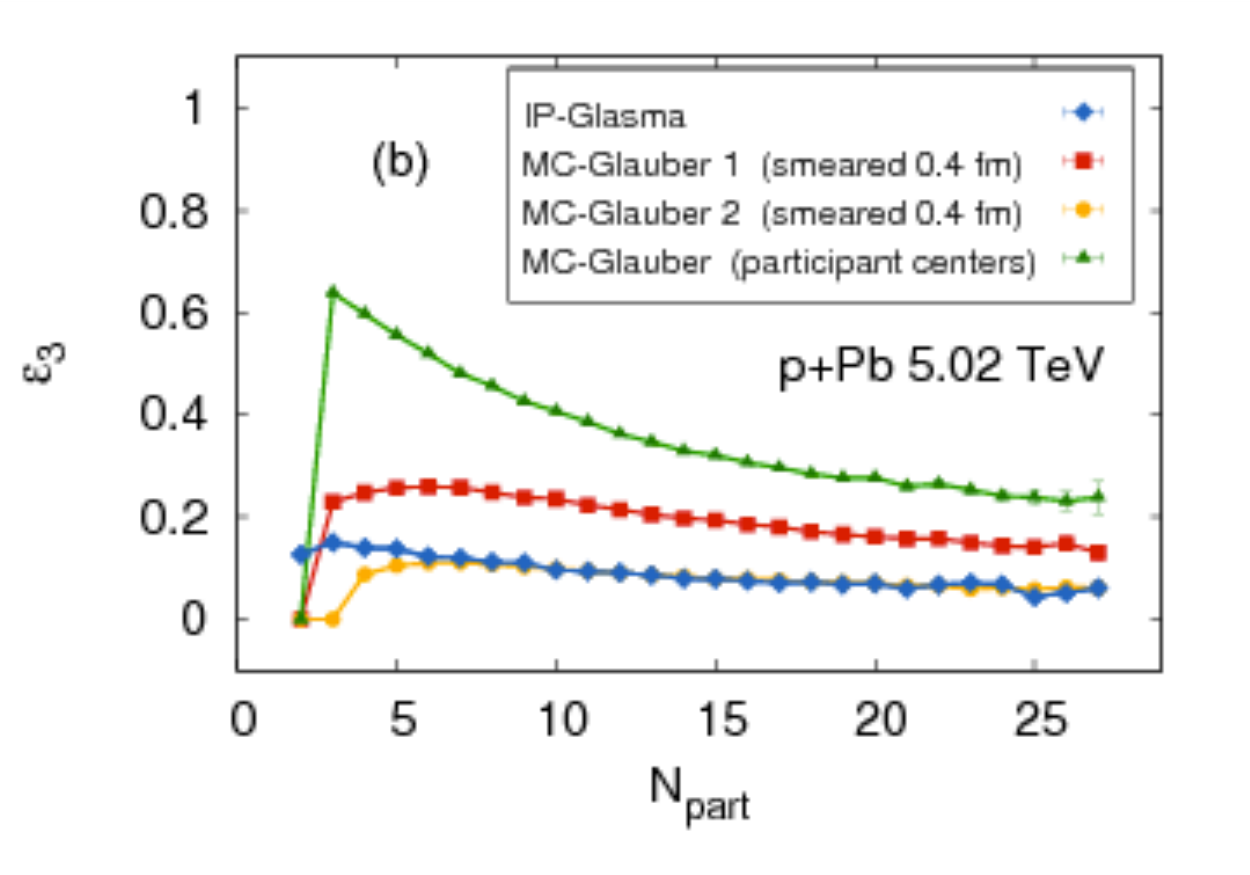}
\end{center}
\caption[a]{Left: various models of the energy density deposition (denoted by red dots) in nucleon-nucleon collisions. In the left plot (a) the energy density is produced at the center of the colliding nucleons even for grazing collisions. The center and right plots (b),(c) correspond to different eccentricities depending on the matter distribution in the nucleon overlap region. For the configuration depicted on the left eccentricity $\epsilon_2=1$, whereas for the configuration in the center $\epsilon_2=0$. Right: the eccentricity harmonic $\epsilon_3$ as a function of the number of wounded nucleons $N_{part}$. In the MC-Glauber (participant centers) model the energy density is deposited in the centers of wounded nucleons (without smearing). Smearing energy densities with the Gaussian distribution (σ0=0.4fm) results in the MC-Glauber 1 model. In the MC-Glauber 2 model the energy density is smeared about the midpoint between colliding nucleons. From ref.~{\protect\cite{Bzdak:2013zma}}.}
\label{fig:epsilon}
\end{figure}

This is shown in \fig{fig:eliptic}, along with the fact at the moment, this state-of-the-art combination of Glasma initial dynamics with viscous hydrodynamics does not allow to distinguish between a constant $\eta/s$ and temperature dependent one. A lot still has to be understood about the nature, scale, magnitude and evolution of the fundamental QCD fluctuations of the initial state. This represents a very exciting avenue for future research, in heavy-ion collisions and more even so in proton-nucleus collisions, for which very little is known. Fig.~\ref{fig:epsilon} shows how poorly things are under control in this case, and that therefore, hydrodynamic fits to p+A data should be interpreted with great care. Indeed, four different models for the energy deposition in binary nucleon-nucleon collisions yield very different eccentricity harmonics in p+A collisions.

\section{Conclusions}

One main conclusion that can be extracted from the studies of the initial stages of heavy ion collisions performed at RHIC and the LHC is that coherence effects are of the uttermost importance for a proper, self-consistent theoretical description of such processes and for their accurate phenomenological description. It is theoretically well established that, in a QCD description, coherence effects arise at small values of Bjorken-$x$ as a consequence of high gluon densities and relate to gluon self-interactions processes that control the small-$x$ evolution of hadronic wave functions and of particle production processes. 

The realization that high-energy QCD scattering is governed by a new, semi-hard scale --the saturation scale-- has provoked a change of paradigm in the field by which observables long-thought to be tractable only through non-perturbative techniques are now amenable to a description in terms of weakly coupled dynamics. However, systems of high gluon densities are still strongly correlated and characterized by non-perturbatively large color-fields $\mathcal{A}\sim 1/g$. These two ideas lay at the basis of the CGC approach, which relies on a separation of degrees of freedom into static color charges and dynamical color fields, and on a renormalization group equation that copes with the arbitrariness in the choice of the scale at which that separation is implemented. This confers the CGC features of an effective theory. The substantial theoretical and phenomenological progress attained over the last two decades has promoted the CGC as the best candidate to characterize the initial stages of nucleus-nucleus collisions. Besides its crucial role as a practical analysis tool, the CGC has also become a useful organizing principle for a variety of observed phenomena, suggesting new ways to look at the data and new measurements.

While no definitive conclusion can be extracted from the analyses of presently available data, it is fair to say that many observables from a variety of collision systems find their natural explanation and a good quantitative description in terms of non-linear dynamics associated to the presence of large gluon densities. Thus, the simultaneous and unified description of small-$x$ data from electron-proton, proton-proton, proton(deuteron)-nucleus and nucleus-nucleus collisions achieved within the CGC framework lends support to the idea that saturation effects are a central ingredient in these collisions. 
This consistent body of phenomenological work validates the universality of the saturation mechanism and provides a strong consistency check of the analyses performed in the field of heavy ion collisions.

Despite the intrinsic technical difficulty of higher-order calculations in the CGC --one should recall that they are performed by expanding on a strong background color field-- progress on the theoretical side has been steadily delivered over the last years.
One main line of research has been the calculation of next-to-leading order corrections, both at the level of the evolution equations and of particle production processes. These corrections turn out to have a strong impact on the leading order results, the most remarkable effect being a significant slowdown of the evolution speed. This has bridged the previous gap between theoretical expectations and data and has opened a new period of phenomenological works based on first-principles calculations. On the negative side, the incipient studies of NLO corrections to particle production in the hybrid formalism indicate that they  overwhelm the LO contribution for large enough transverse momenta, yielding negative cross sections and putting temporarily on hold the hope that they should render the theory closer to the physics of collinear factorization and DGLAP evolution of dilute parton densities. Also, present phenomenological works tend to combine theory ingredients with a varying degree of accuracy, which sets theoretical uncertainties out of control. Very important, the full NLO corrections to some processes now available allow for the use of theoretically well defined fixed-order perturbative schemes. The theoretical understanding of exclusive processes associated to correlations of the produced particles has also been significantly advanced recently.

Another important theoretical directions consist in the study of the system produced after a heavy ion collision, the Glasma, the correlations build up during this stage of the collision, and the study of the thermalization dynamics. Recent works have been able to provide the first compelling indications that the weakly coupled CGC dynamics may lead to the onset of hydrodynamical behavior over the extremely short time scales suggested by hydro analyses. Although further research is needed, this result would imply a major breakthrough in the field and would provide a smooth, well controlled matching with the QGP phase of heavy ion collisions.

The extremely precise data from the RHIC and LHC programs poses enormous challenges to the phenomenological approaches, which must parallel the degree of refinement and exclusiveness reached by experimental data.
Much progress has been attained since the early years of the CGC in order to promote it to a precise and theoretically controlled phenomenological tool. Such progress has been fueled by the richer physic input provided by theoretical advancements and by the use of more sophisticated numerical tools. Thus, the semi-quantitative and relatively simplistic --although extremely insightful-- GBW model of DIS structure functions or the KLN model for hadron multiplicities have been replaced by much more detailed approaches like the AAMQS fits or the rcBK and IP-Glasma Monte Carlo codes. These new tools rely in the use of numerical solutions of the evolution equations at running coupling accuracy and of classical Yang Mills equations combined with the empiric knowledge gained at HERA and a more realistic description of the collision geometry and its fluctuations.

Despite the substantial progress achieved in the field and reported in this review, the ultimate goal of reaching a theoretically solid and complete phenomenological description of the initial stages of heavy ion collisions has not yet been completed. Among the most urgent open problems is the symmetrization of the B-JIMWLK equations to extend their validity --presently restricted to the study of dilute-dense scattering-- to more general, less asymmetric situations. Further, exact --numerical or analytical-- solutions of the B-JIMWLK equations and their precise dynamical content has been barely explored in phenomenological works due to the complexity of the required numerical work. This difficulty has so far been circumvented by the use of the BK equation. It urges to find more efficient numerical works or analytically controlled approximations for their solutions.

Another long-standing issue is the "impact parameter problem", or the present inability of the theory to simultaneously describe the energy evolution and the impact parameter dependence of nuclear wave functions. It is hard to envisage definitive theoretical progress in this direction, since the physics of confinement unavoidably plays an important role. Therefore, some degree of phenomenological modeling for the description of the impact parameter dependence seems unavoidable. Such modeling should be appropriately constrained by empiric information stemming from either the proton-lead run at the LHC or from the proposed facilities for electron-nucleus collisions LHeC and the EIC. Control of the transverse geometry of the collision process is essential for the success of the "HIC Cosmology" program, consisting in the experimental and theoretical study of the transport of initial density fluctuations to the final state of the collision with the goal of extracting the QGP transport parameters.  

Last, but not least, the CGC has no access to the large-$x$ degrees of freedom of QCD scattering, since they are integrated out in the construction of the effective theory. This limits the region of applicability of the CGC and rises the natural question of the precise $x$-value at which the CGC should be applicable. Since the physics is very unlikely to change drastically at some light-cone momentum scale, seeking for a smooth matching of the CGC with the standard leading-twist formalism of collinear factorization and DGLAP evolution is needed to avoid artificial "boundary effects" and, ultimately, to reach an unified and complete roadmap of QCD dynamics in the full kinematic plane. Relaxing the eikonal approximation on which the CGC through the calculation of NLO and kinematic corrections would represent an important first step in that direction.


\section*{Ackowledgements}
The research of Javier.~L. Albacete is funded by a Ram\'on y Cajal fellowship and the grant FPA2010-17915 from Ministerio de Econom\'{\i}a y Competitividad (MINECO).
The authors would like to thank Adrian Dumitru, Heikki M\"{a}ntysaari, Amir Rezaeian and Raju Venugopalan for useful discussions and for providing some of the plots presented in this manuscript.


\itemsep -2pt 
\ifx\mcitethebibliography\mciteundefinedmacro
\PackageError{JHEP-2modM.bst}{mciteplus.sty has not been loaded}
{This bibstyle requires the use of the mciteplus package.}\fi
\providecommand{\href}[2]{#2}
\begingroup\raggedright\endgroup

\end{document}